\documentclass[%
reprint,
superscriptaddress,
amsmath,amssymb,
aps,
]{revtex4-2}
\usepackage{lipsum} 
\usepackage{graphicx}
\usepackage{dcolumn}
\usepackage{bm}
\usepackage{amsmath}
\usepackage{mathrsfs}
\usepackage{appendix}
\usepackage{CJKutf8}  
\usepackage[T1]{fontenc}
\usepackage[hidelinks,breaklinks]{hyperref}

\usepackage[capitalize,nameinlink]{cleveref}
\usepackage{subfigure}

\DeclareMathOperator{\Tr}{Tr}

\def \iu {\mathrm{i}}
\def \d {\mathrm{d}}
\def \derivative#1#2{\frac{\d #1}{\d #2}}
\def \ket#1{|#1\rangle}
\def \bra#1{\langle#1|}
\def \ketbra#1#2{|#1\rangle\! \langle #2|}

\def \average#1{\langle #1\rangle }

\def \coh {\mathrm{coh}}
\def \qtm {\mathrm{qtm}}
\def \cls {\mathrm{cls}}
\def \cq {\mathrm{cq}}

\usepackage[normalem]{ulem} 
\usepackage{xcolor} 

\newcommand\mpwS[1]{
	{\let\helpcmd\sout\parhelp#1\par\relax\relax}
	{} }
\long\def\parhelp#1\par#2\relax{%
	\helpcmd{#1}\ifx\relax#2\else\par\parhelp#2\relax\fi%
}

\newcommand\ylhS[1]{[YH:{\let\helpcmd\sout\parhelp#1\par\relax\relax}] }
\long\def\parhelp#1\par#2\relax{%
	\helpcmd{#1}\ifx\relax#2\else\par\parhelp#2\relax\fi%
}

\newcommand{\edit}[1]{{#1}}
\begin{document}
	
	\begin{CJK*}{UTF8}{gbsn}
	\crefname{appsec}{appendix}{appendices}
	\preprint{APS/123-QED}
	
	\title{On the feasibility of detecting quantum delocalization effects on \edit{relativistic time dilation} in optical clocks
	}
	
	\author{Yanglin Hu (胡杨林)}
	\email{yanglin.hu@u.nus.edu}
	\affiliation{Institut f{\"u}r Theoretische Physik, ETH Zurich, 8093 Z{\"u}rich, Switzerland}
	\affiliation{Centre for Quantum Technologies, National University of Singapore, 117543 Singapore, Singapore}
	\author{Maximilian P. E. Lock}
	\email{maximilian.paul.lock@tuwien.ac.at}
	\affiliation{Atominstitut, Technische Universit{\"a}t Wien, 1020 Vienna, Austria}
	\affiliation{Institute for Quantum Optics and Quantum Information - IQOQI Vienna, Austrian Academy of Sciences, Boltzmanngasse 3, 1090 Vienna, Austria}
	\author{Mischa P. Woods}
	\email{mischa.woods@inria.fr}
 	\affiliation{University Grenoble Alpes, Inria, Grenoble, France}
	\affiliation{Institut f{\"u}r Theoretische Physik, ETH Zurich, 8093 Z{\"u}rich, Switzerland}

	\date{\today}
	
	\begin{abstract}
		We derive the predicted time dilation of delocalized atomic clocks in an optical lattice setup in the presence of a gravitational field to leading order in quantum relativistic corrections. We investigate exotic quantum states of motion whose \edit{relativistic} time dilation is outside of the realm of classical general relativity, finding a regime where $^{24}\mathrm{Mg}$ optical lattice clocks currently in development would comfortably be able to detect this quantum effect (if the technical challenge of generating such states can be met \edit{and the expected accuracy of such clocks can be attained}). We provide a detailed experimental protocol and analyse the effects of noise on our predictions. We also show that the magnitude of our predicted quantum \edit{relativistic} time dilation effect remains just out of detectable reach for the current generation of  $^{87}\mathrm{Sr}$ optical lattice clocks. Our calculations agree with the predicted time dilation of classical general relativity when restricting to Gaussian states. 
		
	\end{abstract}
	
	
	\maketitle
	
		\end{CJK*}
	\section{Introduction}
\edit{Time dilation is one of the key predictions of the special theory of relativity, being central to its experimental validation~\cite{gwinner2005experimental}. It was first observed in 1938 via its contribution to the Doppler shift in the spectrum of hydrogen atoms accelerated to speeds on the order of magnitude of $10^{-3}$c~\cite{ives1938experimental}. Einstein's principle of equivalence between acceleration and gravity implies that the latter also leads to time dilation, a prediction which was first observed on Earth in 1960, over a distance of $22.56$~m~\cite{Pound_1960}. Since then, the incredible accuracy of modern optical clocks has drastically decreased the scales over which time dilation is detectable. In 2010, time dilation was observed between optical clocks moving at relative velocities on the order of $10^{-8}$c, and over a height difference of $33$~cm~\cite{Chou_2010}. Recently, gravitational redshift was observed within a sample of atoms on the single-mm scale~\cite{Bothwell_2022}, leading to the exciting prospect that it may soon be detectable on the length scale of the wavefunction itself. In~\cite{Khandelwal_2020}, a prescription was given for calculating the time dilation in this regime, i.e. for clocks whose velocities and positions are subject to quantum indeterminacy, leading to the prediction of a novel quantum-interference effect. In this work, we study the feasibility of observing this effect in state-of-the-art atomic clocks.}
	
There has been much theoretical work on a quantum theory of spacetime in the last century (see e.g.~\cite{callender2001physics,oriti2009approaches}), but there has been comparatively little experimental progress to guide the way, with the notable exception of certain phenomenological results~\cite{addazi2022quantum}. Given the extreme regimes at which a full theory of quantum gravity is expected to be necessary, it may be fruitful to instead investigate low-energy effects combining relativity and quantum mechanics. There are numerous experiments and phenomena where both quantum and gravitational effects manifest together (see e.g.~\cite{wallace2022quantum} and references therein), which can be described by simply combining quantum theory and relativity (or even Newtonian gravity, e.g.~\cite{colella1975observation}). 
 
 Going a step further, a recent approach consists of quantizing the mass defect associated with the internal states of a moving body, resulting in a coupling between its internal and motional degrees of freedom~\cite{Zych_2017,Zych_2018}. This coupling leads to novel predictions such as a gravitational decoherence~\cite{Pikovski_2015} (not to be confused with the Di\'{o}si-Penrose model~\cite{diosi1987universal,penrose1996gravity}) and time-dilation-induced effects in atom interference experiments~\cite{Zych_2011,Zych_2018,Pikovski_2015}. Moreover, the model has been applied in the Page-Wootters formalism~\cite{Smith_2020}, and is predicted to affect spontaneous emission rates~\cite{Paczos_2022} and atomic spectra~\cite{Grochowski_2021}.
	
 The effect of a quantized mass defect in the time dilation experienced by atomic clocks has been considered in~\cite{Yudin_2018,Haustein_2019,Paige_2020,Martinez-Lahuerta_2022}. In~\cite{Khandelwal_2020}, it was shown that this quantization implies the existence of quantum interference effects in the average time dilation observed for certain states of motion (see also \cite{Smith_2020}). In that analysis, the clocks were freely-falling, in contrast with the spatial confinement of modern state-of-the-art clocks. 
 
 Here we examine the potential for the interference effect to be observed in optical lattice clocks, which are ideal due to both their ultra-high precision \cite{Bloom_2014,Koller_2017,Bothwell_2019}, and the degree of control of their spatial degrees of freedom \cite{Morinaga_1999,Mandel_2003,Forster_2009,Wang_2015,Heinz_2020}. We consider $^{24}\mathrm{Mg}$ optical lattice clocks (and $^{87}\mathrm{Sr}$ in~\cref{apd:sr_protocol}), providing upper bounds on decoherence rates required in order to observe the discrepancy with high confidence. Our numerical results show that the discrepancy could be detectable by a next-generation $^{24}\mathrm{Mg}$ optical lattice clock with a relative accuracy of $10^{-19}$, which are under development and will likely be in operation in the near future \cite{Kulosa_2015,Wu_2020,Fim_2021,Jha_2022} \edit{(assuming that such clocks attain the accuracy that they are expected to)}, opening the door for the quantized-mass-defect model to be subject to experimental scrutiny. \edit{We find that the quantum effect on time dilation is dominated by the motion of the clock in the trap, while the effect on the gravitational contribution to time dilation is negligibly small.}

 We begin by laying out the theoretical model in \cref{sec:Theory}, before describing our proposed experimental protocol in~\cref{sec:Experimental Protocol}, and then finally analyse the effect of noise in \cref{sec:NoiseTolerance}.
	
\section{Theoretical Model} \label{sec:Theory}
    \subsection{\edit{Relativistic quantum clocks}}
	In this section we lay out the theoretical model and its application to atomic clocks. It will set the stage for \cref{sec:Experimental Protocol} where we devise experimental protocols. 

	
	Consider a point-like classical particle with rest mass $m$, position ${\bf x}=(x,y,z)$, and momentum ${\bf p}=(p_x,p_y,p_z)$, subject to some (e.g. optical) potential $U(\bf{x})$. Following from the energy-momentum relation in a weak gravitational field and low-energy limit, the Hamiltonian function in the rest frame of a laboratory on the Earth's surface is given by (see \cref{apd:hamiltonian})
	\begin{align}\label{eqn:Hamiltonian_initial}
	    \begin{split}
	        H =& mc^2 + mgz +\frac{{\bf p}^2}{2m}+U({\bf x})+O(c^{-2}),
	    \end{split}
	\end{align}	
	where $z$ is the particle's height and $g$ is the Earth's surface gravity, and where $O(c^{-n})$ denotes terms proportional to $c^{-n}$, as well as higher orders. The particle's proper time has the line element $\d \tau = \sqrt{g_{\mu\nu}\d x^{\mu} \d x^{\nu}}$, given here by
	\begin{align}
		\d \tau = \left(1+\frac{gz(t)}{c^2}- \frac{({\bf p}^2)(t)}{2m^2c^2}\right)\d t +O(c^{-4}). 
	\end{align}
	After a duration $T$ in the laboratory frame, the particle experiences the proper time
	\begin{align}
		\tau = T + I_0, \label{eqn:proper_time_classical}
	\end{align}
	where
	\begin{align}
		I_0:=\frac{1}{mc^2}\int_0^T \!\d t_1\left(mgz(t_1)- \frac{({\bf p}^2)(t_1)}{2m}\right) . \label{eqn:I_0_alternative}
	\end{align}
    Here $z(t_1)$ and ${\bf p}(t_1)$ are governed by the Poisson bracket
	\begin{align} 
		\frac{\d {\bf p}}{\d t}= - \{H,{\bf p}\},  \quad
		\frac{\d {\bf x}}{\d t}= - \{H,{\bf x}\}. \label{eqn:classical_motion_x_p} 
	\end{align}
    Now let us consider the particle as a point-like clock, with an inner clock degree of freedom denoted by subscript  $\mathrm{c}$, in addition to the above kinematic degrees of freedom, denoted by subscript $\mathrm{k}$. We obtain the quantized Hamiltonian in three steps. First, denoting the internal energy associated with the clock degree of freedom by $H_\mathrm{c}$, we account for the associated mass-defect (i.e. mass-energy equivalence) by making the replacement $mc^2 \to mc^2 + H_\mathrm{c}$.
    Second, we expand the energy-momentum relation, keeping $O(c^{-2})$ and $O(c^{-4})$ terms but neglecting $O(c^{-6})$ terms in the Hamiltonian. Finally, we replace all observables by operators. More details can be found in \cref{apd:hamiltonian}. We thus obtain 
	\begin{align}\label{eqn:Hamiltonian_middle}
		\hat{H} = \hat{H}_\mathrm{k} + \hat{H}_\mathrm{c}+ \frac{\hat{H}_\mathrm{c}}{mc^2}\otimes \hat{V}_\mathrm{k} + \frac{\hat{H}_\mathrm{c}^2}{m^2c^4} \otimes \hat{W}_\mathrm{k},
	\end{align}
	where 
	\begin{align}
		\hat{H}_\mathrm{k} =& mc^2 + mg\hat{z} +\frac{\hat{{\bf p}}^2}{2m} +\hat{U}(\hat{{\bf x}}) +O(c^{-2}) , \label{eqn:Hamiltonian_H_k}\\
		\hat{V}_\mathrm{k} =& mg\hat{z} - \frac{\hat{{\bf p}}^2}{2m}+O(c^{-2}), \label{eqn:Hamiltonian_V_k}\\
		\hat{W}_\mathrm{k} =& \frac{\hat{{\bf p}}^2}{2m}. \label{eqn:Hamiltonian_W_k}
	\end{align}
    \edit{Note that the $O(c^{-2})$ terms in $\hat{H}_{\mathrm{k}}$ includes terms proportional to $\frac{gz}{c^2}$ and $\frac{{\bf p}^2}{mc^2}$, and that the $O(c^{-2})$ terms in $\hat{V}_{\mathrm{k}}$ and the $O(c^0)$ term in $\hat{W}_{\mathrm{k}}$ correspond to the $O(c^{-4})$ terms in \cref{eqn:Hamiltonian_middle}, specifically $\frac{g^2z^2H_\mathrm{c}}{c^4}$, $\frac{{\bf p}^4H_\mathrm{c}}{m^4c^4}$, $\frac{gz{\bf p}^2 H_\mathrm{c}}{mc^4}$, and $\frac{{\bf p}^2H_\mathrm{c}^2}{m^3c^4}$; see \cref{apd:hamiltonian} for details. For brevity, we do not write them explicitly here, and we will later show that these $O(c^{-2})$ terms in $\hat{H}_{\mathrm{k}}$ and $\hat{V}_{\mathrm{k}}$ do not contribute to the lowest order correction to the measured time; see \cref{eqn:I_1_perturbative_argument,eqn:I_2_perturbative_argument}.}
	
    Since the clock degree of freedom couples with the kinematic degrees of freedom, its time is affected by its motion and vice versa. In an optical lattice clock, the clock degree of freedom remains coherent much longer than the kinematic degrees of freedom, see e.g. \cite{Takamoto_2005,Bishof_2011}. We therefore only consider decoherence processes via the kinematic degrees of freedom, modelling the system by the Lindblad equation (see e.g. \cite{Preskill_2019}):
	\begin{align}\label{eqn:evolution}
		\frac{\d}{\d t}\rho = \mathscr{F}_\mathrm{c}(\rho) + \mathscr{F}_\mathrm{k}(\rho) + \mathscr{F}_\mathrm{int}(\rho),
	\end{align}
	where
	\begin{align}
		\mathscr{F}_\mathrm{c}(\rho) & = -\iu [\hat{H}_\mathrm{c},\rho], \\
		\mathscr{F}_\mathrm{k}(\rho) & = -\iu [\hat{H}_\mathrm{k},\rho]+\mathscr{L}_\mathrm{k}(\rho), \\
		\mathscr{F}_\mathrm{int}(\rho) & = - \iu [\frac{1}{mc^2}\hat{H}_\mathrm{c}\otimes \hat{V}_\mathrm{k}+\frac{1}{m^2c^4} \hat{H}_\mathrm{c}^2\otimes \hat{W}_\mathrm{k},\rho], \\
		\mathscr{L}_\mathrm{k} (\rho) & = \frac{1}{2}\sum_i 2\hat{L}_i\rho \hat{L}_i^\dagger-\hat{L}_i^\dagger \hat{L}_i\rho -\rho \hat{L}_i^\dagger \hat{L}_i. \label{eqn:L_terms}
	\end{align}
	Here $\{ L_i\}_i$ are the Lindblad operators describing the effect of environmental noise on the kinematic degree of freedom. Suppose that the initial state $\rho_0$ is a product state between clock degrees of freedom and kinematic degrees of freedom
	\begin{align}
		\rho_0 = \rho_{\mathrm{c},0}\otimes\rho_{\mathrm{k},0}. 
	\end{align}
	We regard the non-correlating parts $\mathscr{F}_c$ and $\mathscr{F}_k$ as the unperturbed evolution and the correlating part $\mathscr{F}_\mathrm{int}$ as the perturbative evolution. The Lindblad equation can be expanded in terms of $\rho=\sum_n\frac{1}{m^{n}c^{2n}}\rho^{(n)}$ where $\rho^{(n)}$ is order $O(c^{-2n})$. This leads to the coupled equations
	\begin{align}
		\frac{\d}{\d t}\rho^{(0)} - \mathscr{F}_\mathrm{c} (\rho^{(0)})-\mathscr{F}_\mathrm{k}(\rho^{(0)}) =\ & 0, \\
		\frac{\d}{\d t}\rho^{(1)}- \mathscr{F}_\mathrm{c}(\rho^{(1)})- \mathscr{F}_\mathrm{k}(\rho^{(1)}) =\ & -\iu [\hat{H}_\mathrm{c} \otimes \hat{V}_\mathrm{k}, \rho^{(0)}], \\
		\frac{\d}{\d t}\rho^{(2)}- \mathscr{F}_\mathrm{c}(\rho^{(2)})- \mathscr{F}_\mathrm{k}(\rho^{(2)}) =\ & -\iu [\hat{H}_\mathrm{c} \otimes \hat{V}_\mathrm{k}, \rho^{(1)}] \nonumber\\
		& -\iu [\hat{H}_\mathrm{c}^2 \otimes \hat{W}_\mathrm{k}, \rho^{(0)}]. 
	\end{align}
	The solution is derived in \cref{apd:perturbative_calculations}. We further introduce $I_1$ and $I_2$ which will allow us to quantify the expectation value and variance of time dilation respectively, when the clock runs for a duration $T$ in the lab frame
	\begin{align}
		I_1 & :=\frac{1}{mc^2}\int_0^T \!\d t_1 \Tr\!\left(\hat{V}_\mathrm{k}[t_1]\rho_{\mathrm{k},0}\right)\label{eqn:I_1_original}\\
		& \ =\frac{1}{mc^2}\int_0^T\!\d t_1 \left(mg\langle \hat{z}[t_1]\rangle-\frac{\langle(\hat{\bf p}^2)[t_1]\rangle}{2m}\right), \label{eqn:I_1_alternative}\\
		I_2 & := \frac{2}{m^2c^4}\int_0^T\!\d t_2\int_0^{t_2}\! \d t_1 \Tr\! \left(\left(\hat{V}_\mathrm{k}[t_2-t_1]\hat{V}_\mathrm{k}\right)\rho_{\mathrm{k}}[t_1]\right), \label{eqn:I_2_original}
	\end{align}	
	where we have reserved square brackets for the evolution of a density matrix $\rho$ and operator $O$ in the interaction picture by 
	\begin{equation}\label{eqn:quantum_motion_rho}
	    \rho[t]:= e^{t\mathscr{F}_\mathrm{k}}(\rho),\qquad   \hat{O}[t]:=e^{t \mathscr{F}^\dagger_\mathrm{k}}(\hat{O}),
	\end{equation}
	respectively. One can easily find the analogy between \cref{eqn:I_0_alternative,eqn:I_1_alternative} by replacing variables such as $z(t_1)$ and ${\bf p}^2(t_1)$ with expectation value of observables such as $\langle \hat{z}[t_1]\rangle $ and $\langle \hat{\bf p}^2[t_1]\rangle $. 

    \subsection{\edit{Relativistic} effects in atomic clocks}
	The clock transition in an atomic clock can be well-modelled by a two-level system whose ground (excited) state is denoted by $\ket{g}$ ($\ket{e}$ respectively) and we thus write the clock Hamiltonian as 
	\begin{align}
		\hat{H}_\mathrm{c} = \frac{1}{2}\hbar\omega_0 \hat{\sigma}_\mathrm{z}.
	\end{align}
	The clock-transition frequency $\omega_0$ of the atom is compared with the frequency $\omega$ of a very stable laser. Any detected difference in frequency is cancelled on the fly by adjusting the laser frequency accordingly. The end result is a high quality time signal from the laser. This comparison may be done via the Ramsey experiment \cite{Ramsey_1950}, for example, whereby the atom initially in its ground state $\ket{g}$ is prepared in the $\ket{g}+\ket{e}$ state by a $\frac{\pi}{2}$-pulse, then allowed to evolve freely for a duration of $T$, before another $\frac\pi2$-pulse is applied and then finally measured in the energy basis. In the absence of relativistic effects, this results in Ramsey fringes described by
	\begin{align}
		\Pr(\ket{e})=\frac{1}{2}\big( 1+p\cos \left((\omega-\omega_0)T\right) \big).\label{eqn:prob_plus}
	\end{align}
	where $\omega_0$ is the centre of the fringes and $p$ (which is usually a function of $\omega-\omega_0$) is their contrast. In order to minimize the variance in the estimate of $\omega_0$, the laser frequency is tuned to the two maximal gradient points either side of the fringe centre at $\omega \approx \omega_0$. (The other maximas of the cosine are reduced by a smaller $p$-value.) Under such optimal conditions, the variance of the estimation of $\omega_0$ is proportional to $T^{-2}p^{-2}$.
	Relativistic effects, however, lead to a shift of the fringe centre to $\tilde{\omega}_0$, and a decreased contrast $\tilde{p}$, and thus an increased variance $\tilde{\sigma}_0^2$.
	Moreover, as we show in \cref{apd:atomic_frequency_standard}, \cref{eqn:prob_plus} still holds but with the replacements 
	\begin{align}
		\omega_0 T & \mapsto \tilde{\omega}_0 T = \omega_0(T+ I_1), \label{eqn:proper_time_frequency_standard}\\
		p & \mapsto \tilde{p} = 1- \frac{\omega_0^2}{2}\left(\Re(I_2)-I_1^2\right),\label{eqn:contrast_frequency_standard}\\
		\sigma_0 & \mapsto \tilde{\sigma}_0 \propto T^{-2}\tilde{p}^{-2}. \label{eqn:VarianceProportionality}
	\end{align}
where we have assumed $p\approx 1$ for $\omega \approx \omega_0$.

In order to find the lowest order relativistic corrections to $\tilde{\omega}_0 T$ and $\tilde{p}$, we rewrite \cref{eqn:I_1_original,eqn:I_2_original}, using $\rho_{\mathrm{k}}^{(n)}$ and $V_{\mathrm{k}}^{(n)}$ to denote the $O(c^{-2n})$ terms in $\rho_{\mathrm{k}}$ and $V_{\mathrm{k}}$, arising from \cref{eqn:Hamiltonian_H_k,eqn:Hamiltonian_V_k} respectively:
	\begin{align}\label{eqn:I_1_perturbative_argument}
	    I_1=\frac{1}{mc^2}\int_0^T \!\d t_1 \Tr\!\left(\Big(\hat{V}_{\mathrm{k}}^{(0)}[t_1]+\frac{\hat{V}_{\mathrm{k}}^{(1)}[t_1]}{mc^2}+...\Big)\rho_{\mathrm{k},0}\right), 
	\end{align}
	and
	\begin{align}\label{eqn:I_2_perturbative_argument}
	    \begin{split}
	        I_2 =&\  \frac{2}{m^2c^4}\int_0^T\!\d t_2 \int_0^{t_2} \!\d t_1 \\ 
	        &\ \Tr\!\left(\Big(\hat{V}_{\mathrm{k}}^{(0)}[t_2-t_1]+\frac{\hat{V}_{\mathrm{k}}^{(1)}[t_2-t_1]}{mc^2}+...\Big)\right.\\
	        &\ \left.\Big(\hat{V}_{\mathrm{k}}^{(0)}+\frac{\hat{V}_{\mathrm{k}}^{(1)}}{mc^2}+...\Big)\Big(\rho_{\mathrm{k}}^{(0)}[t_1]+\frac{\rho_{\mathrm{k}}^{(1)}[t_1]}{mc^2}+...\Big)\right).
	    \end{split}
	\end{align}
	The lowest order of $I_1$ and of $I_2$ only contains $\rho_{\mathrm{k},0}$ and $\hat{V}_{\mathrm{k}}^{(0)}$. \cref{eqn:proper_time_frequency_standard} immediately shows that the lowest order of $\tilde{\omega}_0 T$ contains only $\rho_{\mathrm{k},0}$ and $\hat{V}_{\mathrm{k}}^{(0)}$. In \cref{eqn:contrast_frequency_standard}, the lowest order of $I_1^2$ does not cancel with that of $\Re(I_2)$ in general, as the former is the square of the expectation of an operator and the latter is the expectation of the square of the same operator when the evolution is unitary, as is shown in \cref{apd:perturbative_calculations}. Moreover, \cref{eqn:contrast_frequency_standard} shows that the lowest order of $\tilde{p}$ contains only $\rho_{\mathrm{k}}^{(0)}$ and $\hat{V}_{\mathrm{k}}^{(0)}$. In conclusion, it is sufficient to only take into account $O(c^{0})$ terms in \cref{eqn:Hamiltonian_H_k,eqn:Hamiltonian_V_k} respectively. 
	
	Note that the proper time of a classical clock can be recovered when the kinematic state is Gaussian. As is seen from \cref{eqn:I_0_alternative,eqn:I_1_alternative}, one can obtain \cref{eqn:proper_time_classical} from \cref{eqn:proper_time_frequency_standard} by replacing expectations of operators $\langle \hat{z}[t_1]\rangle$ and $\langle (\hat{\bf p}^2)[t_1]\rangle$ with the classical functions $z(t_1)$ and $({\bf p}^2)(t_1)$. For a pure kinematic state with Gaussian distributed amplitudes specified by initial mean position $\overline{z}(0)$, and mean momentum $\overline{p}(0)$ with variance $\sigma_p^2(0)$, this corresponds to replacing $\langle \hat{z}[t_1]\rangle$ by $\overline{z}(t_1)$ and $\langle (\hat{\bf p}^2)[t_1]\rangle$ by $\overline{p}(t_1)^2+ \sigma_p^2(t_1)$. 

    \subsection{Quantum effects on the \edit{relativistic} time dilation}
	To go beyond the classical relativistic time dilation and examine quantum effects, we will distinguish between two cases, corresponding to two different initial kinematic states: the case of preparing a spatial quantum superposition $\rho_{\mathrm{k},0,\qtm}$, and the case of a spatial classical mixture $\rho_{\mathrm{k},0,\cls}$:
	\begin{align}
		\rho_{\mathrm{k},0,\qtm} & =  \ketbra{\psi}{\psi},\quad \ket{\psi}=\sqrt{K}\left(\cos\theta \ket{\psi_1}+ e^{\iu\phi} \sin\theta \ket{\psi_2}\right)\label{eqn:superposition},
		\\
		\rho_{\mathrm{k},0,\cls} & = \cos^2 \theta \ketbra{\psi_1}{\psi_1} + \sin^2\theta \ketbra{\psi_2}{\psi_2}\label{eqn:mixture}, 
	\end{align}
	where $\ket{\psi_1}$, $\ket{\psi_2}$ are Gaussian states \cite{Schumaker_1986} with different initial mean position and momentum; $K$ is a normalisation constant. Our usage of states with Gaussian-distributed amplitudes is motivated by experimental conditions; we will show in \cref{sec:Experimental Protocol} how such states and their superposition or mixture can be prepared. The two cases will yield different time dilations. We use the notation $I_1\mapsto I_{1,\qtm}$, $I_2 \mapsto I_{2,\qtm}$, $\tilde{\omega}_{0}\mapsto \tilde{\omega}_{0,\qtm}$ and $\tilde{p}\mapsto\tilde{p}_{\qtm}$ when evaluating the case of a quantum superposition, and similarly $I_1\mapsto I_{1,\cls}$, $I_2\mapsto I_{2,\cls}$, $\tilde{\omega}_{0}\mapsto \tilde{\omega}_{0,\cls}$ and $\tilde{p}\mapsto\tilde{p}_{\cls}$ in the case of the classical mixture. The discrepancy between the relativistically shifted frequencies in the two cases is denoted
	\begin{align}
		\tilde{\omega}_{0,\coh} := \tilde{\omega}_{0,\qtm}- \tilde{\omega}_{0,\cls}. 
	\end{align}
	By preparing two atomic clocks in parallel with the two different initial states, one can investigate when this shift is experimentally detectable via \cref{eqn:prob_plus}. Let us first define three quantities 
	\begin{align}
	    \Delta_{1,\coh}& := I_{1,\qtm} - I_{1,\cls}, \label{eqn:delta_1_coh}\\
	    \Delta_{2,\qtm}^2 & := \Re(I_{2,\qtm}) -  I_{1,\qtm}^2, \label{eqn:delta_2_qtm}\\
	    \Delta_{2,\cls}^2 & := \Re(I_{2,\cls}) -  I_{1,\cls}^2. \label{eqn:delta_2_cls} 
	\end{align}
	The quantity $\Delta_{1,\coh}$ is related to the discrepancy in clock times. \edit{Note that because $\rho_{0,\mathrm{k},\qtm}$ and $\rho_{0,\mathrm{k},\cls}$ differ by a constant matrix, the coefficients of the linear contributions of relativistic effects in $I_{1,\qtm}$ and $I_{1,\cls}$ also differ by a constant. Thus their difference, $\Delta_{1,\coh}$, is linear in relativistic effects.} Meanwhile, $\Delta_{2,\qtm}^2$ is related to the contrast $\tilde{p}_{\qtm}$ and thus to the increase in variance of the transition frequency in the quantum superposition case. Similarly, $\Delta_{2,\cls}^2$ is related to $\tilde{p}_{\cls}$ and thus to the increase in variance in the classical mixture case. This correspondence holds not only for atomic clocks, but also for the idealized clocks discussed in \cref{apd:idealised_clock}. Let us now express the frequency discrepancy and the fringe contrasts in terms of these variables. The discrepancy is
    \begin{align}\label{eqn:discrepancy_atomic_clock}
        \tilde{\omega}_{0,\coh} = \frac{\Delta_{1,\coh}}{T}\omega_0, 
    \end{align}
    \edit{including a gravitational term (proportional to the ratio of differences in gravitational potential energy with $mc^{2}$) and a motion term (proportional to the ratio of kinetic energies with $mc^{2}$); see e.g.~\cref{eqn:I_1_alternative}. In the following, we do not explicitly neglect either term, though we will find that the motional term dominates in our setting. The fringe contrasts in~\cref{eqn:prob_plus} in the two cases are then} 
    \begin{align}
        \tilde{p}_{\qtm} & = 1-\frac{\omega_0^2}{2}\Delta_{2,\qtm}^2, \label{eqn:second_why}\\
        \tilde{p}_{\cls} & = 1-\frac{\omega_0^2}{2}\Delta_{2,\cls}^2. \label{eqn:third_why} 
    \end{align} 
    Using relation~\cref{eqn:VarianceProportionality}, the variances satisfy
    \begin{align}
        \tilde{\sigma}_{0,\qtm}^2 & \propto \frac{1}{T^2} +  \frac{\omega_0^2}{T^2} \Delta_{2,\qtm}^2, \\
        \tilde{\sigma}_{0,\cls}^2 & \propto \frac{1}{T^2} + \frac{\omega_0^2}{T^2} \Delta_{2,\cls}^2 .
    \end{align}
    Under the usual assumptions of the addition-of-quadratures rule for standard deviations, we have
    \begin{align}
        \tilde{\sigma}_{0,\coh}^2 = \tilde{\sigma}_{0,\qtm}^2 + \tilde{\sigma}_{0,\cls}^2,
    \end{align}
    thus obtaining 
    \begin{align}\label{eqn:variance_atomic_clock}
        \tilde{\sigma}_{0,\coh}^2 \propto \frac{2}{T^2}  + \frac{\omega_0^2}{T^2}\Delta_{2,\cq}^2,
    \end{align}
    where $\Delta_{2,\cq}^2 = \Delta_{2,\qtm}^2 + \Delta_{2,\cls}^2 $. 
    Because the proportionality factor depends on how the Ramsey interferometry is implemented, we will provide estimates for $\Delta_{2,\cq}^2$ rather than $\tilde\sigma_{0,\coh}^2$ directly. 

    The condition to detect the discrepancy can be expressed as
    \begin{align}\label{eqn:necessary_condition}
        \tilde{\omega}_{0,\coh} \geq \tilde{\sigma}_{0,\coh} ,
    \end{align}
    for atomic clocks. For idealized clocks, interested readers may refer to \cref{apd:idealised_clock} for further discussion.

	\section{Experimental Protocol}\label{sec:Experimental Protocol}
    \subsection{State preparation}
	
	We propose an experimental protocol based on an optical lattice clock, since they afford the highest accuracy, as well as permitting the high level of control required for the preparation of different kinematic states. Such a clock relies on an optical potential to confine the atoms, which depends on their electronic state $\ket{n}$ and position $\hat{\bf x}$, as well as the frequency $\omega$ and the polarization $p$ of the optical field. It is convenient to express the optical potential in the basis $\ket{n}$: 
	\begin{align}
	    \hat{U} = \sum_{n} U_{n}(\hat{\bf x})\ketbra{n}{n}. 
	\end{align}
	Suppose that an atom is in the \edit{electronic} state $\ket{n}$ and that the optical field has one frequency $\omega$ but two polarizations (left $\circlearrowleft$ and right $\circlearrowright$ circularly polarized light), then the optical potential that the atom feels is given by~\cite{Derevianko_2011,Ludlow_2015}, 
	\begin{align}
		U_{n}(\hat{\bf x}) = \frac{1}{4}\sum_{p\in\{\circlearrowleft,\circlearrowright\}} \alpha_{n}(\omega,p) |E_p(\hat{\bf x})|^2, \label{eqn:general_optical_pot}
	\end{align}
	where $\alpha_{n}(\omega,p)$ is the polarizability of the atom and is dependent on $n$, $\omega$ and $p$. In a one-dimensional optical lattice, the optical potential is a one-dimensional standing wave:
	\begin{align}\label{eqn:potential well before expnasion}
		U_n(\hat{\bf x}) = -U_{n,\max} \cos^2k\hat{z}. 
	\end{align}
	Atoms are usually confined in a deep optical lattice in order to reduce the recoil shift and the Doppler shift and increase the accuracy \cite{Hobson_2016}. Therefore, the potential can be expanded around a minimum point, giving the harmonic oscillator approximation,
	\begin{align}\label{eqn:pot_expanded}
		U_n(\hat{x})\approx \frac{1}{2}m \omega_{n,\mathrm{z}}^2\hat{z}^2 - U_{n,\mathrm{max}},
	\end{align}
	where $\omega_{n,\mathrm{z}}=\sqrt{\frac{2U_{n,\max}}{m}}k$. The optical lattice is also aligned parallel to the gravitational field, which suppresses the hopping between sites and improves the accuracy \cite{Lemonde_2005}. 
	
	We choose coherent states $\ket{\alpha}$ and $\ket{-\alpha}$ (which are Gaussian~\cite{Schumaker_1986}) for $\ket{\psi_1}$ and $\ket{\psi_2}$ in the spatial superposition and the spatial mixture, i.e.~\cref{eqn:superposition} and~\cref{eqn:mixture}. There are two reasons to choose coherent states here. One is that they are the most classical choice in the sense that they have a non-negative Wigner function and saturate the position-momentum uncertainty relation. More importantly, they can be prepared easily in the experiment. With the harmonic oscillator approximation, the ground state of the deep optical lattice is the vacuum state. Coherent states can be prepared by displacing the vacuum state, for example with a state-dependent optical lattice \cite{Bloch_2005}. \edit{An atom prepared in such a state will oscillate, resulting in time dilation due to both motion and gravitational redshift.} Note that since the optical potential depends on the electronic state, atoms in two different electronic states can be subject to different optical lattices. Suppose that there are two electronic states $\ket{g}$ and $\ket{e}$ for which the polarizability in \cref{eqn:general_optical_pot} satisfies
	\begin{align}
		\alpha_{\ket{g}}(\omega ,\sigma^+)& > \alpha_{\ket{e}}(\omega ,\sigma^+), \\
		\alpha_{\ket{g}}(\omega ,\sigma^-)& < \alpha_{\ket{e}}(\omega ,\sigma^-),
	\end{align}
	and consider two coinciding optical lattices induced by lasers with $\sigma^\pm$ polarizations respectively. An appropriate phase modulation of the $\sigma^{\pm}$ lasers can displace the optical lattices in two opposite directions. Atoms in $\ket{g}$ and $\ket{e}$ are mainly subject to optical lattices induced by $\sigma^{+}$ and $\sigma^{-}$-polarized lasers respectively, and thus move along with the corresponding optical lattice in two opposite directions. In such a state-dependent optical lattice, the spatial ground states induced by the $\sigma^{\pm}$-polarized lasers are the respective coherent states $\ket{\pm \alpha}$ of the original optical lattice. 

    To prepare the superposition of coherent states given in~\cref{eqn:superposition}, one can first prepare atoms in the spatial ground state and a superposition of $\ket{g}$ and $\ket{e}$, then state-dependently displace and replace the lattice. After that, the desired stated is achieved by measuring in the $\{\ket{g}+\ket{e}, \ket{g}-\ket{e}\}$ basis and post-selecting the $\ket{g}-\ket{e}$ outcome. To prepare the mixture, one can follow the same method, except measure in the $\{\ket{g},\ket{e}\}$ basis and do not post-select. 
	
	Another important technique is electron shelving \cite{Nagourney_1986}. Suppose that there are a stable energy eigenstate $\ket{g}$ and an unstable energy eigenstate $\ket{f}$, and that $\ket{f}$ spontaneously decays to $\ket{g}$. Let an atom be in $\ket{g}$. We induce the transition between $\ket{g}$ and $\ket{f}$ with a strong laser. The atom will jump up to $\ket{f}$ due to the laser but will soon jump down to $\ket{g}$ due to the spontaneous decay of $\ket{f}$ repeatedly.  In each cycle, the atom scatters one photon. Scattered photons lead to two results. One is the possibility to detect these photons, from which we infer that the atom is in $\ket{g}$. Another is the transfer of the kinematic momentum and energy from photons to the atom, from which the atom can gain enough energy to escape from the optical lattice. We will make use of the first to measure the state of the atom and the second to remove atoms in a given state from the optical lattice. Both applications can be found in \cite{Westergaard_2010}. 
	
    In practice, there is usually no direct transition between $\ket{g}$ and $\ket{e}$. In that case, a Raman transition via a third level, is required for the transition between $\ket{g}$ and $\ket{e}$. Interested readers may refer to \cite{Bateman_2010} for a general description for the Raman transition with one and multiple intermediate states. 

    \subsection{A protocol for detecting the quantum discrepancy}
    
	\begin{figure}[htbp!]
		\centering
		\includegraphics[scale=0.25]{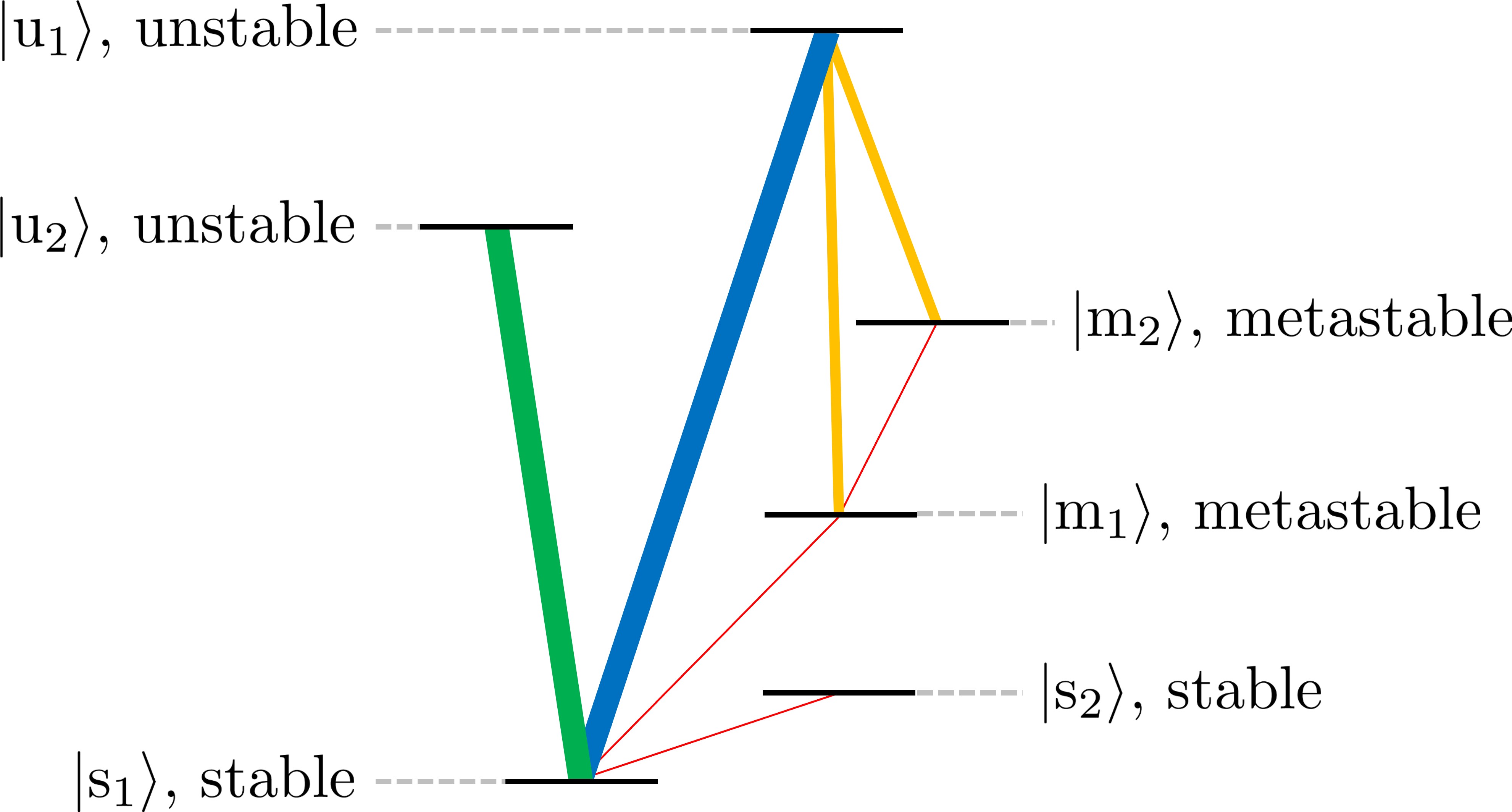}
		\caption{The energy diagram of the desired atom in our general protocol. The line width corresponds to the transition strength. The line colors indicate the relative energies of the transitions e.g. the blue line has a higher frequency than the green line.}
		\label{fig:energy_diagram_general}
	\end{figure}

    \begin{figure*}[htbp!]
	\centering
	\includegraphics[scale=0.3]{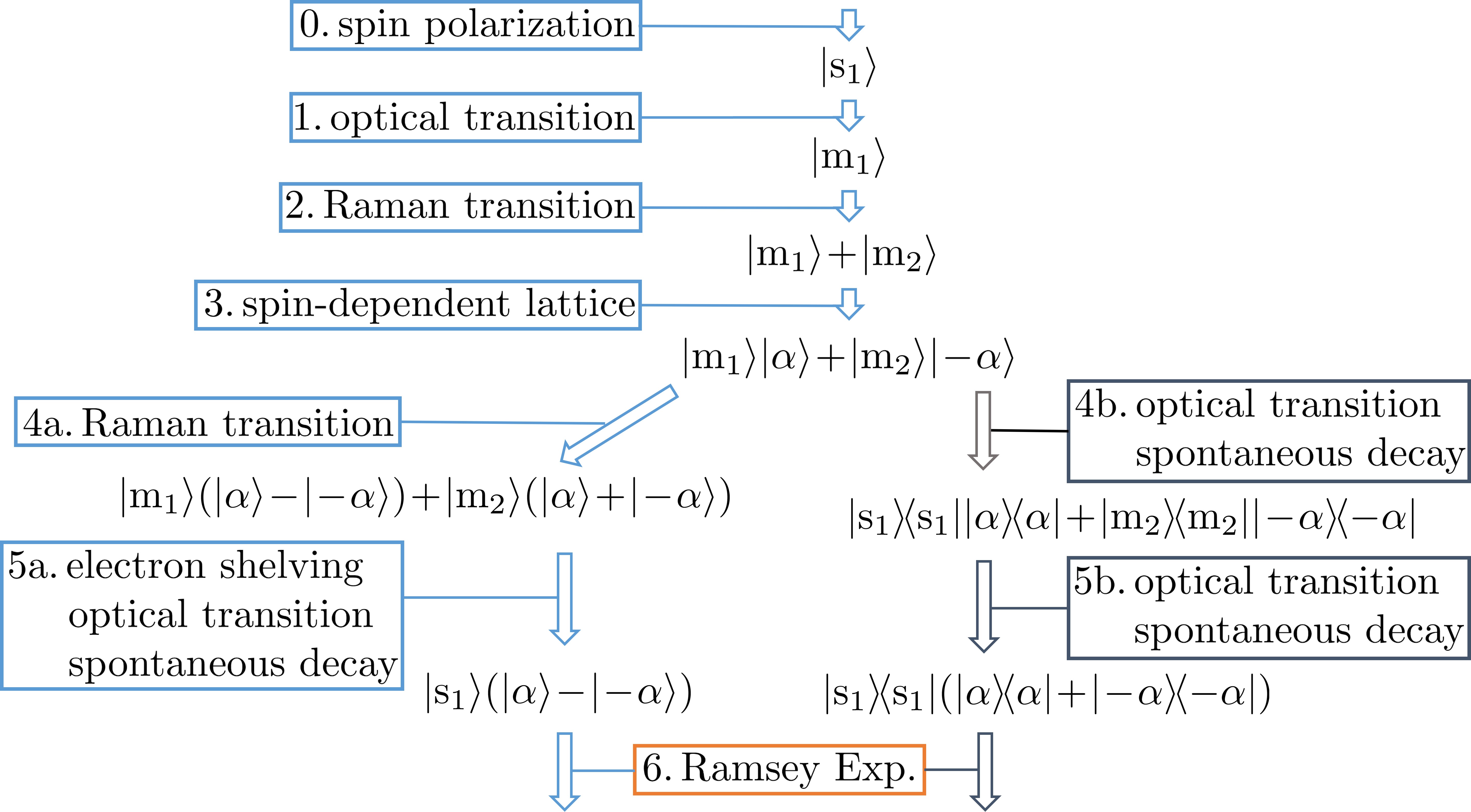}
	\caption{The scheme of our protocol. Steps 4a and 5a corresponds to a clock with a superposition as input and Step 4b and Step 5b to a clock with a mixture as input.}\label{fig:protocol}
    \end{figure*}
    
	We now describe our proposed experimental protocol in general, before applying it in the specific case of a $^{24}\mathrm{Mg}$ clock in the following section (and a $^{87}\mathrm{Sr}$ clock in \cref{apd:sr_protocol}). The atom is required to have an energy structure as in \cref{fig:energy_diagram_general}. Two stable states denoted by $\ket{\mathrm{s}_{1,2}}$ are used for the clock transition. Preferably, the transition between $\ket{\mathrm{s}_{1}}$ and $\ket{\mathrm{s}_2}$ is forbidden, which means that the natural linewidth of the transition is narrow. Two meta-stable states denoted by $\ket{\mathrm{m}_{1,2}}$ are used for the state-dependent optical lattice. We require that transitions between $\ket{\mathrm{m}_{1}}$ and $\ket{\mathrm{m}_2}$ and $\ket{\mathrm{s}_1}$ and $\ket{\mathrm{m}_1}$ are possible. Since all states involved in these two transitions are either stable or meta-stable, they are likely to be weak. In addition, two unstable states denoted by $\ket{\mathrm{u}_{1,2}}$ are used for measurements, post-selection and spontaneous decay. Strong transitions between $\ket{\mathrm{s}_1}$ and $\ket{\mathrm{u}_1}$ and $\ket{\mathrm{s}_1}$ and $\ket{\mathrm{u}_2}$ are needed. Moreover, weak transitions between $\ket{\mathrm{m}_1}$ and $\ket{\mathrm{u}_1}$ and $\ket{\mathrm{m}_2}$ and $\ket{\mathrm{u}_1}$ are also required.

	Our protocol is shown in \cref{fig:protocol}, and proceeds as follows:
    \edit{
    \begin{itemize}
        \item \emph{Step 0}, atoms are loaded into optical lattices induced by lasers at the magic wavelength with $\sigma^{\pm}$ and initialized to stable states $\ket{\mathrm{s}_{1}}$. 
        \item \emph{Step 1}, a $\pi$-pulse excites atoms from the stable state $\ket{\mathrm{s}_1}$ to the meta-stable state $\ket{\mathrm{m}_{1}}$. 
        \item \emph{Step 2}, a $\frac{\pi}{2}$-pulse induces the transition between meta-stable states $\ket{\mathrm{m}_{1}}$ and $\ket{\mathrm{m}_2}$ and prepares the superposition $\ket{\mathrm{m}_{1}}+\ket{\mathrm{m}_{2}}$. 
        \item \emph{Step 3}, we adiabatically displace optical lattices induced by $\sigma^\pm$ toward opposite directions to realize state-dependent optical lattices and diabatically replace them after that, yielding $\ket{\mathrm{m}_{1}}\ket{\alpha}+\ket{\mathrm{m}_{2}}\ket{-\alpha}$. 
    \end{itemize}
    }
    
	In the following, we split \emph{Step 4} and \emph{Step 5} into two cases: case \emph{a} for the kinematic superposition in \cref{eqn:superposition}, 
    \edit{
    \begin{itemize}
        \item \emph{Step 4a}, another $\frac{\pi}{2}$-pulse induces the transition between $\ket{\mathrm{m}_1}$ and $\ket{\mathrm{m}_2}$ and prepares $\ket{\mathrm{m}_1}(\ket{\alpha}-\ket{-\alpha})+\ket{\mathrm{m}_2}(\ket{\alpha}+\ket{-\alpha})$.
        \item \emph{Step 5a}, we transit atoms in $\ket{\mathrm{m}_1}$ to $\ket{\mathrm{s}_1}$ via $\ket{\mathrm{u}_1}$ and clear atoms in $\ket{\mathrm{m}_2}$ by electron shelving. To be concrete, we first apply a laser inducing the transition from $\ket{\mathrm{m}_2}$ to $\ket{\mathrm{u}_1}$ and a strong laser inducing the transition between $\ket{\mathrm{s}_1}$ and $\ket{\mathrm{u}_2}$ to realize electron shelving. That is, atoms in $\ket{\mathrm{m}_2}$ first transit to $\ket{\mathrm{u}_1}$, quickly decay to $\ket{\mathrm{s}_1}$ via spontaneous emission and then cycle between $\ket{\mathrm{s}_1}$ and $\ket{\mathrm{u}_2}$. During electron shelving, atoms originally in $\ket{\mathrm{m}_2}$ are cleared while atoms in $\ket{\mathrm{m}_1}$ remain unchanged. After that, a $\pi$-pulse then excites atoms in $\ket{\mathrm{m}_1}$ to $\ket{\mathrm{u}_1}$. These atoms quickly decay to $\ket{\mathrm{s}_1}$ via spontaneous emission and we obtain the spatial superposition $\ketbra{\mathrm{s}_1}{\mathrm{s}_1}\otimes(\ket{\alpha}-\ket{-\alpha})(\bra{\alpha}-\bra{-\alpha})$. 
    \end{itemize}
    and case \emph{b} for the mixture in \cref{eqn:mixture}, 
    \begin{itemize}
        \item \emph{Step 4b}, atoms in $\ket{\mathrm{m}_1}$ are excited to $\ket{\mathrm{u}_1}$ by a $\pi$-pulse and then spontaneously decay to $\ket{\mathrm{s}_1}$.
        \item \emph{Step 5b}, atoms in $\ket{\mathrm{m}_2}$ are excited to $\ket{\mathrm{u}_1}$ by another $\pi$-pulse and again spontaneously decay to $\ket{\mathrm{s}_1}$. These steps result in the spacial mixture $\ketbra{\mathrm{s}_1}{\mathrm{s}_1}\otimes(\ketbra{\alpha}{\alpha} +\ketbra{-\alpha}{-\alpha})$.
    \end{itemize}
    Both cases follow the same last step:
    \begin{itemize}
        \item \emph{Step 6}, we perform the ordinary interrogation and detection procedure of an optical lattice clock. A typical interrogation procedure consists of a Ramsey experiment to compare the frequency of the laser and the clock transition, in which atoms freely evolve for some duration $T$ in the laboratory frame. A typical detection procedure consists of a measurement in the basis of $\ket{\mathrm{s}_1}$ and $\ket{\mathrm{s}_2}$ with electron shelving.
    \end{itemize}
    }
    
    \subsection{Application to a $^{24}\mathrm{Mg}$ clock}
    Clock-specific protocols can be divided into two categories: optical lattice clocks based on bosons, e.g. $^{24}\mathrm{Mg}$ \cite{Wu_2020,Fim_2021,Jha_2022} and $^{88}\mathrm{Sr}$ \cite{Akatsuka_2010}, and on fermions, e.g. $^{87}\mathrm{Sr}$ \cite{Takamoto_2005} respectively. We consider $^{24}\mathrm{Mg}$ and $^{87}\mathrm{Sr}$ optical lattice clocks to illustrate typical optical lattice clocks with bosons and fermions respectively. Since $^{24}\mathrm{Mg}$ are currently in development~\cite{Kulosa_2015,Wu_2020,Fim_2021,Jha_2022} and yield the most promising results as far as detectability is concerned, we present them here, while $^{87}\mathrm{Sr}$ are relegated to~\cref{apd:sr_protocol}.

    \begin{figure}[htbp!]
		\centering
		\includegraphics[scale=0.25]{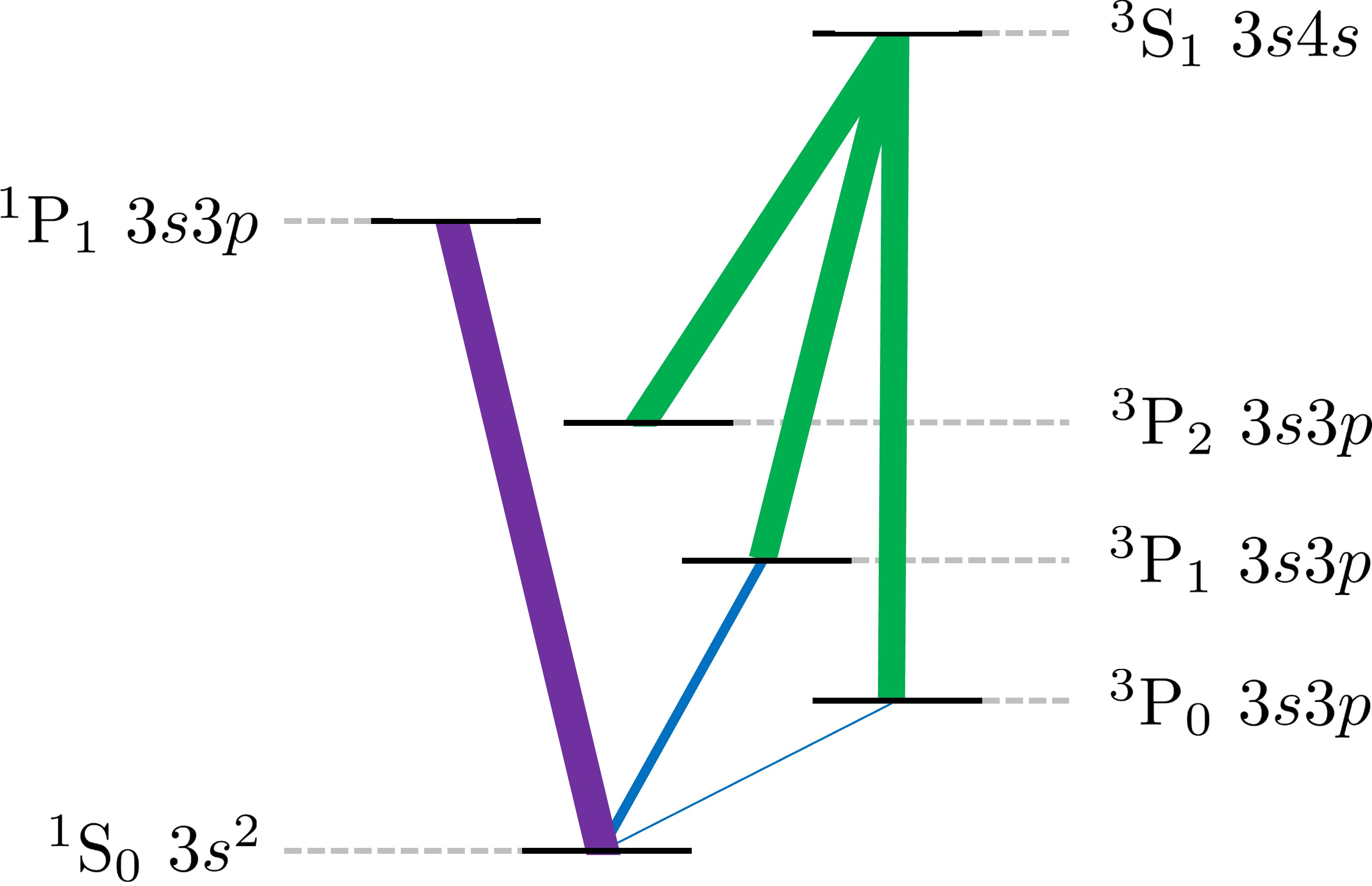}
		\caption{The energy diagram of the $^{24}\mathrm{Mg}$ atom. Each state is labeled by its electronic state $ ^{2S+1} L_{J} $. The line color corresponds the transition light color, and the width of the line corresponds to the strength of the transition. }\label{fig:simplified_energy_diagram_Mg}
	\end{figure}
    
    \cref{fig:simplified_energy_diagram_Mg} shows the relevant energy diagram of $^{24}\mathrm{Mg}$ which is bosonic and has zero nuclear spin (and therefore no hyperfine splitting). This results in a simple electronic structure but at the same time a difficulty in finding enough states for all operations because of the limited number of states. $^1\mathrm{S}_0$ is used as the lower stable state $\ket{\mathrm{u}_1}$. In \emph{Step 0} to \emph{Step 5}, we treat $^3\mathrm{P}_0$ and $^3\mathrm{P}_2$ as meta-stable states $\ket{\mathrm{m}_1}$ and $\ket{\mathrm{m}_2}$ respectively. In \emph{Step 4} and \emph{Step 5}, we treat $^3\mathrm{P}_1$ and $^1\mathrm{P}_1$ as unstable states $\ket{\mathrm{u}_1}$ and $\ket{\mathrm{u}_2}$. In \emph{Step 6}, $^3\mathrm{P}_0$ is used as the upper stable state $\ket{\mathrm{s}_2}$. The transition between $^1\mathrm{S}_0$ and $^3\mathrm{P}_2$ (the clock transition), though strictly speaking forbidden, is induced applying by a strong magnetic field \cite{Hobson_2016,Fim_2021,Jha_2022}. Transitions between $^3\mathrm{P}_2$ and $^3\mathrm{P}_0$, $^3\mathrm{P}_2$ and $^3\mathrm{P}_1$ as well as $^3\mathrm{P}_0$ and $^3\mathrm{P}_1$ are realized by Raman transitions via $^3\mathrm{S}_1$. Although the natural linewidth of $^3\mathrm{P}_1$ is rather narrow, an atom in $^3\mathrm{P}_1$ does spontaneously decay to $^1\mathrm{S}_0$ in a timescale much smaller than the lifetime of $^{3}\mathrm{P}_0$. Besides, due to the use of $^3\mathrm{P}_0$ as $\ket{\mathrm{m}_1}$ in \emph{Step 0} to \emph{Step 5} and $\ket{\mathrm{s}_2}$ in \emph{Step 6}, we must additionally clear atoms in $^3\mathrm{P}_0$ by electron shelving between \emph{Step 5} and \emph{Step 6}. This is possible by pumping atoms from $^3\mathrm{P}_0$ to $^3\mathrm{S}_1$ while re-pumping atoms from $^3\mathrm{P}_1$ and $^3\mathrm{P}_2$ to $^3\mathrm{S}_1$.  

    \subsection{Numerical analysis}\label{sec:FreeEvolution}
    We now evaluate and numerically simulate our results for the experimental protocols with the aim of obtaining predictions for the discrepancy and the increase in variance in \cref{eqn:delta_1_coh,eqn:delta_2_qtm,eqn:delta_2_cls}. For an optical atomic clock, the transition frequency between the ground state and the excited state has to be a constant with respect to different spatial positions (in the absence of time-dilation effects). Thus the clock works at the magic wavelength, whereby the energy shift due to the external electric fields of both energy levels are the same to leading order \cite{Takamoto_2005}. We thus replace $U_{n,\max}$ and $\omega_{n,\mathrm{z}}=\sqrt{\frac{2U_{n,\max}}{m}}k$ with $U_{\max}$ and $\omega_{\mathrm{z}}=\sqrt{\frac{2U_{\max}}{m}}k$ respectively. We write \cref{eqn:pot_expanded} in terms of creation and annihilation operators:
	\begin{align}
		\hat{a} &:= \frac{1}{\sqrt{2}z_\mathrm{s}} \Big(\hat{z}+ \frac{g}{\omega_\mathrm{z}^2}+\frac{\iu}{m\omega_\mathrm{z}}\hat{p}_\mathrm{z}\Big), \label{eqn:annihilation_op}\\
		\hat{a}^\dagger &:= \frac{1}{\sqrt{2}z_\mathrm{s}} \Big(\hat{z}+ \frac{g}{\omega_\mathrm{z}^2}-\frac{\iu}{m\omega_\mathrm{z}}\hat{p}_\mathrm{z}\Big),\label{eqn:creation_op}
	\end{align}
	where $z_\mathrm{s} := \sqrt{\frac{\hbar}{m\omega_\mathrm{z}}}$. We can then derive an expression for the total Hamiltonian in \cref{eqn:Hamiltonian_middle}, for the case of relativistic atoms in a one-dimensional optical lattice by substituting \cref{eqn:pot_expanded,eqn:creation_op,eqn:annihilation_op} into \cref{eqn:Hamiltonian_H_k,eqn:Hamiltonian_V_k,eqn:Hamiltonian_W_k}, thus obtaining
	\begin{align}
		\hat{H}_\mathrm{k} &= \hbar\omega_\mathrm{z} (\hat{a}^\dagger \hat{a} +\frac{1}{2}) -\frac{mg^2}{2\omega_\mathrm{z}^2}, \\
		\frac{\hat{V}_\mathrm{k}}{mc^2}&= C_\mathrm{g}(\hat{a}+\hat{a}^\dagger) -C_\mathrm{r}+ C_\mathrm{k}(\hat{a}^2+\left.{\hat{a}}^\dagger\right.^2-2\hat{a}^\dagger \hat{a} -1), \\
		\frac{\hat{W}_\mathrm{k}}{m^2c^4}&= -C_\mathrm{g}(\hat{a}^2+\left.\hat{a}^\dagger\right.^2-2\hat{a}^\dagger \hat{a} -1),
	\end{align}
	where $C_\mathrm{g} := \frac{gz_\mathrm{s}}{\sqrt{2}c^2}$, $C_\mathrm{r}:=\frac{g^2}{\omega_\mathrm{z}^2c^2}$ and \edit{$C_\mathrm{k}:=\frac{\hbar \omega_\mathrm{z}}{4mc^2}$}. 
	
	We first consider the noiseless case by setting the Lindblad operators $\{ L_i\}$ to zero in~\cref{eqn:L_terms}, and examine the effect of noise later, in~\cref{sec:NoiseTolerance}. The time evolution of polynomials of creation and annihilation operators in the noiseless case is derived in~\cref{apd:free_evolution}. Then, 
	\begin{align}
	    \begin{split}
		    \frac{\hat{V}_\mathrm{k}[t]}{mc^2} =&\ C_\mathrm{g}(\hat{a} e^{-i\omega_\mathrm{z} t} +\textnormal{h.c.}) -C_\mathrm{r}-C_\mathrm{k}\\
		    &+C_\mathrm{k}\left((\hat{a}^2 e^{-i2\omega_\mathrm{z} t} +\textnormal{h.c.})-2\hat{a}^\dagger \hat{a} \right). 
		\end{split}
	\end{align}
	With $\hat{V}_\mathrm{k}[t]$ in hand, we can compute $I_1$ and $I_2$ according to \cref{eqn:I_1_original,eqn:I_2_original} \edit{by setting 
    \begin{align}
        \rho_{{\rm k},0,\qtm} & = \frac{1}{1+C_{\rm i}} (\cos\theta \ket{\alpha} + e^{\iu\phi}\sin\theta\ket{-\alpha}) \nonumber\\
        &\quad \quad \quad \quad \cdot (\cos\theta \bra{\alpha} + e^{-\iu\phi}\sin\theta\bra{-\alpha}),\\
        \rho_{{\rm k},0,\cls} & = \cos^2\theta \ketbra{\alpha}{\alpha} +\sin^2\theta \ketbra{-\alpha}{-\alpha}. 
    \end{align}
    where $\alpha=\alpha_0e^{\iu \varphi}$ and $C_\mathrm{i} := e^{-2\alpha_0^2}\sin2\theta\cos\phi$.} We split $I_1$ into $I_{1,\rm l}$ which is an integration of non-oscillating terms and $I_{1,\rm o}$ which is an integration of oscillating terms (see~\cref{apd:integration_free}). The ratio between them is given by
	\begin{align}
		\frac{I_{1,\rm l}}{ I_{1,\rm o }} \propto \omega_\mathrm{z}T .
	\end{align}
	For our protocol, typical parameters are the frequency of the harmonic oscillator $\omega_\mathrm{z}=10^5$~Hz and the interrogation time of the Ramsey experiment $T=1$~s, and thus $\omega_\mathrm{z} T\approx 10^5$ \cite{Hobson_2016}. Therefore, we can only keep $I_{1,\rm l}$:
	\begin{align}\label{eqn:I_1_l_free_evolution}
		I_{1} \approx-\left(C_\mathrm{k}(2\average{\hat{a}^\dagger \hat{a} }+1) +C_\mathrm{r}\right)T, 
	\end{align}
	where $\average{\,\cdot\,}= \Tr(\,\cdot\, \rho_{\mathrm{k},0})$. Similarly, we split $I_2$ into $I_{2,\rm q}$ which is an double integration of non-oscillating terms and $I_{2,\rm o}$ which is a double integration of oscillating terms (see~\cref{apd:integration_free}). The ratio between them is 
	\begin{align}
		\frac{I_{2,\rm q}}{I_{2,\rm o}} \propto \omega_\mathrm{z} T. 
	\end{align}
	Again we can only keep $I_{2,\rm q}$ and obtain
	\begin{align}\label{eqn:I_2_1_free_evolution}
		\begin{split}
			I_{2} \approx &\ 4C_\mathrm{k}^2 \average{\left.\hat{a}^\dagger\right.^2\hat{a}^2 } T^2+\left(C_\mathrm{r} + C_\mathrm{k}\right)^2T^2  \\
			&+4C_\mathrm{k}\left(C_\mathrm{r} + 2C_\mathrm{k}\right) \average{\hat{a}^\dagger\hat{a} }T^2. 
		\end{split}
	\end{align}
	Plugging in initial states $\rho_{\mathrm{k},0,\qtm}$ in \cref{eqn:superposition} and $\rho_{\mathrm{k},0,\cls}$ in \cref{eqn:mixture}, one finds
	\begin{align}
		\begin{split}\label{eqn:I_1_qtm_free_evolution}
			I_{1,\qtm} =&\ -\left(C_\mathrm{r} + C_\mathrm{k} \right)T- 2C_\mathrm{k}\frac{1-C_\mathrm{i}}{1+C_\mathrm{i}}\alpha_0^2 T,
		\end{split}\\
		\begin{split}\label{eqn:I_2_qtm_free_evolution}
			I_{2,\qtm} =&\ \left(C_\mathrm{r} + C_\mathrm{k}\right)^2T^2 +4C_\mathrm{k}^2 \alpha_0^4T^2\\
			& + 4C_\mathrm{k}\left(C_\mathrm{r} + 2C_\mathrm{k}\right)\frac{1-C_\mathrm{i}}{1+C_\mathrm{i}} \alpha_0^2 T^2, 
		\end{split}
	\end{align} 
	and
	\begin{align}
		\begin{split}\label{eqn:I_1_cls_free_evolution}
			I_{1,\cls} =&\ -\left(C_\mathrm{r} + C_\mathrm{k} \right)T - 2C_\mathrm{k}\alpha_0^2T, 
		\end{split}\\
		\begin{split}\label{eqn:I_2_cls_free_evolution}
			I_{2,\cls} =&\ \left(C_\mathrm{r} + C_\mathrm{k}\right)^2T^2 +4C_\mathrm{k}^2 \alpha_0^4T^2\\
			&+ 4C_\mathrm{k}\left(C_\mathrm{r} + 2C_\mathrm{k}\right)  \alpha_0^2 T^2. 
		\end{split}
	\end{align}
	Recall the discrepancy $\Delta_{1,\coh}$ and the increase in variance in the quantum superposition case and the classical mixture case $\Delta_{\mathrm{2,qtm}}^2$ and $\Delta_{\mathrm{2,cls}}^2$ respectively defined in \cref{eqn:delta_1_coh,eqn:delta_2_qtm,eqn:delta_2_cls}. We obtain
	\begin{align}
		\Delta_{1,\coh} &=  4C_\mathrm{k}\frac{C_\mathrm{i}}{1+C_\mathrm{i}}\alpha_0^2 T, \label{eqn:delta_1_free}\\
		\Delta_{2,\qtm}^2 &= 16C_\mathrm{k}^2\frac{C_i}{(1+C_i)^2} \alpha_0^4T^2+ 4C_\mathrm{k}^2\frac{1-C_\mathrm{i}}{1+C_\mathrm{i}} \alpha_0^2 T^2, \label{eqn:delta_2_qtm_free}\\
		\Delta_{2,\cls}^2 &=  4C_\mathrm{k}^2 \alpha_0^2 T^2. \label{eqn:delta_2_cls_free}
	\end{align}

\edit{Here, we highlight that the discrepancy is dominated by the motional rather than the gravitational time dilation, as $\Delta_{1,\coh}$ is proportional to $C_{\rm k}$ coming from $\frac{\hat{{\bf p}}^2}{2m}$ in $V_{\rm k}$ rather than $C_{\rm g}$ coming from $mg\hat{z}$ in $V_{\rm k}$. This is because, as the position of the atom oscillates, the gravitational time dilation also oscillates, leading to a negligible contribution. Ignoring motion in the $xy$-plane, the average time-dilation effect is then determined by the quantum version of the $v^{2}/c^{2}$ term appearing in the classical Lorentz factor, i.e.
\begin{align} \label{eCLSvc}
    \left\langle\frac{\hat{p}_{z}^2(t)}{m^{2}c^{2}}\right\rangle_\mathrm{cls} = 4 C_\mathrm{k} \left(\alpha_0^2 \cos(2\omega_{\rm z} t - 2\varphi) + \alpha_0^2 + \frac{1}{2}\right),
\end{align}
for the classical mixture, and 
\begin{align} \label{eQTMvc}
\left\langle\frac{\hat{p}_{z}^2(t)}{m^{2}c^{2}}\right\rangle_\mathrm{qtm} = \left\langle\frac{\hat{p}_{z}^2(t)}{m^{2}c^{2}}\right\rangle_\mathrm{cls} - 8 \alpha_0^2 C_\mathrm{k} \frac{C_{\rm i}}{1+C_{\rm i}}
\end{align}
for the quantum superposition. This is the origin of the discrepancy accumulated over time, according to Eq.~\eqref{eqn:delta_1_free}, and its magnitude is determined by $\alpha_0^{2}C_\mathrm{k}\frac{C_{\rm i}}{1+C_{\rm i}}$. Recalling that $C_\mathrm{k}=\frac{\hbar \omega_\mathrm{z}}{4mc^2}$ and $C_{\rm i} = e^{-2\alpha_0^2}\sin2\theta\cos\phi$, we see that $\phi=\pi$ and $\theta=\frac{\pi}{4}$ are optimal choices for maximising the discrepancy. Furthermore, one must choose a value of $\alpha_0$ that is, of course, nonzero, but small enough that the exponential decrease of $C_{\rm i}$ (corresponding to the decrease in interference due to the decreasing overlap between components of the superposition) does not render the discrepancy negligible.}
    
\begin{figure}[htbp!]
		\centering
		\includegraphics[width=\columnwidth]{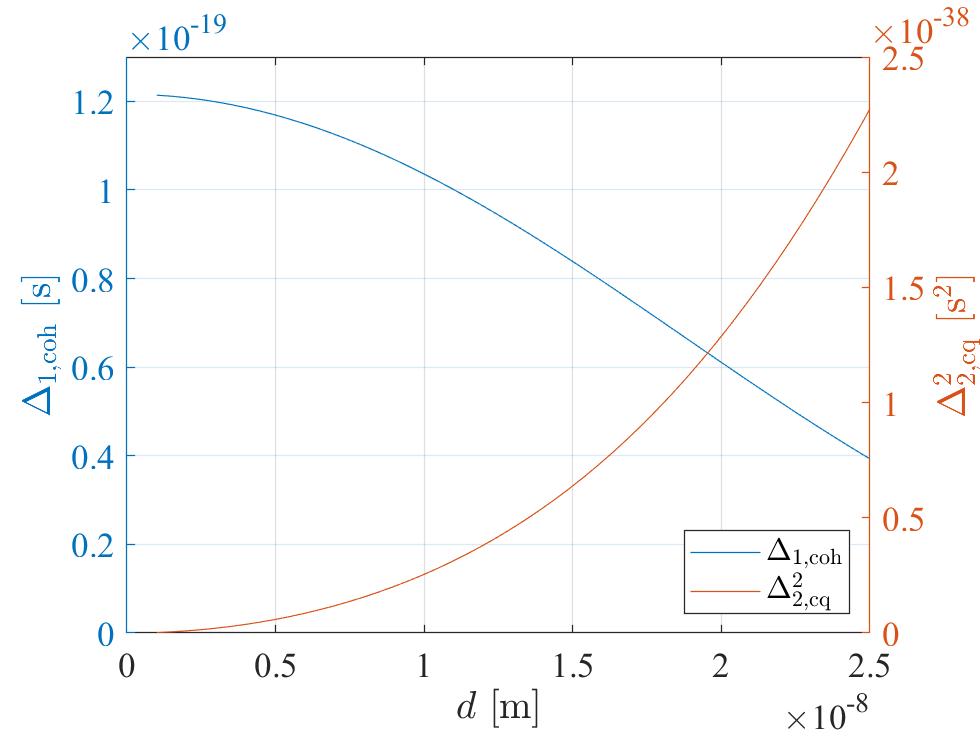}
		\caption{The discrepancy $\Delta_{1,\coh}$ between the quantum superposition case and the classical mixture case and the increase in variance $\Delta_{2,\qtm}^2+\Delta_{2,\cls}^2$ of this discrepancy (see \cref{eqn:discrepancy_atomic_clock,eqn:variance_atomic_clock}) versus the displacement $d$ respectively for a $^{24}\mathrm{Mg}$ clock. The parameters are $\lambda_{\mathrm{Mg}}=468$~nm, $U_{\max,\mathrm{Mg}}=300 E_{\mathrm{r},\mathrm{Mg}}$, $\phi=\pi$, \edit{$\theta=\frac{\pi}{4}$} and $T=1$~s. \edit{Note that $d= \sqrt{2}z_\mathrm{s} \alpha$, i.e. $d=10$~nm corresponds to $\alpha=0.395$ for the $^{24}{\rm Mg}$ clock here.}}
		\label{fig:quantum_modification_displacement_Mg}
	\end{figure}

    Recalling \cref{eqn:discrepancy_atomic_clock,eqn:variance_atomic_clock}, we now calculate $\Delta_{1,\coh}$ and $\Delta_{2,\cq}^2$ for both $^{24}\mathrm{Mg}$ and $^{87}\mathrm{Sr}$ optical lattice clocks. A $^{24}\mathrm{Mg}$ optical lattice clock works at the magic wavelength $\lambda_{\mathrm{Mg}}=468$~nm \cite{Kulosa_2015}. The trap depth is chosen as $U_{\max,\mathrm{Mg}} = 300 E_{\mathrm{r},\mathrm{Mg}}$, where the recoil energy of $^{24}\mathrm{Mg}$ is $E_{\mathrm{r},\mathrm{Mg}}=\frac{2\pi^2 \hbar^2 }{m_{\mathrm{Mg}}\lambda_{\mathrm{Mg}}^{2} }$. The interrogation time is set to $T=1$~s. The displacement $d$ is the half of the \edit{maximal} distance between centers of two coherent states $\ket{\pm\alpha}$ \edit{where $\alpha$ and $d$ are related by $d= \sqrt{2}z_\mathrm{s} \alpha$, and $d=10$~nm corresponds to $\alpha=0.395$ for the $^{24}{\rm Mg}$ clock here}.   \cref{fig:quantum_modification_displacement_Mg} illustrates $\Delta_{1,\coh}$ and $\Delta_{2,\cq}^2$ with respect to the displacement $d$ for the $^{24}\mathrm{Mg}$ optical lattice clock. The figure for the $^{87}\mathrm{Sr}$ optical lattice clock can be found in \cref{apd:sr_protocol}. 
	
	It can be seen from \cref{fig:quantum_modification_displacement_Mg} for $^{24}\mathrm{Mg}$ (and \cref{fig:quantum_modification_displacement_Sr} in~\cref{apd:sr_protocol} for $^{87}\mathrm{Sr}$) that the discrepancy is significant enough to be detected and that the increase in variance is still tolerable at a small displacement. This suggests that the discrepancy is in principle detectable. $^{87}\mathrm{Sr}$ clocks have a relative accuracy of $10^{-18}$ to $10^{-19}$ \cite{Bothwell_2019}, and the next-generation $^{24}\mathrm{Mg}$ clocks are expected to exceed $^{87}\mathrm{Sr}$ clocks in accuracy due to the avoidance of certain systematic effects such as the AC Stark shift by room-temperature blackbody radiation \cite{Kulosa_2015} or the vector and tensor lattice shifts \cite{Hobson_2016} as well as a shortened dead time between two subsequent measurements \cite{Hobson_2016}. The relative discrepancy for the $^{87}\mathrm{Sr}$ clock is $10^{-21}$ (see~\cref{apd:sr_protocol}), which seems not practical to detect with state-of-art $\mathrm{Sr}$ clocks. However, the relative discrepancy for the $^{24}\mathrm{Mg}$ clock is of $10^{-19}$, and is therefore in principle detectable in the next generation $^{24}\mathrm{Mg}$ clocks. \edit{The observed time dilation in this case corresponds to an average value of $\frac{v^{2}}{c^{2}}$ on the order of $10^{-19}$ in both the classical and quantum cases, according to Eqs.~\eqref{eQTMvc} and~\eqref{eCLSvc}. For comparison, the time dilation measured in~\cite{Chou_2010} corresponds to $\frac{v^2}{c^2}\sim 10^{-16}$.}
	
    \section{Noise tolerance} \label{sec:NoiseTolerance}
    The effect of noise can also be taken into account within our scheme. We will consider the effects of amplitude and phase damping, as well as diffusion~\cite{Gottesman_2001}. 
 
Amplitude damping can be modelled by the Lindblad operator
	\begin{align}
		\hat{L}_{\rm a} = \sqrt{\Gamma_\mathrm{a}} \hat{a}, 
	\end{align}
	where $\Gamma_\mathrm{a}$ is the decay rate of the amplitude damping channel. A detailed derivation for the time evolution of polynomials of the creation and annihilation operator can be found in \cref{apd:amplitude_damping_channel}. Then, 
	\begin{align}
		\hat{V}_\mathrm{k}[t] =&\ C_\mathrm{g}(\hat{a} e^{-i\omega_\mathrm{z} t} +\textnormal{h.c.})e^{-\frac{\Gamma_\mathrm{a}}{2} t} -C_\mathrm{r}-C_\mathrm{k}\nonumber\\
		&+C_\mathrm{k}\left((\hat{a}^2 e^{-i2\omega_\mathrm{z} t} +\textnormal{h.c.})-2\hat{a}^\dagger \hat{a} \right)e^{-\Gamma_\mathrm{a} t}. 
	\end{align}
	We again compute $I_1$. We consider the regime where $T\simeq \Gamma_\mathrm{a}^{-1}\,\edit{\gg}\, \omega_\mathrm{z}^{-1}$, i.e. where the decay rate is much slower than timescale associated with the trap frequency, and moreover, because we apply the measurement when the effect of decoherence becomes non-negligible but not overwhelming, we further assume $T \approx \Gamma_\mathrm{a}^{-1}$. In this case, $I_1$ includes a damped oscillating term and a constant term, which we collectively call $I_{1,\mathrm{o}}$,  and a linear term and a slowly decaying term, which we collectively call $I_{1,\mathrm{l}} $. We will split $I_1$ into $I_{1,\rm l}$ which is an integration of non-oscillating terms and $I_{1,\rm, o}$ which is an integration of oscillating terms. (see~\cref{apd:integration_amplitude}).  The ratio is given by
	\begin{align}
		\frac{I_{1,\mathrm{l}}}{ I_{1,\mathrm{o}} } \propto \omega_{\rm z} T. 
	\end{align}
	By our assumption, we only keep $I_{1,\rm l}$. Thus we get
	\begin{align}
		I_{1} =-\left(C_\mathrm{k} +C_\mathrm{r}\right)T-2C_\mathrm{k}\average{\hat{a}^\dagger \hat{a} }\frac{1-e^{-\Gamma_\mathrm{a} T}}{\Gamma_{\mathrm{a}}}. 
	\end{align}
	We can also compute $I_2$. Here $I_2$ is decomposed into two parts, $I_{2,\rm q}$ which is a double integration of non-oscillating terms and $I_{2,\rm o}$ which is a double integration of oscillating terms (see~\cref{apd:integration_amplitude}).  The ratio between them is 
	\begin{align}
            \frac{I_{2,\rm q}}{I_{2,\rm o}} \propto \omega_{\rm z} T.
	\end{align}
	We again only keep $I_{2,\rm q}$ and obtain
	\begin{align}
		\begin{split}
			I_{2}=&\ 4C_\mathrm{k}^2 \average{\left.\hat{a}^\dagger\right.^2 \hat{a}^2 }\frac{1-2e^{-\Gamma_\mathrm{a} T}+e^{-2\Gamma_\mathrm{a} T}}{\Gamma_\mathrm{a}^2} \\
			&\ + 4C_\mathrm{k}\left(C_\mathrm{r} + C_\mathrm{k}\right) \average{\hat{a}^\dagger\hat{a} }\frac{T(1-e^{-\Gamma_\mathrm{a}T})}{\Gamma_\mathrm{a}} \\
			&\ +8C_\mathrm{k}^2\average{\hat{a}^\dagger\hat{a} }\frac{1-e^{-\Gamma_\mathrm{a}T}-\Gamma_\mathrm{a}T e^{-\Gamma_\mathrm{a}T}}{\Gamma_\mathrm{a}^2}\\
			&\ +\left(C_\mathrm{r} + C_\mathrm{k}\right)^2T^2 . 
		\end{split}
	\end{align}
	Plugging in initial states $\rho_{\mathrm{k},0,\qtm}$ in \cref{eqn:superposition} and $\rho_{\mathrm{k},0,\cls}$ in \cref{eqn:mixture}, we obtain
	\begin{align}
		\begin{split}
			I_{1,\qtm} =&\ -\left(C_\mathrm{k} +C_\mathrm{r}\right)T\\
			&\ -2C_\mathrm{k} \frac{1-C_\mathrm{i}}{1+C_\mathrm{i}}\alpha_0^2 \frac{1-e^{-\Gamma_\mathrm{a} T}}{\Gamma_{\mathrm{a}}},
		\end{split}\\
		\begin{split}
			I_{2,\qtm} =&\ 4C_\mathrm{k}^2 \alpha_0^4\frac{1-2e^{-\Gamma_\mathrm{a} T}+e^{-2\Gamma_\mathrm{a} T}}{\Gamma_\mathrm{a}^2} \\
			&\ + 4C_\mathrm{k}\left(C_\mathrm{r} + C_\mathrm{k}\right) \frac{1-C_\mathrm{i}}{1+C_\mathrm{i}}\alpha_0^2\frac{T(1-e^{-\Gamma_\mathrm{a}T})}{\Gamma_\mathrm{a}} \\
			&\ + 8C_\mathrm{k}^2\frac{1-C_\mathrm{i}}{1+C_\mathrm{i}}\alpha_0^2\frac{1-e^{-\Gamma_\mathrm{a}T}-\Gamma_\mathrm{a}T e^{-\Gamma_\mathrm{a}T}}{\Gamma_\mathrm{a}^2}\\
			&\ +\left(C_\mathrm{r} + C_\mathrm{k}\right)^2T^2, 
		\end{split}
	\end{align}
	and
	\begin{align}
		\begin{split}
			I_{1,\cls} =&\ -\left(C_\mathrm{k} +C_\mathrm{r}\right)T-2C_\mathrm{k} \alpha_0^2 \frac{1-e^{-\Gamma_\mathrm{a} T}}{\Gamma_{\mathrm{a}}},   
		\end{split}\\
		\begin{split}
			I_{2,\cls} =&\ 4C_\mathrm{k}^2 \alpha_0^4\frac{1-2e^{-\Gamma_\mathrm{a} T}+e^{-2\Gamma_\mathrm{a} T}}{\Gamma_\mathrm{a}^2} \\
			&\ + 4C_\mathrm{k}\left(C_\mathrm{r} + C_\mathrm{k}\right) \alpha_0^2\frac{T(1-e^{-\Gamma_\mathrm{a}T})}{\Gamma_\mathrm{a}} \\
			&\ + 8C_\mathrm{k}^2\alpha_0^2\frac{1-e^{-\Gamma_\mathrm{a}T}-\Gamma_\mathrm{a}T e^{-\Gamma_\mathrm{a}T}}{\Gamma_\mathrm{a}^2}\\
			&\ +\left(C_\mathrm{r} + C_\mathrm{k}\right)^2T^2 . 
		\end{split}
	\end{align}
	Therefore
	\begin{align}
		\begin{split}
			\Delta_{1,\coh} =&\   4C_\mathrm{k}\frac{C_\mathrm{i}}{1+C_\mathrm{i}}\alpha_0^2 \frac{1-e^{-\Gamma_\mathrm{a} T}}{\Gamma_{\mathrm{a}}}, \label{eqn:delta_1_amplitude}
		\end{split}\\
		\begin{split}
			\Delta_{2,\qtm}^2= &\ 16C_\mathrm{k}^2\frac{C_i}{(1+C_i)^2} \alpha_0^4\frac{1-2e^{-\Gamma_\mathrm{a} T}+e^{-2\Gamma_\mathrm{a} T}}{\Gamma_\mathrm{a}^2}\\
			&\ + 8C_\mathrm{k}^2\frac{1-C_\mathrm{i}}{1+C_\mathrm{i}} \alpha_0^2 \frac{1-e^{-\Gamma_\mathrm{a}T}-\Gamma_\mathrm{a}T e^{-\Gamma_\mathrm{a}T}}{\Gamma_\mathrm{a}^2}, \label{eqn:delta_2_qtm_amplitude}
		\end{split} \\
		\begin{split}
			\Delta_{2,\cls}^2  =&\ 8C_\mathrm{k}^2 \alpha_0^2 \frac{1-e^{-\Gamma_\mathrm{a}t}-\Gamma_\mathrm{a}T e^{-\Gamma_\mathrm{a}T}}{\Gamma_\mathrm{a}^2}. \label{eqn:delta_2_cls_amplitude}
		\end{split}
	\end{align}
	From the above equations, we conclude that the amplitude damping channel will set an upper bound to the expectation value and the variance of the quantum modification. Our results show that the amplitude damping sets an upper bound of the discrepancy which is proportional to $\Gamma_\mathrm{a}^{-1}$. For a $^{24}\mathrm{Mg}$ optical lattice clock with the same parameters defined previously, one should ensure $\Gamma_\mathrm{a}\lesssim 1$~Hz in order for the discrepancy to reach $10^{-19}$.

	Phase damping can be modelled by the Lindblad operator
	\begin{align}
		\hat{L}_{\rm p} = \sqrt{\Gamma_\mathrm{p}} \hat{a}^\dagger \hat{a}.
	\end{align} 
	We again consider the regime where $ T \simeq \Gamma_\mathrm{p}^{-1} \gg \omega_\mathrm{z}^{-1}$. Applying a similar method to that of the amplitude damping channel, our derivation  (see~\cref{apd:integration_phase}) shows that in this regime,  it results in the same equations as \cref{eqn:delta_1_free,eqn:delta_2_qtm_free,eqn:delta_2_cls_free}. Therefore, the phase damping does not adversely affect the ability to detect the discrepancy.
	
    The diffusion is considered similarly by setting
	\begin{align}
		\hat{L}_{\rm d1} = \sqrt{\Gamma_\mathrm{d}} \hat{a}, \qquad
		\hat{L}_{\rm d2} = \sqrt{\Gamma_\mathrm{d}} \hat{a}^\dagger. 
	\end{align}
    As before it is assumed that $\omega_\mathrm{z}^{-1} \ll \Gamma_{\rm d}^{-1} \simeq T$. With exactly the same routine (see~\cref{apd:integration_diffusion}), we obtain
    \begin{align}
    \begin{split}\label{eqn:delta_1_diffusion}
        \Delta_{1,\mathrm{coh}}  =&  4C_{\rm k} \frac{C_{\rm i}}{1+C_{\rm i}}\alpha_0^2 T,
    \end{split}   \\
    \begin{split}\label{eqn:delta_2_qtm_diffusion}
        \Delta_{2,\mathrm{qtm}}^2= &16C_{\rm k}^2\frac{C_{\rm i}}{1+C_{\rm i}^2} \alpha_0^4 T^2 + \frac{8}{3} C_{\rm k}^2 \frac{1-C_{\rm i}}{1+C_{\rm i}} \alpha_0^2 \Gamma_{\mathrm{d}} T^3  \\
        & + 4C_{\rm k}^2 \frac{1-C_{\rm i}}{1+C_{\rm i}} \alpha_0^2 T^2 + \frac{4}{3}C_{\rm k}^2 \Gamma_{\mathrm{d}} T^3 + \frac{2}{3} C_{\rm k}^2 \Gamma_{\mathrm{d}}^2 T^4, 
    \end{split}\\
    \begin{split}\label{eqn:delta_2_cls_diffusion}
        \Delta_{2,\mathrm{cls}}^2  =&  4C_{\rm k}^2 \alpha_0^2 T^2 + \frac{8}{3}C_{\rm k}^2 \alpha_0^2\Gamma_{\mathrm{d}} T^3 \\
        & + \frac{4}{3} C_{\rm k}^2 \Gamma_{\mathrm{d}} T^3 + \frac{2}{3} C_{\rm k}^2 \Gamma_{\mathrm{d}}^2 T^4. 
    \end{split}
\end{align}
    
    Our results show that the diffusion increases the variance which is related to ${\rm poly}(\Gamma_{\rm d}T) \cdot T^2$.   For a $^{24}\mathrm{Mg}$ optical lattice clock with the same parameters defined previously, the ratio between the increase in variance and the original variance is $\sim 10^{-8}$ for $\Gamma_{\rm d}\sim 1$~Hz, which shows the accuracy of the clock does not decrease significantly.
	
\begin{figure}[htbp!]
		\centering
            
            \subfigure[Discrepancy]{
            \includegraphics[width=0.9\columnwidth]{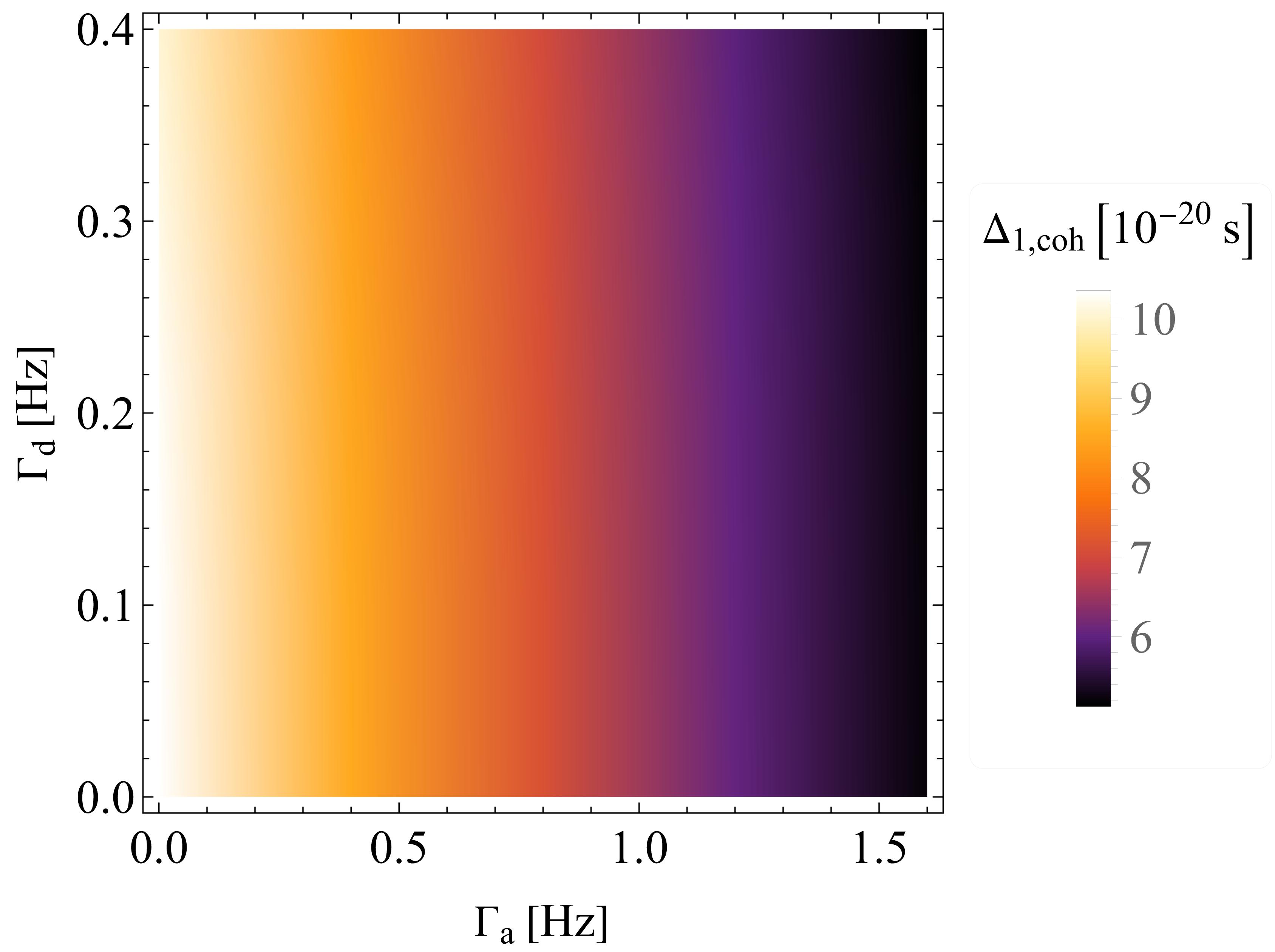}
            }
            \subfigure[Increase in variance]{
            \includegraphics[width=0.9\columnwidth]{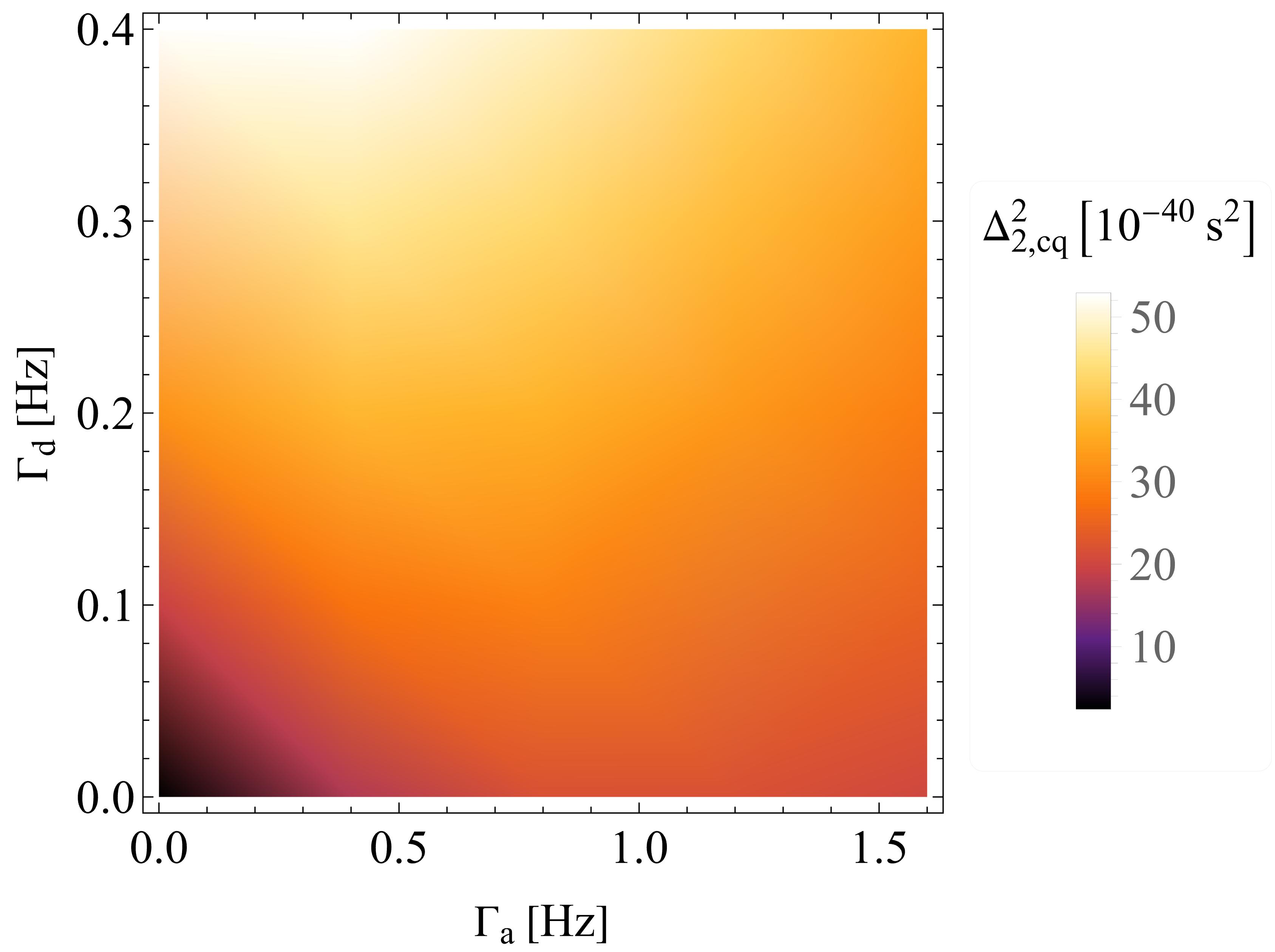}
            }
		\caption{ (a) The discrepancy $\Delta_{1,\coh}$ and (b) the increase in variance $\Delta_{2,\cq}^2$ (see \cref{eqn:discrepancy_atomic_clock,eqn:variance_atomic_clock}) versus decay rate and the diffusion rate $\Gamma_a$ and $\Gamma_d$ respectively for a $^{24}\mathrm{Mg}$ clock, where the parameters are chosen as $\lambda_{\mathrm{Mg}}=468$~nm, $U_{\max,\mathrm{Mg}} = 300 E_{\mathrm{r},\mathrm{Mg}}$, $T=1$~s, $\phi=\pi$, \edit{$\theta=\frac{\pi}{4}$} and $d=10$~nm \edit{($\alpha=0.395$)}. }
		\label{fig:discrepancy_variance_atomic_Mg}
	\end{figure}

	We also compute the discrepancy and the increase in variance with damping and diffusion for the $^{24}\mathrm{Mg}$ optical lattice clock. The parameters are again set to $\lambda_{\mathrm{Mg}}=468$~nm, $U_{\max,\mathrm{Mg}} = 300 E_{\mathrm{r},\mathrm{Mg}}$, $T=1$~s, $\phi=\pi$, \edit{$\theta=\frac{\pi}{4}$} and $d=10$~nm \edit{($\alpha = 0.395$)}. The computation was performed with Mathematica, where the library QULIB was used \cite{Landi_2021}. We do not include dephasing because it alters neither the discrepancy nor the variance. Results are plotted in~\cref{fig:discrepancy_variance_atomic_Mg}. 
	
	\section{Discussion}
	Our approach is founded on the widely-used quantized-mass-defect model \cite{Lammerzahl_1995,Pikovski_2015,Zych_2017,Zych_2018,Yudin_2018,Haustein_2019,Paige_2020,Smith_2020,Martinez-Lahuerta_2022}, which assumes a certain formulation of the Einstein equivalence principle, i.e. equivalence between the rest, inertial and gravitational internal energy, before quantizing the relevant quantities to operators. This particular combination of quantum theory and relativity has never been tested, and may not agree with the predictions of a ``correct'' theory of quantum gravity. Our experimental protocol would function as a test of the quantized-mass-defect model, with its prediction on time dilation of a quantum clock, allowing it to be falsified in the case that the experiment does not agree with the theory.

    Our approach probes this model for two different types of initial clock states: one for which no quantum effects manifest in the clock time (i.e. the classical-mixture states), and one for which they do (i.e. the quantum superposition case). In the former case, the quantized-mass-defect model gives rise to a purely classical relativistic time dilation effect, but in the latter the model results in an interplay of a purely relativistic effect with a purely quantum one, namely the interference of matter waves. Nonetheless, general relativity alone cannot realistically be expected to make a prediction for time dilation for the case of quantum states with highly non-classical features since it is outside the purview of the model; proper time (and thus time-dilation between observers) is a function of well-defined spacetime trajectories, which do not exist in quantum theory (see e.g.~\cite{lock2017relativistic,lock2019quantum}).

While we have shown that the predicted accuracy of the next generation of optical clocks should be able to detect the quantum effect in the time dilation \edit{(assuming that such clocks attain the accuracy that they are expected to) \cite{Kulosa_2015,Wu_2020,Fim_2021,Jha_2022}}, we have assumed that the technical challenge of generating the necessary exotic states of motion can be achieved without sacrificing this accuracy, for example due to imperfect process fidelity, or systematic errors induced by the delocalization of the clock. This is a nontrivial assumption, and represents a serious technical challenge. Nonetheless, the in-principle possibility of detecting this result tells us that we are very close to a regime where we can observe quantum effects in general relativistic time dilation, and may serve as a motivation to the resolution of this challenge.


\edit{
We have found that, despite the initial superposition is across different gravitational potentials, the quantum effect on the clock's gravitational time dilation is outweighed by the quantum effect on its time dilation due to motion. Relativistic time dilation due to an optical clock oscillating in a trapping potential has been observed in~\cite{Chou_2010}, and one can thus consider the effect as a quantum modification of that time dilation, but with one clock superposed in two places instead of two different clocks. To have the gravitational effect dominate instead of the kinematical effect, one might consider preventing the oscillation by having the two components of the superposition trapped in different potentials. However, this requires that the two components are in different electronic states, preventing them from interfering. The resulting time dilation would thus be indistinguishable from a mixture, with the quantum discrepancy relying on the fact that the two components interfere with each other. It remains to be seen whether the oscillatory motion can be prevented in some other manner which does not destroy the interference, and thus the quantum effect on the gravitational component of the time dilation.}

\section{Conclusion}
	
	We have studied the feasibility of detecting quantum interference effects in the time dilation experienced by an optical lattice clock. Our numerical analysis found that state-of-art or next-generation optical lattice clocks may be capable of detecting the discrepancy between the quantum superposition case and the classical mixture case, using either $^{87}\mathrm{Sr}$ or $^{24}\mathrm{Mg}$. Although the electronic structure of $^{87}\mathrm{Sr}$ is more convenient, our work demonstrates that state-of-the-art $^{87}\mathrm{Sr}$ clocks  have a lower accuracy and a smaller interference effect compared to next-generation $^{24}\mathrm{Mg}$ clocks, rendering such an experiment more difficult. Fortunately, in spite of its unfavorable electronic structure, next-generation $^{24}\mathrm{Mg}$ clocks are predicted by our work to have a higher accuracy and a larger discrepancy, and it is possible that the discrepancy can be detected. If such experiments are carried out, one can compare the experiment results and the predictions of current phenomenological quantum relativistic theories, providing a much-needed experimental signpost on the path to a theory which fully incorporates quantum mechanics \edit{into our understanding of spacetime}.
 	
	
	\begin{acknowledgments}
    Y.H. is funded by the National Research Foundation, Singapore and A*STAR under its CQT Bridging Grant.  
	M.P.E.L. acknowledges support from ERC-2021-COG 101043705 ``Cocoquest'' as well as the John Templeton Foundation (Grant 62423).
	M.P.W. was supported by an Ambizione fellowship from the Swiss National Science Foundation (grant No.~PZ00P2\_179914) in addition to the NCCR QSIT. The opinions expressed in this publication are those of the authors and do not necessarily reflect the views of the John Templeton Foundation.
	\end{acknowledgments}

	\bibliography{ClocksTimDim}
	
	\onecolumngrid
	\appendix 
	\section*{Appendices}
	
	\section{Hamiltonian in the Non-Relativistic Limit}\label{apd:hamiltonian}

	In this section, we derive our system Hamiltonian in the low velocity and weak field limit. Here we follows the routine in \cite{Zych_2017}. We restrict ourselves to a static spacetime with the metric denoted by $g_{\mu\nu}$ with signature $(+---)$. In that case, we have $g^{00}=(g_{00})^{-1}$, and recall that $g_{0i}=g_{i0}=0$ and $g_{ij}=g_{ji}$ for $i,j=1,2,3$. Consider that a point particle of mass \edit{$m_\mathrm{rest}c^2$} with a clock degree of freedom follows a world line $x^\mu(t)$ and has a four-momentum $p_\mu(t)$. In the rest frame of the particle, the metric is $g_{\mu\nu}'$ and the particle has a four-momentum $p_{\mu}'$. In that case, we have $g'\ ^{00}=1$. The scalar product of the four-momentum is coordinate-invariant
    \begin{align}
        p_\mu p^\mu = p'_\mu p'^\mu. 
    \end{align}
    Noting the rest energy $\edit{m_\mathrm{rest}c^2}=p'_0c$ and the energy $H=p_0c$, we obtain
    \begin{equation}\label{eqn:energy_momentum}
        H = \sqrt{-g_{00}\left(c^2p_jp^j-m_{\mathrm{rest}}^2c^4\right)}. 
    \end{equation}
    We now restrict ourselves to a weak field and low energy. We apply the post-Newtonian approximation. Consider the frame of an observer at rest at infinity. Let the coordinate be $\tilde{x}^{\mu}=(t,{\bf r})$, the metric is given by
    \begin{align}
        \tilde{g}_{00} & = 1+\frac{2\Phi}{c^2}+\frac{2\Phi^2}{c^4}, \\
        \tilde{g}_{ij} & = -\delta_{ij} \left(1-\frac{2\Phi}{c^2}\right). 
    \end{align}
    where $\Phi = -\frac{GM}{r}$ and $r=|{\bf r}|$. It can be transformed to the frame of an observer at rest at $r=r_0$ on the earth via
    \begin{align}
         g_{00} &= \left(1-\frac{\Phi_0}{c^2}\right)^2 \tilde{g}_{00}, \\
         g_{ij} &= \tilde{g}_{ij},
    \end{align}
    where $\Phi_0 = -gr_0$ is the gravitational potential on the earth and $g = \frac{GM}{r_0^2}$ is the gravitational acceleration. Now we expand the Hamiltonian near a point at $r=r_0$ on the earth and let the $z$-axis be parallel to the radius at this point. We further add the potential $U$ the clock is subject to into the Hamiltonian, which is an approach used in many works~\cite{Bloch_2005,Howl_2019,Haustein_2019,Martinez-Lahuerta_2022,Yudin_2018}. The resulting Hamiltonian is 
	\begin{align}\label{eqn:H}
	    H = m_{\mathrm{rest}}c^2 + m_{\mathrm{rest}}gz +\frac{{\bf p}^2}{2m_\mathrm{rest}}+U+ O(c^{-2}),
	\end{align}
	where $O(c^{-2})$ terms consist of $\frac{m_{\mathrm{rest}}g^2z^2}{c^2}$, $\frac{{\bf p}^4}{m_{\mathrm{rest}}^3c^2}$, $\frac{gz {\bf p}^2}{m_{\mathrm{rest}}c^2}$ and other higher order terms. As we show in \cref{eqn:I_1_perturbative_argument,eqn:I_2_perturbative_argument} and arguments thereafter, $O(c^{-2})$ terms do not contribute to the leading order of our results, and hence they are not written explicitly here. \edit{Now, assuming that the particle includes some internal structure, namely a clock degree of freedom with corresponding energy $H_{\mathrm{c}}$, we can decompose $m_{\mathrm{rest}}c^2$ to include the mass defect corresponding to this energy~\cite{Zych_2017}}, i.e.
	\begin{align}\label{eqn:H_rest}
	    \edit{m_{\mathrm{rest}}c^2} = mc^2\left(1 + \frac{H_{\mathrm{c}}}{mc^2}\right).
	\end{align}
	We also refer to terms proportional to $\frac{mg^2z^2}{c^2}$, $\frac{{\bf p}^4}{m^3c^2}$, $\frac{gz{\bf p}^2}{mc^2}$, $\frac{gzH_{\mathrm{c}}}{c^2}$ and $\frac{{\bf p}^2H_\mathrm{c}}{m^2c^2}$ as $O(c^{-2})$ terms, $\frac{g^2z^2H_\mathrm{c}}{c^4}$, $\frac{{\bf p}^4H_\mathrm{c}}{m^4c^4}$, $\frac{gz{\bf p}^2 H_\mathrm{c}}{mc^4}$, and $\frac{{\bf p}^2H_\mathrm{c}^2}{m^3c^4}$ as $O(c^{-4})$ terms, and other higher order terms as $O(c^{-6})$. We preserve $O(c^{-2})$ terms and $O(c^{-4})$ terms but ignore $O(c^{-6})$ terms while expanding \cref{eqn:H} with \cref{eqn:H_rest}. By re-arranging all the terms according to their common factors $\frac{H_\mathrm{c}}{mc^2}$ and $\frac{H_\mathrm{c}^2}{m^2c^4}$, one obtains 
    \begin{equation}
        H = mc^2+ H_{\mathrm{k}}+H_{\mathrm{c}}+\frac{H_{\mathrm{c}}}{mc^2} V_\mathrm{k}+\frac{H_\mathrm{c}^2}{m^2c^4} W_\mathrm{k}, \label{eqn:at this stage}
    \end{equation}
    where 
    \begin{align}
        H_{\mathrm{k}} &= mgz + \frac{{\bf p}^2}{2m} + U+ O(c^{-2}),\label{eqn:H_k}\\
        V_\mathrm{k} &= mgz-\frac{{\bf p}^2}{2m} + O(c^{-2}),\label{eqn:V_k}\\
        W_\mathrm{k} &= \frac{{\bf p}^2}{2m}.\label{eqn:W_k}
    \end{align}
 One should note that $O(c^{-2})$ terms in \cref{eqn:H} only result in $O(c^{-2})$ terms in \cref{eqn:H_k,eqn:V_k,eqn:W_k} after the expansion. Therefore, we are not neglecting $O(c^{-2})$ and $O(c^{-4})$ terms in~\cref{eqn:at this stage}. Instead, we keep all these terms compactly in $H_\mathrm{k}$, $V_\mathrm{k}$ and $W_\mathrm{k}$ in our calculation. Moreover, arguments in \cref{eqn:I_1_perturbative_argument,eqn:I_2_perturbative_argument} show that it is sufficient to only consider the explicitly written terms in~\cref{eqn:at this stage}. To quantize the framework, we replace observables with operators and obtain the quantized Hamiltonian
    \begin{equation}\label{eqn:hamiltonian}
        \hat{H} = \hat{H}_{\mathrm{k}}+\hat{H}_{\mathrm{c}}+\frac{\hat{H}_{\mathrm{c}}}{mc^2}\otimes \hat{V}_\mathrm{k}+\frac{\hat{H}_\mathrm{c}^2}{m^2c^4}\otimes \hat{W}_\mathrm{k}, 
    \end{equation}
    where $\hat{H}_\mathrm{k}$, $\hat{V}_\mathrm{k}$ and $\hat{W}_\mathrm{k}$ corresponds to those in \cref{eqn:H_k,eqn:V_k,eqn:W_k} with observables replaced by operators.

	\section{Perturbative Calculations}\label{apd:perturbative_calculations} 
	
	In this section, we are using the notation of \cref{eqn:quantum_motion_rho},
	i.e. square brakets are reserved for time evolution of density matrices $\rho[t]$ in the Schrodinger's picture and that of operators $\hat{O}[t]$ in the Heisenberg's picture. Consider an inhomogeneous linear differential equation of a density matrix $X[t]$
	\begin{align}\label{eqn:ini condition X}
		\derivative{}{t}X[t]-\mathcal{F}_{\mathrm{c}}\left(X[t]\right)-\mathcal{F}_{\mathrm{k}}\left(X[t]\right)= f(t), 
	\end{align}
	with the initial condition
	\begin{align}
		X[0] = 0. 
	\end{align}
	The above inhomogeneous linear differential equation can be solved with Green's function method, see e.g. \cite{Riley_2006}. We follow the same routine but slightly modify the formula. We first compute the function $G(t,t')$ which satisfies the homogeneous linear differential equation
	\begin{align}
		\derivative{G}{t}(t,t') - \mathcal{F}_{\mathrm{c}}\left(G(t,t')\right)-\mathcal{F}_{\mathrm{k}}\left(G(t,t')\right)= 0,
	\end{align}
	with the initial condition
	\begin{align}
		G(t',t') = f(t').
	\end{align}
	Then the general solution of the original equation is given by
	\begin{align}\label{eqn:eqn:X solved}
		X[t] = \int_0^t G(t,t') \d t' .
	\end{align}
	Now we rewrite \cref{eqn:evolution} into 
	\begin{align}
	    \derivative{\rho}{t} - \mathcal{F}_{\mathrm{c}}(\rho)-\mathcal{F}_{\mathrm{k}}(\rho)=- \frac{\iu}{mc^2}[\hat{H}_{\mathrm{c}}\otimes \hat{V}_{\mathrm{k}},\rho] -\frac{\iu}{m^2c^4} [\hat{H}_{\mathrm{c}}^2\otimes \edit{\hat{W}_{\mathrm{k}}},\rho]. 
	\end{align}
	Then we expand $\rho$ into $\sum_n \frac{1}{m^{n} c^{2n}}\rho^{(n)}$. The zeroth order density matrix satisfies the homogeneous equation
	\begin{align}
		\derivative{\rho^{(0)}}{t} - \mathcal{F}_{\mathrm{c}}(\rho^{(0)})-\mathcal{F}_{\mathrm{k}}(\rho^{(0)})=0, 
	\end{align}
	with the initial condition
	\begin{align}
		\rho^{(0)}[0]=\rho_{\mathrm{c},0}\otimes \rho_{\mathrm{k},0}. 
	\end{align}
	The solution is 
	\begin{align}
		\rho^{(0)}[t]=e^{\mathcal{F}_{\mathrm{c}} t} (\rho_{\mathrm{c},0})\otimes e^{\mathcal{F}_{\mathrm{k}} t} (\rho_{\mathrm{k},0}). 
	\end{align}
	The first and second order density matrix satisfies 
	\begin{align}
		\derivative{\rho^{(1)}}{t}  - \mathcal{F}_{\mathrm{c}}(\rho^{(1)})-\mathcal{F}_{\mathrm{k}}(\rho^{(1)})=&-\iu[\hat{H}_{\mathrm{c}}\otimes \hat{V}_\mathrm{k}, \rho^{(0)}], 
	\end{align}
	and
	\begin{align}
		\derivative{\rho^{(2)}}{t} - \mathcal{F}_{\mathrm{c}}(\rho^{(2)})-\mathcal{F}_{\mathrm{k}}(\rho^{(2)}) =-\iu[\hat{H}_{\mathrm{c}}\otimes \hat{V}_\mathrm{k}, \rho^{(1)}]-\iu[\hat{H}_{\mathrm{c}}^2\otimes \hat{W}_\mathrm{k}, \rho^{(0)}],
	\end{align}
	with the initial condition
	\begin{align}
		\rho^{(1)}[0]=\rho^{(2)}[0]=0. 
	\end{align}
	The solution is 
	\begin{align}
		\rho^{(1)}[t] =& -\iu \hat{H}_{\mathrm{c}}\rho_\mathrm{c}^{(0)}[t]\otimes \int_0^t \d t_1  e^{\mathcal{F}_\mathrm{k} (t-t_1)} \left(\hat{V}_\mathrm{k}  \rho_\mathrm{k}^{(0)}[t_1] \right) + \textnormal{h.c.},
	\end{align}
	and
	\begin{align}
		\rho^{(2)}[t] = &\ \hat{H}_{\mathrm{c}}\rho_\mathrm{c}^{(0)}[t]\hat{H}_{\mathrm{c}}\otimes \int_0^t  \d t_2 \int_0^{t_2} \d t_1 e^{\mathcal{F}_\mathrm{k} (t-t_2)}\left(\hat{V}_\mathrm{k} e^{\mathcal{F}_\mathrm{k} (t_2-t_1)} \left( \rho_\mathrm{k}^{(0)}[t_1]\hat{V}_\mathrm{k}\right)\right)+\textnormal{h.c}.\nonumber\\
		&\ -\hat{H}_{\mathrm{c}}^2\rho_\mathrm{c}^{(0)}[t]\otimes \int_0^{t} \d t_2 \int_0^{t_2} \d t_1 e^{\mathcal{F}_\mathrm{k} (t-t_2)}\left(\hat{V}_\mathrm{k} e^{\mathcal{F}_\mathrm{k} (t_2-t_1)}\left(\hat{V}_\mathrm{k} \rho_\mathrm{k}^{(0)}[t_1]\right)\right)+\textnormal{h.c.} \nonumber\\
		&\ -\iu \hat{H}_{\mathrm{c}}^2\rho_\mathrm{c}^{(0)}[t]\otimes \int_0^t \d t_1  e^{\mathcal{F}_\mathrm{k} (t-t_1)} \left(\hat{W}_\mathrm{k}  \rho_\mathrm{k}^{(0)}[t_1] \right) +\textnormal{h.c.}, 
	\end{align}
	where we have used that the evolution of the clock state is unitary. By taking the partial trace, we obtain the evolution of the reduced density matrix of the clock degree of freedom
	\begin{align}\label{eqn:rho_first_order}
		\frac{1}{mc^2}\rho_\mathrm{c}^{(1)}[t] = -\iu  \hat{H}_{\mathrm{c}}\rho_\mathrm{c}^{(0)}[t] I_1 +\textnormal{h.c.}, 
	\end{align}
	and
	\begin{align}\label{eqn:rho_second_order}
		\frac{1}{m^2c^4}\rho_\mathrm{c}^{(2)}[t] = &\ \frac{1}{2}\left(\hat{H}_{\mathrm{c}}\rho_\mathrm{c}^{(0)}[t]\hat{H}_{\mathrm{c}}
		-\hat{H}_{\mathrm{c}}^2\rho_\mathrm{c}^{(0)}[t]\right)I_2-\iu \hat{H}_{\mathrm{c}}^2\rho_\mathrm{c}^{(0)}[t] I_2'+ \textnormal{h.c.},
	\end{align}
	where
	\begin{align}
		I_1& =  \frac{1}{mc^2}\int_0^t \d t_1 \Tr_\mathrm{k}\left(\hat{V}_\mathrm{k} \rho_\mathrm{k}^{(0)}[t_1]\right), \label{eqn:I_1_apd}\\
		I_2& =  \frac{2}{m^2c^4}\int_0^{t} \d t_2 \int_0^{t_2} \d t_1\Tr_\mathrm{k}\left(\hat{V}_\mathrm{k} e^{\mathcal{F}_\mathrm{k} (t_2-t_1)} \left(\hat{V}_\mathrm{k} \rho_\mathrm{k}^{(0)}[t_1]\right)\right), \label{eqn:I_2_apd} \\
		I_2'& = \frac{1}{m^2c^4} \int_0^t \d t_1 \Tr_\mathrm{k}\left(\hat{W}_\mathrm{k} \rho_\mathrm{k}^{(0)}[t_1]\right) \label{eqn:I_2_prime_apd}. 
	\end{align}
	One should note that $I_1$ and thus $\frac{1}{mc^2}\rho_{\mathrm{c}}^{(1)}$ are of order $\frac{1}{mc^2}$. Similarly, $I_2$ and thus $\frac{1}{m^2c^4}\rho_{\mathrm{c}}^{(2)}$ are of order $\frac{1}{m^2c^4}$. Another useful equality is
	\begin{align}
		\Tr_\mathrm{k}\!\left(\hat{V}_\mathrm{k} e^{\mathcal{F}_\mathrm{k} (t_2-t_1)} \left(\hat{V}_\mathrm{k} \rho_\mathrm{k}^{(0)}[t_1]\right)\right)^* 	= \Tr_\mathrm{k}\!\left(\hat{V}_\mathrm{k} e^{\mathcal{F}_\mathrm{k} (t_2-t_1)} \left( \rho_\mathrm{k}^{(0)}[t_1]\hat{V}_\mathrm{k}\right)\right). 
	\end{align}
	When the time evolution is unitary, we have 
	\begin{align}
	    I_1 &=\frac{1}{mc^2} \Tr_{\mathrm{k}}\!\left(\left(\int_{0}^{t} \d t_1 \hat{V}_{\mathrm{k}}[t_1]\right) \rho_{\mathrm{k},0}\right), \\
	    \Re(I_2) &= \frac{1}{m^2c^4} \Tr_{\mathrm{k}}\!\left(\left(\int_0^t \d t_1 \hat{V}_{\mathrm{k}}[t_1]\right)^2 \rho_{\mathrm{k},0}\right), 
	\end{align}
	with which we conclude $\Re(I_2)$ does not cancel with $I_1^2$ in general. 
	
	\section{Atomic Frequency Standard}\label{apd:atomic_frequency_standard}
	Here we provide a basic description of an atomic frequency standard which suffices for our purposes. A more detailed description can be found in \cite{Riehle_2006}. The Hamiltonian of a two-level atom is given by 
	\begin{align}
		\hat{H}_\mathrm{c} = \frac{1}{2} \omega_0 \hat{\sigma}_\mathrm{z}. 
	\end{align}
    In this section, we only consider a perfect Ramsey experiment in which only relativistic effects are present, ignoring other effects such as decay, collision, etc. Before considering the relativistic case, we first review the case in which the atom is at rest. The atomic frequency standard compares the laser frequency and the transition frequency, which can be performed by the Ramsey experiment. Let $\ket{\psi_{\mathrm{c}}^{(0)}(t)}$ and $\rho_\mathrm{c}^{(0)}[t]=\ketbra{\psi_{\mathrm{c}}^{(0)}(t)}{\psi_{\mathrm{c}}^{(0)}(t)}$ describe the zeroth order time evolution of the atomic state (which is also the time evolution without relativistic effects). For simplicity, we assume that all laser pulses are short but strong such that laser pulses can change the atomic state immediately. We use $t^-$ and $t^+$ to denote the time before and after the pulse at $t$, respectively. Initially the atom is prepared in 
	\begin{align}
		\ket{\psi_{\mathrm{c}}^{(0)}(0^-)}=\ket{\psi_{\mathrm{c},0}}=\ket{g}.
	\end{align}
	Applying a $\frac{\pi}{2}$-pulse at $t=0$, the state of the atom is 
	\begin{align}
		\ket{\psi_{\mathrm{c}}^{(0)}(0^+)}=\frac{1}{\sqrt{2}}(\ket{g}+\ket{e}).
	\end{align}
	After a period of free evolution, the state of the atom at time $t$ in the rotating frame becomes
	\begin{align}\label{eqn:atomic_frequency_property}
		\ket{\psi_{\mathrm{c}}^{(0)}(t)}=\frac{1}{\sqrt{2}}(\ket{g}+e^{-\iu (\omega_0 -\omega )t}\ket{e}).
	\end{align}
	Now we apply a $\frac{\pi}{2}$-pulse at time $t=T$. The state of the atom is 
	\begin{align}
	    \ket{\psi_{\mathrm{c}}^{(0)}(T^-)}=\frac{1}{\sqrt{2}}(\ket{g}+e^{-\iu (\omega_0-\omega)T}\ket{e}), 
	\end{align}
	before the pulse and 
	\begin{align}
		\ket{\psi_{\mathrm{c}}^{(0)}(T^+)}=\iu\sin\frac{(\omega_0-\omega)T}{2}\ket{g}+ \cos \frac{(\omega_0-\omega)T}{2}\ket{e}, 
	\end{align}
	after the pulse. Finally, a measurement in the energy basis is performed. The probability of $\ket{e}$ is given by
	\begin{align}
		\Pr[\ket{e}] = \frac{1}{2}\left(1+\cos(\omega-\omega_0)T\right).
	\end{align}
	This corresponds to the ideal Ramsey experiment with maximal contrast. In reality, experiments can suffer from noise, which results in a smaller contrast $p$
	\begin{align}
		\Pr[\ket{e}] = \frac{1}{2}\left(1+p \cos(\omega-\omega_0)T\right).
	\end{align}
	In principle, we can keep $\omega$ close to $\omega_0$ by continuously maximizing $\Pr[\ket{e}]$ with respect to $\omega$. In practice, in order to improve the accuracy, we will measure the maxima gradient points on both sides on the maxima and take their average. The slope on these points are 
	\begin{align}
	    \frac{\d}{\d \omega } \Pr[\ket{e}] = \frac{Tp}{2},
	\end{align}
	which means that the variance $\sigma_0^2$ of $\omega_0$ depends on the variance $\sigma_{\ket{e}}^2$ of $\Pr[\ket{e}]$ by
	\begin{align}
	    \sigma_{0}^2 = \frac{4}{T^2p^2} \sigma_{\ket{e}}^2,
	\end{align}
    $\sigma_{\ket{e}}^2$, however, depends on the experimental condition of the atomic frequency standard, which is out of the scope of this paper. Therefore, we will instead write
	\begin{align}
	    \sigma_0^2\propto \frac{1}{T^2p^2}. 
	\end{align}
	Now consider that the two-level atom moves and experiences a gravitation field. We preserve the expansion of $\rho_{\mathrm{c}}[t]=\sum_n \frac{1}{m^nc^{2n}} \rho_{\mathrm{c}}^{(n)}[t]$ up to the order of $O(c^{-4})$. We again apply $\frac{\pi}{2}$-pulses at $t=0$ and $t=T$ respectively. Substituting \cref{eqn:rho_first_order,eqn:rho_second_order,eqn:atomic_frequency_property} into $\rho_{\mathrm{c}}[t]$, we obtain the state of the atom at time $t$
	\begin{align}\label{eqn:atomic_frequency_comparison}
		\rho_\mathrm{c}[t] \approx &\  \tilde{p}\ket{\tilde{\psi}_{\mathrm{c}}}\!\bra{\tilde{\psi}_{\mathrm{c}}} +(1-\tilde{p})\mathbb{I}_{\mathrm{c}},
	\end{align}
	where
	\begin{align}
		\ket{\tilde{\psi}_\mathrm{c}(t)} = \frac{1}{\sqrt{2}}(\ket{g}+ e^{-\iu (\tilde{\omega}_0-\omega)t} \ket{e}), 
	\end{align}
	and 
	\begin{align}
		\tilde{\omega}_0 t &= \omega_0(t+I_1), \\
		\tilde{p} & = 1-\frac{\omega_0^2}{2}\left((\Re(I_2)-I_1^2\right). 
	\end{align}
	Similarly, the lowest order of $\tilde{\omega}_0$ contains $\frac{V_\mathrm{k}}{mc^2}$, and the lowest order of $\tilde{p}$ contains $\frac{V_\mathrm{k}^2}{m^2c^4}$; it is sufficient to keep $O(c^{0})$ terms and omit $O(c^{-2})$ terms in  $H_\mathrm{k}$ and $V_\mathrm{k}$. By comparison between \cref{eqn:atomic_frequency_property,eqn:atomic_frequency_comparison}, the process with relativistic effects can be seen as a mixture of the process without relativistic effects but with modified frequency $\tilde{\omega}_0$ with a probability of $\tilde{p}$, and a completely depolarizing process with a probability of $1-\tilde{p}$. Therefore, the probability of $\ket{e}$ up to the order of $O(c^{-4})$ is given by 
	\begin{align}  
		\tilde{\Pr}[\ket{e}] =\frac{1}{2}\left(1+ \tilde{p}\cos((\tilde{\omega}_0-\omega)T)\right) . 
	\end{align}
	In this case, the variance $\tilde{\sigma}_0$ of $\tilde{\omega}_0$ is 
    \begin{align}
        \tilde{\sigma}_0^2\propto \frac{1}{T^2\tilde{p}^2}.
    \end{align}
	Comparing variances of perfect experiments with still atoms in a flat space and with moving atoms in a curved space which are proportional to $T^{-2}$ and $T^{-2}\edit{\tilde{p}}$ respectively, the increase in variance is
	\begin{align}
	    \tilde{\sigma}_0^2 - \sigma_0^2 \propto \frac{\omega_0^2}{T^2} \left((\Re(I_2)-I_1^2\right).
	\end{align}
	Now we distinguish between the quantum superposition case and the classical mixture case. We denote the quantities in the quantum superposition case with $\tilde{\omega}_0\mapsto \tilde{\omega}_{0,\qtm}$, $\tilde{p}\mapsto\tilde{p}_{\qtm}$ , $I_1\mapsto I_{1,\qtm}$ and $I_{2,\qtm}$ and in classical mixture case with $\tilde{\omega}_0\mapsto \tilde{\omega}_{0,\cls}$, $\tilde{p}\mapsto \tilde{p}_{\cls}$, $I_1\mapsto I_{1,\cls}$ and $I_{2,\cls}$. Let the discrepancy $\tilde{\omega}_{0,\coh}$ between two cases be 
	\begin{align}
	    \tilde{\omega}_{0,\coh}=\tilde{\omega}_{0,\qtm}-\tilde{\omega}_{0,\cls}. 
	\end{align}
	The discrepancy and the contrasts can be written in terms of $\Delta_{1,\coh}$, $\Delta_{2,\qtm }^2 $ and $\Delta_{2,\cls}^2$ in \cref{eqn:delta_1_coh,eqn:delta_2_qtm,eqn:delta_2_cls} as 
	\begin{align}
	    \tilde{\omega}_{\coh}&= \frac{\omega_0}{T} \Delta_{1,\coh}, \\
        \tilde{p}_{\qtm} &= 1-\frac{\omega_0^2}{2}\Delta_{2,\qtm}^2, \\
        \tilde{p}_{\cls} &= 1-\frac{\omega_0^2}{2}\Delta_{2,\cls}^2, 
	\end{align}
    from which \cref{eqn:discrepancy_atomic_clock,eqn:variance_atomic_clock} can be derived, as is shown in the main text. 

\section{Idealized Clocks}\label{apd:idealised_clock}
	An idealized clock in the non-relativistic limit is defined by its commutator between two observables, a time operator $T_\mathrm{c}$ and the Hamiltonian $H_\mathrm{c}$, and the clock state itself $\ket{\psi_\textup{Ideal}(t)}$,
	\begin{align}
		-\iu[\hat{T}_\mathrm{c},\hat{H}_\mathrm{c}] \ket{\psi_\textup{Ideal}(t)} = \ket{\psi_\textup{Ideal}(t)},
	\end{align}
	for all time $t$ and $\ket{\psi_\textup{Ideal}(t)}$, the time evolved initial state according to the clock Hamiltonian $H_\mathrm{c}$. We will make use of the property that 
	\begin{align}\label{eqn:idealised_clock_property}
	    e^{\iu \hat{H}_\mathrm{c} t } \hat{T}_{\mathrm{c}} e^{-\iu \hat{H}_\mathrm{c} t } = \hat{T}_{\mathrm{c}} + t \mathbb{I}.
	\end{align}	
	The idealized clock can be mimicked quite well by a quasi-ideal clock. For more details of both idealized clocks and quasi-ideal clocks, readers may refer to \cite{Woods_2019,Khandelwal_2020}. We consider the expectation value $\average{\hat{T}_{\rm c}}(t) = \Tr(\hat{T}_{\rm c} \rho_{\rm c}[t])$ and the variance $\sigma_{\rm c}^2(t) = \Tr(\hat{T}_{\rm c}^2 \rho_{\rm c}[t])-\Tr(\hat{T}_{\rm c} \rho_{\rm c}[t])^2 $ of the time operator. For a confined idealized clock described by \cref{eqn:Hamiltonian_middle}, we expand $\rho_{\rm c}[t]=\sum_{n}\frac{1}{m^nc^{2n}}\rho_{\rm c}^{(n)}[t]$ and preserve terms up to the order of $O(c^{-4})$. Substituting \cref{eqn:rho_first_order,eqn:rho_second_order} into $\average{\hat{T}_{\mathrm{c}}}$ and $\sigma_\mathrm{c}^2$ and making use of \cref{eqn:idealised_clock_property}, we obtain the expectation value and the variance 
	\begin{align}
		\average{\hat{T}_\mathrm{c}}(t) &= \average{\hat{T}_\mathrm{c}}(0)+t+I_1, \label{eqn:proper_time_ideal_clock}\\
		\sigma_\mathrm{c}^2(t) &= \sigma_\mathrm{c}^2(0)+ \Re(I_2) - I_1^2 + \left(\Im(I_2)+2I_2'\right) \left(\average{[\hat{T}_\mathrm{c},\hat{H}_\mathrm{c}]}(0)-2\average{\hat{T}_\mathrm{c}}(0)\average{\hat{H}_\mathrm{c}}(0)\right). \label{eqn:variance_ideal_clock}
	\end{align}
	where $I_1$, $I_2$ and $I_2'$ are defined in \cref{eqn:I_1_apd,eqn:I_2_apd,eqn:I_2_prime_apd}, and $\average{\hat{O}}(0)=\Tr(\hat{O}\rho_{\mathrm{c},0})$. Since the lowest order of $\average{\hat{T}_{\rm c}}$ contains $\frac{\hat{V}_\mathrm{k}}{mc^2}$, and the lowest order of $\sigma_{\mathrm{c}}^2$ contains $\frac{\hat{V}_\mathrm{k}^2}{m^2c^4}$ and $\frac{\hat{W}_\mathrm{k}}{m^2c^4}$, it is sufficient to keep $O(c^{0})$ terms and omit $O(c^{-2})$ terms in  $\hat{H}_\mathrm{k}$, $\hat{V}_\mathrm{k}$ and $\hat{W}_\mathrm{k}$. Now we distinguish between two cases with different initial kinematic state. One is the case of preparing a spatial quantum superposition $\rho_{\mathrm{c},\qtm}$ in \cref{eqn:superposition}, and we replace $\average{\hat{T}_{\rm c}}\mapsto \average{\hat{T}_{\rm c}}_{\qtm}$, $\sigma_{\mathrm{c}}^2\mapsto \sigma_{\mathrm{c},\qtm}^2$, $I_1\mapsto I_{1,\qtm}$, $I_2\mapsto I_{2,\qtm}$ and $I_2'\mapsto I_{2,\qtm}'$. The other is the case of preparing a spatial classical mixture $\rho_{0,\cls}$ in \cref{eqn:mixture}, and we replace $\average{\hat{T}_{\rm c}}\mapsto \average{\hat{T}_{\rm c}}_{\cls}$, $\sigma_{\mathrm{c}}^2\mapsto\sigma_{\mathrm{c},\cls}^2$, $I_1\mapsto I_{1,\cls}$, $I_2\mapsto I_{2,\cls}$ and $I_2'\mapsto I_{2,\cls}'$. We will denote the discrepancy $\average{\hat{T}_{\mathrm{c}}}$ between these two cases and the variance $\sigma_{\coh}^2$ of the discrepancy by 
	\begin{align}
		\average{\hat{T}_\mathrm{c}}_\coh(t)&=  \average{\hat{T}_{\mathrm{c}}}_{\qtm}(t)-\average{\hat{T}_{\mathrm{c}}}_{\cls}(t), \\
		\sigma_{\coh}^2 &= \sigma_{\mathrm{c},\qtm}^2 + \sigma_{\mathrm{c},\cls}^2. 
	\end{align}
	We also explicitly write down $\average{\hat{T}_\mathrm{c}}_{\coh}$ and $\sigma_{\coh}^2$
	\begin{align}
		\average{\hat{T}_\mathrm{c}}_\coh &= \Delta_{1,\coh},\\
		\sigma_{\coh}^2 &= \Delta_{2,\qtm}^2 + \Delta_{2,\cls}^2+ 2\average{\hat{T}_\mathrm{c}^2}(0) - 2\average{\hat{T}_\mathrm{c}}(0)^2  + \left(\average{\{\hat{T}_\mathrm{c}, \hat{H}_{\mathrm{c}}\}}(0) - 2 \average{\hat{T}_\mathrm{c}}(0)\average{\hat{H}_\mathrm{c}}(0)\right)(\Sigma_{2,\qtm}+\Sigma_{2,\cls}).
	\end{align}
	where $\Sigma_{2,\qtm}=\Im (I_{2,\qtm} ) + 2I_{2,\qtm}'$,   $\Sigma_{2,\cls}=\Im (I_{2,\cls} ) + 2I_{2,\cls}'$, and $\Delta_{1,\coh}$, $\Delta_{2,\qtm }^2 $ and $\Delta_{2,\cls}^2$ are given in \cref{eqn:delta_1_coh,eqn:delta_2_qtm,eqn:delta_2_cls}. The variance $\sigma_{\coh}^2$ can be decomposed into a clock-state-independent term $\sigma_{\coh,\mathrm{i}}^2$ and a clock-state-dependent term $\sigma_{\coh,\mathrm{d}}^2$ given by
	\begin{align}
		\sigma_{\coh,\mathrm{i}}^2 &= \Delta_{2,\qtm}^2  + \Delta_{2,\cls}^2, \\
		\sigma_{\coh,\mathrm{d}}^2 &= \sigma_{\coh}^2 -\sigma_{\coh,\mathrm{i}}^2.
	\end{align}
    We regard the clock-state-independent term $\sigma_{\coh,\mathrm{i}}^2$ as a more general term of the discrepancy of relativistic time dilation than the clock-state-dependent term $\sigma_{\coh,\mathrm{d}}^2$, and thus pay more attention on the former. 
    
\begin{figure}[htbp!]
	\centering
        \subfigure[Discrepancy vs Standard deviation]{\includegraphics[width=0.5\columnwidth]{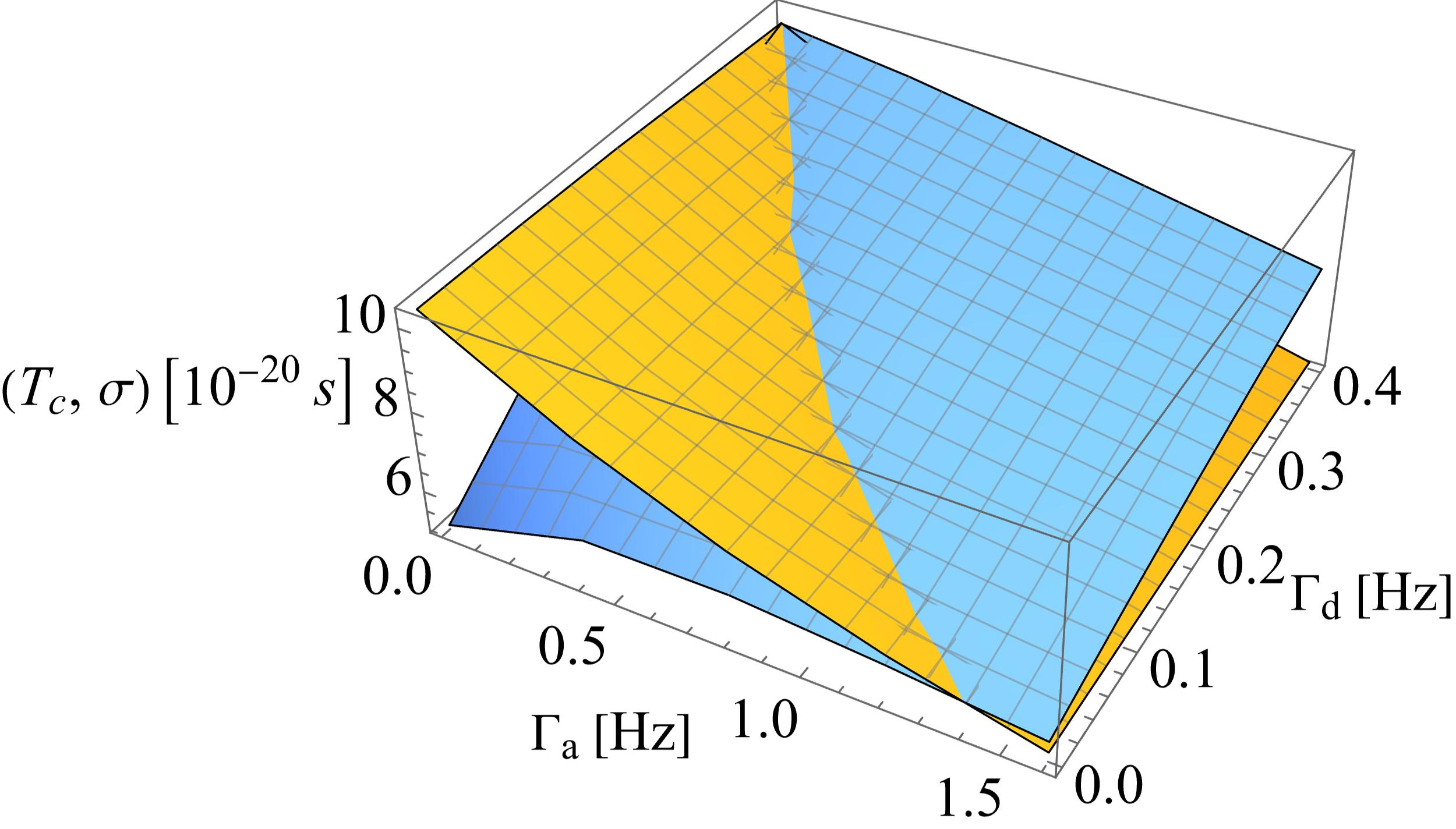}}
        \quad 
        \subfigure[Discrepancy minus Standard Deviation]{\includegraphics[width=0.4\columnwidth]{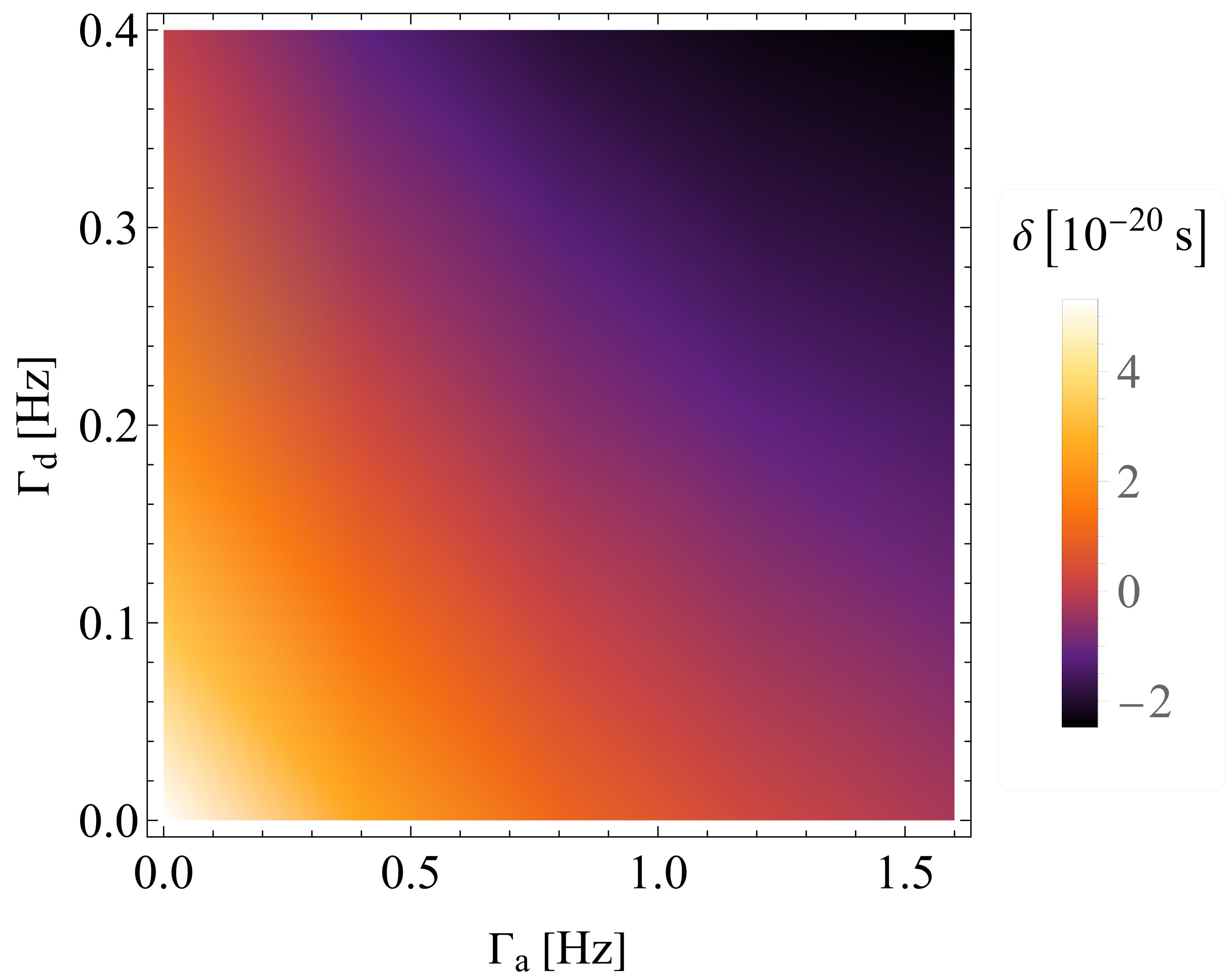}
        }
        \caption{(a) The discrepancy $\langle\hat{T}_{\rm c}\rangle$ (orange) and the standard deviation $\sigma_{\coh,{\rm i}}$ (blue) for an ideal clock. The discrepancy is detectable when the orange surface is above the blue one. (b) The difference between the discrepancy and the standard deviation $\delta=\langle \hat{T}_{\rm c}\rangle -\sigma_{\coh, {\rm i}}$ for an ideal clock. The discrepancy is detectable when $\delta$ is positive. The parameters are the same as the clock in the main text, i.e. $\lambda_{\mathrm{Mg}}=468$~nm, $U_{\max,\mathrm{Mg}} = \edit{300} E_{\mathrm{r},\mathrm{Mg}}$, $T=1$~s, $\phi=\pi$, \edit{$\theta=\frac{\pi}{4}$} and $d=10$~nm \edit{($\alpha=0.395$)}. }
	\label{fig:discrepancy_vs_standard_deviation_Mg}
    \end{figure}

    We compute and plot the discrepancy $\langle\hat{T}_{\rm c}\rangle_{\coh}$ and the clock-state-independent standard deviation $\sigma_{\coh,{\rm i}}$ with respect to decoherence rate for the idealized clock with the same parameters as in the main text, i.e. $\lambda_{\mathrm{Mg}}=468$~nm, $U_{\max,\mathrm{Mg}} = \edit{300} E_{\mathrm{r},\mathrm{Mg}}$, $T=1$~s, $\phi=\pi$, \edit{$\theta=\frac{\pi}{4}$} and $d=10$~nm \edit{($\alpha=0.395$)}, in \cref{fig:discrepancy_vs_standard_deviation_Mg}. 
    
    \section{Protocol for \texorpdfstring{$^{87}\mathrm{Sr}$}{TEXT} Optical Lattice Clock}
    \label{apd:sr_protocol}
\begin{figure}[htbp!]
	\centering
        \includegraphics[width=0.4\linewidth]{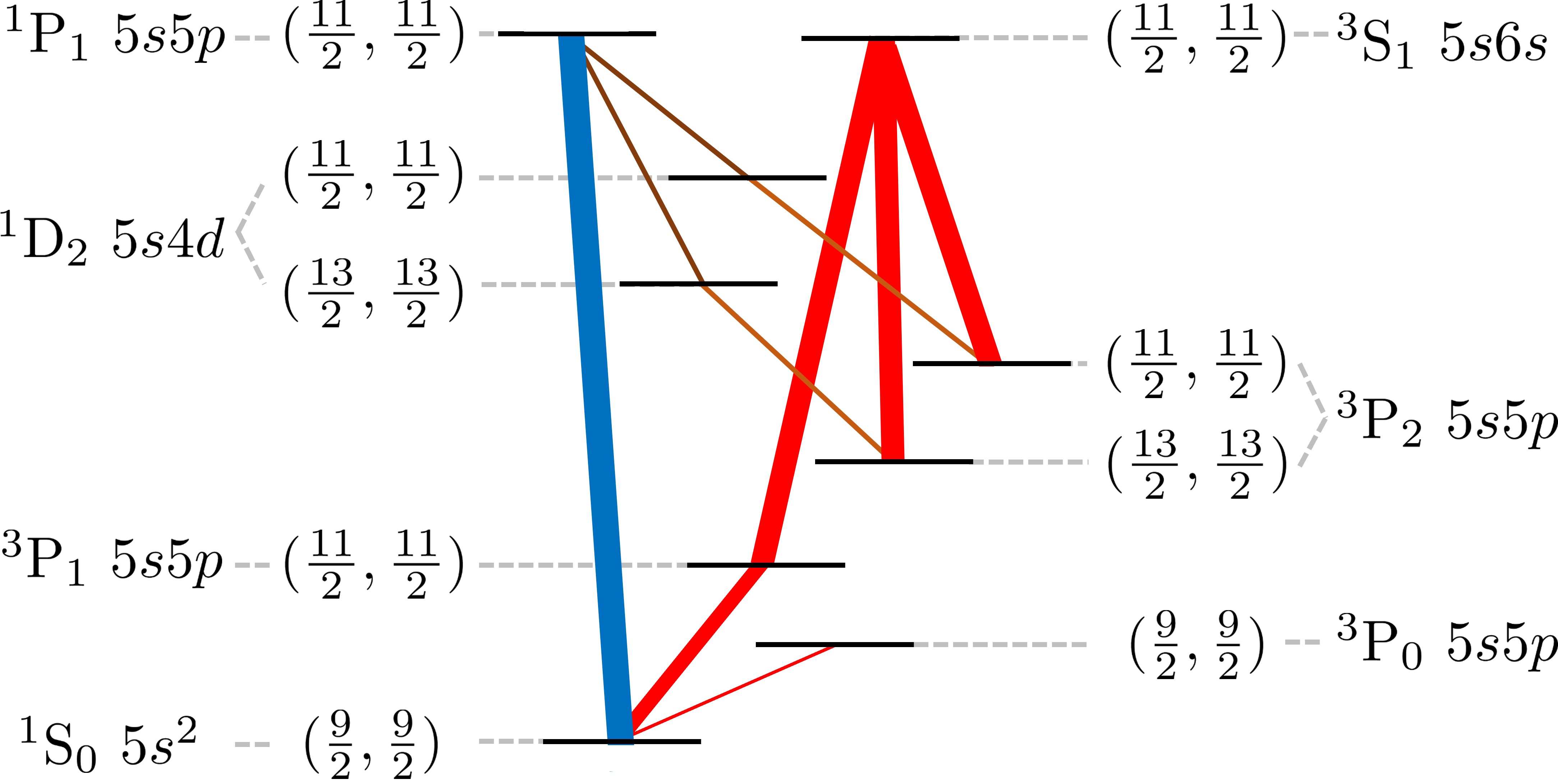}
        \caption{The energy diagram of the $^{87}\mathrm{Sr}$ atom. Each state is labeled by both its electronic state $ ^{2S+1} L_{J} $ and its hyperfine state $(F,m_F)$. The line color corresponds to the transition light color and the line width corresponds to the transition strength.}
	\label{fig:simplified_energy_diagram_Sr}
    \end{figure}

    The relevant energy diagram of the $^{87}\mathrm{Sr}$ is shown in \cref{fig:simplified_energy_diagram_Sr} \cite{Hobson_2016,Derevianko_2011,NIST_2013,Sansonetti_2010}. $^{87}\mathrm{Sr}$ is a fermion. It has a non-zero nuclear spin, and therefore a much more complicated electronic structure than $^{24}\mathrm{Mg}$ due to the hyperfine splitting. This feature complicates experimental procedures, but it also provides enough states for operations such as electron shelving and cooling. 
	
    Let us first describe how to adapt the general protocol into the $^{87}\mathrm{Sr}$ protocol. $^{87}\mathrm{Sr}$ atoms have to be identified with not only fine states but also hyperfine states. An additional magnetic field has to be applied in order to have a well-defined hyperfine state. We mention that the direction of the magnetic field is not necessarily the same as the direction of the optical lattice. The $\ket{F=\frac{9}{2}, m_F=\frac{9}{2}}$ states $^{1}\mathrm{S}_{0}$ and $^{3}\mathrm{P}_{0}$ are used as stable states $\ket{\mathrm{u}_1}$ and $\ket{\mathrm{u}_2}$ respectively. The $\ket{F =\frac{11}{2},m_F=\frac{11}{2}}$ and $\ket{F=\frac{13}{2}, m_F=\frac{13}{2}}$ states of $^{3}\mathrm{P}_{2}$ are used as meta-stable states $\ket{\mathrm{m}_1}$ and $\ket{\mathrm{m}_2}$ respectively. The $\ket{F=\frac{11}{2}, m_F=\frac{11}{2}}$ state of $^{1}\mathrm{P}_{1}$ is used as the unstable state $\ket{\mathrm{u}_1}$ and any hyperfine state of $^{1}\mathrm{P}_{1}$ can be used as the unstable state $\ket{\mathrm{u}_2}$. 
	
    The transitions between these states are specified by not only the frequency but also the polarization of the laser, in order to transit between correct hyperfine states. We will denote the right and left handed circular polarized laser with the wave vector parallel to the magnetic field by the $\sigma_{R}^{\pm}$ laser respectively. One should not confuse $\sigma_{R}^{\pm}$ which describes the laser inducing the transition with $\sigma^{\pm}$ which describes the laser inducing the optical lattice. We will denote the linear polarized laser with the wave vector perpendicular to the magnetic field by the $\pi$ laser. The Raman transition from $\ket{F=\frac{9}{2}, m_F=\frac{9}{2}}$ of $^1\mathrm{S}_0$ to $\ket{F=\frac{13}{2}, m_F=\frac{13}{2}}$ of $^{3}\mathrm{P}_2$ can be realized by a laser between $\ket{F=\frac{9}{2}, m_F=\frac{9}{2}}$ of $^1\mathrm{S}_0$ and $\ket{F=\frac{11}{2}, m_F=\frac{11}{2}}$ of $^3\mathrm{P}_1$ with $\sigma_R^{+}$ polarization, a laser between $\ket{F=\frac{11}{2}, m_F=\frac{11}{2}}$ of $^3\mathrm{P}_1$ and $\ket{F=\frac{11}{2}, m_F=\frac{11}{2}}$ of $^3\mathrm{S}_1$ with $\pi$ polarization and a laser between $\ket{F=\frac{13}{2}, m_F=\frac{13}{2}}$ of $^3\mathrm{P}_2$ and $\ket{F=\frac{11}{2}, m_F=\frac{11}{2}}$ of $^3\mathrm{S}_1$ with $\sigma_R^-$ polarization. The Raman transition between two hyperfine states of $^3\mathrm{P}_2$ can be realized by a laser between $\ket{F=\frac{13}{2}, m_F=\frac{13}{2}}$ of $^3\mathrm{P}_2$ and $\ket{F=\frac{11}{2}, m_F=\frac{11}{2}}$ of $^3\mathrm{S}_1$ with $\sigma_R^-$ polarization and a laser between $\ket{F=\frac{11}{2}, m_F=\frac{11}{2}}$ of $^3\mathrm{P}_2$ and $\ket{F=\frac{13}{2}, m_F=\frac{13}{2}}$ of $^3\mathrm{P}_2$ with $\pi$ polarization. The Raman transition from $\ket{F=\frac{13}{2}, m_F=\frac{13}{2}}$ of $^3\mathrm{P}_2$ to $\ket{F=\frac{11}{2}, m_F=\frac{11}{2}}$ of $^1\mathrm{P}_1$ can be realized by a laser between $\ket{F=\frac{13}{2}, m_F=\frac{13}{2}}$ of $^3\mathrm{P}_2$ and $\ket{F=\frac{13}{2}, m_F=\frac{13}{2}}$ of $^1\mathrm{D}_2$ with $\pi$ polarization and a laser between $\ket{F=\frac{13}{2}, m_F=\frac{13}{2}}$ of $^1\mathrm{D}_2$ and $\ket{F=\frac{11}{2}, m_F=\frac{11}{2}}$ of $^1\mathrm{P}_1$ with $\sigma_R^{-}$ polarization. Similarly, the Raman transition from $\ket{F=\frac{11}{2}, m_F=\frac{13}{2}}$ of $^3\mathrm{P}_2$ to $\ket{F=\frac{11}{2}, m_F=\frac{11}{2}}$ of $^1\mathrm{P}_1$ can be realized by a laser between $\ket{F=\frac{11}{2}, m_F=\frac{11}{2}}$ of $^3\mathrm{P}_2$ and $\ket{F=\frac{11}{2}, m_F=\frac{11}{2}}$ of $^1\mathrm{D}_2$ with $\pi$ polarization and a laser between $\ket{F=\frac{11}{2}, m_F=\frac{11}{2}}$ of $^1\mathrm{D}_2$ and $\ket{F=\frac{11}{2}, m_F=\frac{11}{2}}$ of $^1\mathrm{P}_1$ with $\pi$ polarization. $\ket{F=\frac{11}{2}, m_F=\frac{11}{2}}$ of $^1\mathrm{P}_1$ can only spontaneously decay to $\ket{F=\frac{9}{2}, m_F=\frac{9}{2}}$ of $^1\mathrm{S}_0$, which keeps atoms spin-polarized, i.e. in the same hyperfine state. Thus $\ket{F=\frac{11}{2}, m_F=\frac{11}{2}}$ of $^1\mathrm{P}_1$ is the only suitable hyperfine state of $^1\mathrm{P}_1$ for $\ket{\mathrm{u}_1}$. As for $\ket{\mathrm{u}_2}$, all hyperfine states are suitable for measurements because we do not need to keep atoms spin-polarized after measurements. 
    
	Atoms loaded into an optical lattice can stay in arbitrary hyperfine state. Therefore, we have to additionally spin-polarize atoms in \emph{Step 0}, as is described in \cite{Takamoto_2006}. Suppose that initially atoms are in arbitrary hyperfine states $\ket{F=\frac{9}{2},m_F}$ state of $^{1}\mathrm{S}_{0}$. Atoms can be spin-polarized into $\ket{F=\frac{9}{2}, m_{F} =\frac{9}{2}}$ of $^{1}\mathrm{S}_0$ as follow. A horizontal magnetic field is applied to atoms. A laser with circular polarization $\sigma_R^+$ transit atoms from $\ket{F=\frac{9}{2},m_F}$ of $^1\mathrm{S}_0$ to $\ket{F=\frac{9}{2},m_F+1}$ of $^{3}\mathrm{P}_1$. Meanwhile, atoms in $\ket{F=\frac{9}{2},m_F+1}$ of $^{3}\mathrm{P}_1$ spontaneously decay to $\ket{F=\frac{9}{2},m_F}$, $\ket{F=\frac{9}{2},m_F+1}$ and $\ket{F=\frac{9}{2},m_F+2}$ of $^{1}\mathrm{S}_0$. $m_F$ is thus increased by $1$ in each cycle on average until atoms reach $\ket{F=\frac{9}{2},m_F=\frac{9}{2}}$ of $^1\mathrm{S}_0$, which is the dark state of the transition. Similarly, atoms can be spin-polarized into $\ket{F=\frac{9}{2},m_F=-\frac{9}{2}}$ of $^1\mathrm{S}_0$ by reversing the polarisation of the laser. 
	
\begin{figure}[htbp!]
		\centering
		\includegraphics[width=0.5\linewidth]{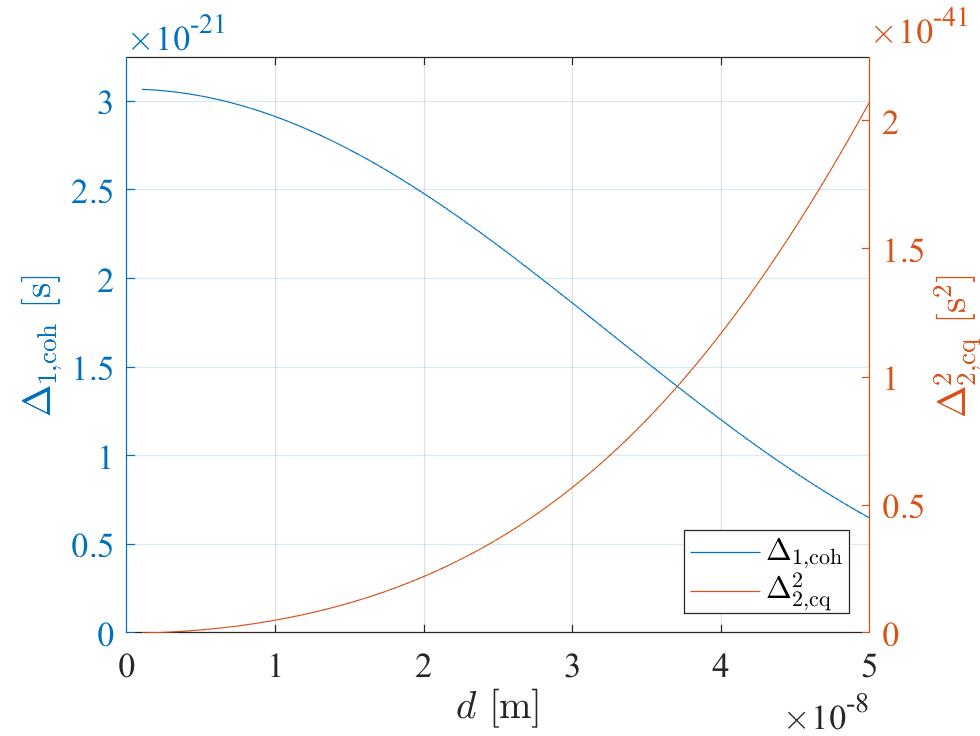}
		\caption{The discrepancy $\Delta_{1,\coh}$ between the quantum superposition case and the classical mixture case and the increase in variance  $\Delta_{1,\coh}$ of the quantum superposition case and the classical mixture case versus the displacement $d$  respectively for a $^{87}\mathrm{Sr}$ clock. The parameters are $\lambda_{\mathrm{Sr}}=813$~nm, $U_{\max,\mathrm{Sr}}=300 E_{\mathrm{r},\mathrm{Sr}}$, $\phi=\pi$, \edit{$\theta=\frac{\pi}{4}$} and $T=1$~s. \edit{Note that $d=\sqrt{2}z_\mathrm{s} \alpha$, and $d=10$~nm corresponds to $\alpha= 0.227$ for the $^{87}{\rm Sr}$ clock here.} }
		\label{fig:quantum_modification_displacement_Sr}
	\end{figure}

	A $^{87}\mathrm{Sr}$ optical lattice clock has a magic wavelength at $\lambda_{\mathrm{Sr}}=813$~nm \cite{Takamoto_2005}. The the trap depth is set to $U_{\max,\mathrm{Sr}} = 300E_{\mathrm{r},\mathrm{Sr}}$, where $E_{\mathrm{r},\mathrm{Sr}}=\frac{2\pi^2\hbar^2}{m_{\mathrm{Sr}}\lambda_{\mathrm{Sr}}^2}$ is the recoil energy of $^{87}\mathrm{Sr}$. The interrogation time is assumed to be $T=1$~s. We also set $\phi = \pi $ and \edit{$\theta =\frac{ \pi}{4}$}. The relation between $\Delta_{1,\coh}$ and $\sqrt{\Delta_{2,\qtm}^2+\Delta_{2,\cls}^2}$ with respect to the displacement $d$ can be found in \cref{fig:quantum_modification_displacement_Sr}. \edit{Again, $d=\sqrt{2}z_\mathrm{s} \alpha$, and $d=10$~nm corresponds to $\alpha= 0.227$ for the $^{87}{\rm Sr}$ clock here.}

	\section{Evolution of Operators under Noise} \label{apd:operator_evolution}
	\subsection{Free Evolution}\label{apd:free_evolution}
    The dual of the Lindblad equation for the amplitude damping channel can be written as
	\begin{align}
		\derivative{\hat{A}}{t} = \iu \omega_\mathrm{z} [\hat{a}^\dagger \hat{a}, \hat{A}]. 
	\end{align}
	Due to the unitarity of free evolution, we have
	\begin{align}\label{eqn:free_a_adagger}
		(\left.\hat{a}^\dagger\right.^m \hat{a}^n)[t] = (\hat{a}^\dagger[t])^m (\hat{a}[t])^n,
	\end{align}
    where
	\begin{align}
		\hat{a}[t] = \hat{a}e^{-\iu \omega_{\mathrm{z}} t}. 
	\end{align}
	The solution of polynomials of the creation and annihilation operator is given by
	\begin{align}
		(\left.\hat{a}^\dagger\right.^m \hat{a}^n)[t] = \left.\hat{a}^\dagger\right.^m \hat{a}^n e^{\iu(m-n)\omega_\mathrm{z} t }. 
	\end{align}
	\subsection{Amplitude Damping Channel}\label{apd:amplitude_damping_channel}
	The dual of the Lindblad equation of the amplitude damping channel can be written as
	\begin{align}
		\derivative{\hat{A}}{t} = \iu \omega_\mathrm{z} [\hat{a}^\dagger \hat{a}, \hat{A}]+\frac{ \Gamma_\mathrm{a}}{2} ([\hat{a}^\dagger,\hat{A}]\hat{a} -\hat{a}^\dagger [\hat{a},\hat{A}]). 
	\end{align}
	The expectation value and the variance only include polynomials of creation and annihilation operators in the normal order of up to the fourth order. Therefore, we consider the time evolution of polynomials up to the fourth order. For the first order, we assume that 
	\begin{align}
		\hat{a}[t] = f_{11}(t) \hat{a}. 
	\end{align}
	We then get
	\begin{align}
		\derivative{f_{11}}{t} \hat{a} = -\iu \omega_\mathrm{z}  f_{11} \hat{a} - \frac{\Gamma_\mathrm{a}}{2} f_{11} \hat{a}, 
	\end{align}
	and therefore
	\begin{align}
		\hat{a}[t] = \hat{a}e^{-\iu\omega_\mathrm{z}  t -\frac{\Gamma_\mathrm{a}}{2} t}. 
	\end{align}
	For the second order, we assume that 
	\begin{align}
		(\hat{a}^2)[t] &= f_{22}(t) \hat{a}^2, \\
		(\hat{a}^\dagger\hat{a})[t] &= f_{21}(t) \hat{a}^\dagger \hat{a}. 
	\end{align}
	We then get
	\begin{align}
		\derivative{f_{22}}{t} \hat{a}^2 &= -2\iu \omega_\mathrm{z}  f_{22} \hat{a}^2 - \Gamma_\mathrm{a} f_{22} \hat{a}^2, \\
		\derivative{f_{21}}{t} \hat{a}^\dagger \hat{a} &= - \Gamma_\mathrm{a} f_{21} \hat{a}^\dagger \hat{a}, 
	\end{align}
	and therefore
	\begin{align}
		(\hat{a}^2)[t] &= \hat{a}^2e^{-2\iu\omega_\mathrm{z}  t -\Gamma_\mathrm{a} t},\\
		(\hat{a}^\dagger \hat{a})[t] &= \hat{a}^\dagger \hat{a} e^{-\Gamma_\mathrm{a} t}. 
	\end{align}
	As for the third order, we assume that 
	\begin{align}
		(\hat{a}^3)[t] &= f_{33}(t) \hat{a}^2, \\
		({\hat{a}^\dagger} \hat{a}^2)[t] &= f_{32}(t) {\hat{a}^\dagger} \hat{a}^2. 
	\end{align}
	We then get
	\begin{align}
		\derivative{f_{33}}{t} \hat{a}^3 &= -3\iu \omega_\mathrm{z}  f_{33} \hat{a}^3 - \frac{3}{2}\Gamma_\mathrm{a} f_{33} \hat{a}^3 ,\\
		\derivative{f_{32}}{t} {\hat{a}^\dagger} \hat{a}^2 &= -\iu \omega_\mathrm{z}  f_{32} {\hat{a}^\dagger} \hat{a}^2 -\frac{3}{2} \Gamma_\mathrm{a} f_{32}{\hat{a}^\dagger} \hat{a}^2 ,
	\end{align}
	and therefore
	\begin{align}
		(\hat{a}^3)[t] &= \hat{a}^3e^{-3\iu\omega_\mathrm{z}  t -\frac{3}{2}\Gamma_\mathrm{a} t}, \\
		({\hat{a}^\dagger} \hat{a}^2)[t] &= {\hat{a}^\dagger} \hat{a}^2 e^{-\iu\omega_\mathrm{z}  t-\frac{3}{2}\Gamma_\mathrm{a} t}. 
	\end{align}
	Finally, for the fourth order 
	\begin{align}
		(\hat{a}^4)[t] &= f_{44}(t) \hat{a}^2, \\
		({\hat{a}^\dagger} \hat{a}^3)[t]  &= f_{43}(t) {\hat{a}^\dagger} \hat{a}^3, \\
		(\left.\hat{a}^\dagger\right.^2 \hat{a}^2)[t]  &= f_{42}(t) \left.\hat{a}^\dagger\right.^2 \hat{a}^2.  
	\end{align}
	We then get
	\begin{align}
		\derivative{f_{44}}{t} \hat{a}^4 &= -4\iu \omega_\mathrm{z}  f_{44} \hat{a}^4 - 2\Gamma_\mathrm{a} f_{44} \hat{a}^4, \\
		\derivative{f_{43}}{t} {\hat{a}^\dagger} \hat{a}^3 &= -2\iu \omega_\mathrm{z}  f_{43} {\hat{a}^\dagger} \hat{a}^3 -2 \Gamma_\mathrm{a} f_{43}{\hat{a}^\dagger} \hat{a}^3, \\
		\derivative{f_{42}}{t} \left.\hat{a}^\dagger\right.^2 \hat{a}^2 &= -2 \Gamma_\mathrm{a} f_{42}\left.\hat{a}^\dagger\right.^2 \hat{a}^2,
	\end{align}
	and therefore
	\begin{align}
		(\hat{a}^4)[t] &= \hat{a}^4e^{-4\iu\omega_\mathrm{z}  t -2\Gamma_\mathrm{a} t},\\
		({\hat{a}^\dagger} \hat{a}^3)[t] &= {\hat{a}^\dagger} \hat{a}^3 e^{-2\iu\omega_\mathrm{z}  t-2\Gamma_\mathrm{a} t}, \\
		(\left.\hat{a}^\dagger\right.^2 \hat{a}^2)[t] &= \left.\hat{a}^\dagger\right.^2 \hat{a}^2 e^{-2\Gamma_\mathrm{a} t}. 
	\end{align}
	By observing the above solutions, we can find the general solution of polynomials of the creation and annihilation operator
	\begin{align}\label{eqn:amplitude_a_adagger}
		(\left.\hat{a}^{\dagger}\right.^m \hat{a}^n)[t] = \left.\hat{a}^{\dagger}\right.^m \hat{a}^n e^{\iu(m-n)\omega_\mathrm{z} t  - \frac{1}{2}(m+n) \Gamma_\mathrm{a} t}. 
	\end{align}
	\subsection{Phase Damping Channel}\label{apd:phase_damping_channel}
	The dual of the Lindblad equation for the phase damping channel is given by
	\begin{align}
		\derivative{\hat{A}}{t} = \iu \omega_\mathrm{z} [\hat{a}^\dagger \hat{a}, \hat{A}]-\frac{ \Gamma_\mathrm{p}}{2} [\hat{a}^\dagger \hat{a}, [\hat{a}^\dagger \hat{a},\hat{A}]]. 
	\end{align}
	For the first order, we assume that 
	\begin{align}
		\hat{a}[t] = f_{11}(t) \hat{a}.
	\end{align}
	We then get
	\begin{align}
		\derivative{f_{11}}{t} \hat{a} = -\iu \omega_\mathrm{z}  f_{11} \hat{a} - \frac{\Gamma_\mathrm{p}}{2} f_{11} \hat{a}, 
	\end{align}
	and therefore
	\begin{align}
		\hat{a}[t] = \hat{a}e^{-\iu\omega_\mathrm{z}  t -\frac{\Gamma_\mathrm{p}}{2} t}. 
	\end{align}
	For the second order, we assume that 
	\begin{align}
		(\hat{a}^2)[t] &= f_{22}(t) \hat{a}^2, \\
		(\hat{a}^\dagger \hat{a})[t] &= f_{21}(t) \hat{a}^\dagger \hat{a}. 
	\end{align}
	We then get
	\begin{align}
		\derivative{f_{22}}{t} \hat{a}^2 &= -2\iu \omega_\mathrm{z}  f_{22} \hat{a}^2 -2\Gamma_\mathrm{p} f_{22} \hat{a}^2, \\
		\derivative{f_{21}}{t} \hat{a}^\dagger \hat{a} &= 0 ,
	\end{align}
	and therefore
	\begin{align}
		(\hat{a}^2)[t] &= \hat{a}^2e^{-2\iu\omega_\mathrm{z}  t -2\Gamma_\mathrm{p} t},\\
		(\hat{a}^\dagger \hat{a})[t] &= \hat{a}^\dagger \hat{a}.  
	\end{align}
	As for the third order,  
	\begin{align}
		(\hat{a}^3)[t] &= f_{33}(t) \hat{a}^3, \\
		(\hat{a}^\dagger \hat{a}^2)[t] &= f_{32}(t) \hat{a}^\dagger \hat{a}^2 .
	\end{align}
	We then get
	\begin{align}
		\derivative{f_{33}}{t} \hat{a}^3 &= -3\iu \omega_\mathrm{z}  f_{33} \hat{a}^3 -\frac{9}{2}\Gamma_\mathrm{p} f_{33} \hat{a}^3, \\
		\derivative{f_{32}}{t} \hat{a}^\dagger \hat{a}^2 &= -\iu \omega_\mathrm{z}  f_{32}\hat{a}^\dagger  \hat{a}^2 -\frac{\Gamma_\mathrm{p}}{2} f_{32} \hat{a}^\dagger  \hat{a}^2, 
	\end{align}
	and therefore
	\begin{align}
		(\hat{a}^3)[t] &= \hat{a}^3e^{-3\iu\omega_\mathrm{z}  t -\frac{9}{2}\Gamma_\mathrm{p} t},\\
		(\hat{a}^\dagger \hat{a}^2)[t] &= \hat{a}^\dagger \hat{a}^2  e^{-\iu\omega_\mathrm{z}  t -\frac{\Gamma_\mathrm{p} }{2}t}.
	\end{align}
	Finally, for the fourth order,  
	\begin{align}
		(\hat{a}^4)[t] &= f_{44}(t) \hat{a}^4, \\
		(\hat{a}^\dagger \hat{a}^3)[t] =&\ f_{43}(t) \hat{a}^\dagger \hat{a}^3, \\ 
		(\left.\hat{a}^\dagger\right.^2 \hat{a}^2)[t] &= f_{42}(t) \left.\hat{a}^\dagger\right.^2  \hat{a}^2. 
	\end{align}
	We then get
	\begin{align}
		\derivative{f_{44}}{t} \hat{a}^4 &= -4\iu \omega_\mathrm{z}  f_{44} \hat{a}^4 -8\Gamma_\mathrm{p} f_{44} \hat{a}^4, \\
		\derivative{f_{43}}{t} {\hat{a}^\dagger} \hat{a}^3 &= -2\iu \omega_\mathrm{z}  f_{43}{\hat{a}^\dagger}  \hat{a}^3 -2\Gamma_\mathrm{p} f_{43}{\hat{a}^\dagger}  \hat{a}^3,\\
		\derivative{f_{42}}{t} \left.\hat{a}^\dagger\right.^2 \hat{a}^2 &= 0, 
	\end{align}
	and therefore
	\begin{align}
		(\hat{a}^4)[t] &= \hat{a}^4e^{-4\iu\omega_\mathrm{z}  t -8\Gamma_\mathrm{p} t},\\
		(\hat{a}^\dagger \hat{a}^3)[t] &= \hat{a}^\dagger \hat{a}^3 e^{-2\iu\omega_\mathrm{z}  t -2\Gamma_\mathrm{p}t}, \\
		(\left.\hat{a}^\dagger\right.^2 \hat{a}^2)[t] &= \left.\hat{a}^\dagger \right.^2 \hat{a}^2. 
	\end{align}
	By observation, we conclude that 
	\begin{align}\label{eqn:phase_a_adagger}
		(\left.\hat{a}^\dagger\right.^m \hat{a}^n)[t]=\left.\hat{a}^\dagger\right.^m \hat{a}^n e^{\iu(m-n)\omega_\mathrm{z} t -\frac{1}{2}(m-n)^2\Gamma_\mathrm{p} t}. 
	\end{align}
	\subsection{Diffusion Channel}\label{apd:diffusion_channel}
	The dual of the Lindblad equation for the diffusion damping channel is given by
	\begin{align}
		\derivative{\hat{A}}{t} = \iu \omega_\mathrm{z} [\hat{a}^\dagger \hat{a}, \hat{A}]-\frac{ \Gamma_\mathrm{d}}{2} ([\hat{a},[\hat{a}^\dagger,\hat{A}]]+[\hat{a}^\dagger,[\hat{a},\hat{A}]]). 
	\end{align}
	For the first order, we assume that 
	\begin{align}
		\hat{a}[t] = f_{11}(t) \hat{a}.
	\end{align}
	We then get
	\begin{align}
		\derivative{f_{11}}{t} \hat{a} = -\iu \omega_\mathrm{z}  f_{11} \hat{a} ,
	\end{align}
	and therefore
	\begin{align}
		\hat{a}[t] = \hat{a}e^{-\iu\omega_\mathrm{z}  t }.
	\end{align}
	For the second order, we assume that 
	\begin{align}
		(\hat{a}^2)[t] &= f_{22}(t) \hat{a}^2, \\
		(\hat{a}^\dagger \hat{a})[t] &= f_{21}(t) \hat{a}^\dagger \hat{a} + g_{21}(t) .
	\end{align}
	We then get
	\begin{align}
		\derivative{f_{22}}{t} \hat{a}^2 &= -2\iu \omega_\mathrm{z}  f_{22} \hat{a}^2, \\
		\derivative{f_{21}}{t} \hat{a}^\dagger \hat{a}+\derivative{g_{21}}{t} &= \Gamma_{\mathrm{d}} f_{21},
	\end{align}
	and therefore
	\begin{align}
		(\hat{a}^2)[t] &= \hat{a}^2e^{-2\iu\omega_\mathrm{z}  t },\\
		(\hat{a}^\dagger \hat{a})[t] &= \hat{a}^\dagger \hat{a} + \Gamma_\mathrm{d} t.
	\end{align}
	As for the third order,  
	\begin{align}
		(\hat{a}^3)[t] &= f_{33}(t) \hat{a}^3, \\
		(\hat{a}^\dagger \hat{a}^2)[t] &= f_{32}(t) \hat{a}^\dagger \hat{a}^2 + g_{32}(t) \hat{a}.
	\end{align}
	We then get
	\begin{align}
		\derivative{f_{33}}{t} \hat{a}^3 &= -3\iu \omega_\mathrm{z}  f_{33} \hat{a}^3, \\
		\derivative{f_{32}}{t} \hat{a}^\dagger \hat{a}^2 +\derivative{g_{32}}{t} \hat{a} &= -\iu \omega_\mathrm{z}  f_{32}\hat{a}^\dagger  \hat{a}^2 -\iu \omega_\mathrm{z}  g_{32}  \hat{a} +2\Gamma_\mathrm{d} f_{32} \hat{a},
	\end{align}
	and therefore
	\begin{align}
		(\hat{a}^3)[t] &= \hat{a}^3e^{-3\iu\omega_\mathrm{z}  t },\\
		(\hat{a}^\dagger \hat{a}^2)[t] &= \hat{a}^\dagger \hat{a}^2  e^{-\iu\omega_\mathrm{z}  t} + 2\Gamma_{\mathrm{d}}t \hat{a} e^{-\iu\omega_\mathrm{z}  t}. 
	\end{align}
	Finally, for the fourth order,  
	\begin{align}
		(\hat{a}^4)[t] &= f_{44}(t) \hat{a}^4, \\
		(\hat{a}^\dagger \hat{a}^3)[t] &= f_{43}(t) \hat{a}^\dagger \hat{a}^3 + g_{43}(t) \hat{a}^2,\\ 
		(\left.\hat{a}^\dagger\right.^2 \hat{a}^2)[t] &= f_{42}(t) \left.\hat{a}^\dagger\right.^2  \hat{a}^2 + g_{42}(t) \hat{a}^\dagger \hat{a} + h_{42}(t) .
	\end{align}
	We then get
	\begin{align}
		\derivative{f_{44}}{t} \hat{a}^4 &= -4\iu \omega_\mathrm{z}  f_{44} \hat{a}^4, \\
		\derivative{f_{43}}{t} {\hat{a}^\dagger} \hat{a}^3+\derivative{g_{43}}{t} \hat{a}^2 &= -2\iu \omega_\mathrm{z}  f_{43}{\hat{a}^\dagger} \hat{a}^3 -2\iu \omega_\mathrm{z}  g_{43}  \hat{a}^2+3\Gamma_{d} f_{43}\hat{a}^2,\\
		\derivative{f_{42}}{t} \left.\hat{a}^\dagger\right.^2 \hat{a}^2+\derivative{g_{42}}{t} {\hat{a}^\dagger} \hat{a} &+\derivative{h_{42}}{t} = 4 \Gamma_{d} f_{42}\hat{a}^\dagger \hat{a} +\Gamma_{\mathrm{d}} g_{42},
	\end{align}
	and therefore
	\begin{align}
		(\hat{a}^4)[t] &= \hat{a}^4e^{-4\iu\omega_\mathrm{z}  t }, \\
		(\hat{a}^\dagger \hat{a}^3)[t] &= \hat{a}^\dagger \hat{a}^3 e^{-2\iu\omega_\mathrm{z}  t} +3\Gamma_{\mathrm{d}} t \hat{a}^2 e^{-2\iu\omega_\mathrm{z}  t}, \\
		(\left.\hat{a}^\dagger\right.^2 \hat{a}^2)[t] &= \left.\hat{a}^\dagger\right.^2 \hat{a}^2+ 4\Gamma_{\mathrm{d}} t \hat{a}^\dagger \hat{a} + 2\Gamma_{\mathrm{d}}^2 t^2 .
	\end{align}
	By observation, we conclude that 
	\begin{align}\label{eqn:diffusion_a_adagger}
		(\left.\hat{a}^\dagger\right.^m \hat{a}^n)[t]=&\left(\sum_{k}\frac{m!n!}{k!(m-k)!(n-k)!}
		\Gamma_{\mathrm{d}}^kt^k\left.\hat{a}^\dagger\right.^{m-k} \hat{a}^{n-k}\right)\cdot e^{\iu(m-n)\omega_\mathrm{z}  t }. 
	\end{align}
	
    \section{Integration}\label{apd:integration}
    In this section, we explicitly compute integrals $I_1$ and $I_2$ and explain how we deal with different terms in $I_1$ and $I_2$. As is explained in~\cref{sec:Theory}, we compute $I_1$ and $I_2$ following \cref{eqn:I_1_original,eqn:I_2_original} by keeping $\hat{V}_{\rm k}$ and $\rho_{\rm k}$ to the leading order ($O(c^{0})$). We also use the interaction picture where the operator changes but the state does not (to the leading order) 
    \begin{align}
        I_1 & = \frac{1}{mc^2} \int_0^T \d t_1 \Tr\left(\hat{V}_{\rm k}[t_1]\rho_{\rm k,0}\right), \label{eqn:I_1_appendix}\\
        I_2 & = \frac{2}{m^2c^4} \int_0^T \d t_2 \int_0^{t_2} \d t_1 \Tr\left( \left(\hat{V}_{\rm k}[t_2-t_1]\hat{V}_{\rm k}\right)\![t_1] \rho_{\rm k,0}\right), \label{eqn:I_2_appendix}
    \end{align}
    where 
    \begin{align}
        \frac{\hat{V}_{\rm k}}{mc^2} = C_{\rm g} (\hat{a}+\hat{a}^\dagger) - C_{\rm r} + C_{\rm k}(\hat{a}^2 + \left.\hat{a}^\dagger\right.^2 - 2\hat{a}^\dagger \hat{a} -1). \label{eqn:V_k_appendix}
    \end{align}
    In order to compute the integration, we compute the time evolution of $\hat{V}_{\rm k}[t_1]$ and $\left(\hat{V}_{\rm k}[t_2-t_1] \hat{V}_{\rm k}\right)\![t_1]$ by making use of the time evolution of polynomials of $\hat{a}^\dagger$ and $\hat{a}$, which are discussed in~\cref{apd:operator_evolution}. We first put $\hat{V}_{\rm k}$ into anti-normal ordering (creation operators on the left and annihilation operators on the right). Observing that $\hat{V}_{\rm k}$ contains quadratic polynomials of $\hat{a}$ and $\hat{a}^\dagger$, we then obtain $\hat{V}_{\rm k}[t_1]$ and $\hat{V}_{\rm k}[t_2-t_1]$ by evolving polynomials of $\hat{a}$ and $\hat{a}^\dagger$ accordingly. We again put $\hat{V}_{\rm k}[t_2-t_1]\hat{V}_{\rm k}$ into anti-normal ordering. Observing that $\hat{V}_{\rm k}[t_2-t_1]\hat{V}_{\rm k}$ contains quartic polynomials of $\hat{a}$ and $\hat{a}^\dagger$, we finally obtain $\left(\hat{V}_{\rm k}[t_2-t_1]\hat{V}_{\rm k}\right)\![t_1]$ by evolving polynomials of $\hat{a}$ and $\hat{a}^\dagger$ accordingly. $\hat{V}_{\rm k}[t_1]$ and $\left(\hat{V}_{\rm k}[t_2-t_1] \hat{V}_{\rm k}\right)\![t_1]$ both contain oscillating terms with respect to at least one of $t_1$ or $t_2$ and non-oscillating terms. After integration, the former is small compared to the latter. 
    \subsection{Free Evolution}\label{apd:integration_free}
    For free evolution, the time evolution of relevant operators are 
    \begin{align}
        \hat{V}_{\rm k}[t_1] & =  (C_{\rm g}\hat{a} e^{-\iu \omega_{\rm z} t_1 }+ C_{\rm k} \hat{a}^2 e^{- \iu 2 \omega_{\rm z} t_1 }  + \textnormal{h.c.})  \nonumber \\
        & \ \ \ \ - \left(C_{\rm r} + C_{\rm k}(  2\hat{a}^\dagger \hat{a} + 1)\right), \label{eqn:V_k_free}\\
        \left(\hat{V}_{\rm k}[t_2-t_1] \hat{V}_{\rm k}\right)\![t_1] & = (C_{\rm g}\hat{a} e^{-\iu \omega_{\rm z} t_2 }+ C_{\rm k} \hat{a}^2 e^{- \iu 2 \omega_{\rm z} t_2 }  + \textnormal{h.c.}) (C_{\rm g}\hat{a} e^{-\iu \omega_{\rm z} t_1 }+ C_{\rm k} \hat{a}^2 e^{- \iu 2 \omega_{\rm z} t_1 }  + \textnormal{h.c.}) \nonumber \\ 
        & \ \ \ \ - (C_{\rm g}\hat{a} e^{-\iu \omega_{\rm z} t_2 }+ C_{\rm k} \hat{a}^2 e^{- \iu 2 \omega_{\rm z} t_2 }  + \textnormal{h.c.}) \left(C_{\rm r} +C_{\rm k} (2\hat{a}^\dagger \hat{a} + 1)\right)  \nonumber\\
        & \ \ \ \ - \left(C_{\rm r} +C_{\rm k} (2\hat{a}^\dagger \hat{a} + 1)\right) (C_{\rm g}\hat{a} e^{-\iu \omega_{\rm z} t_1 } + C_{\rm k} \hat{a}^2 e^{- \iu 2 \omega_{\rm z} t_1 }  + \textnormal{h.c.})  \nonumber\\
        & \ \ \ \ + \left(C_{\rm r} +C_{\rm k} (2\hat{a}^\dagger \hat{a} + 1)\right)^2, \label{eqn:V_k_V_k_free}
    \end{align}
    where we have used \cref{eqn:free_a_adagger} for the evolution of $\hat{a}$ and $\hat{a}^\dagger$. We can split $\hat{V}_{\rm k}[t_1]$ in \cref{eqn:V_k_free} into oscillating terms (first line) whose integration over $t_1$ is denoted by $I_{1,\rm o}$ and non-oscillating terms (second line) whose integration over $t_1$ is denoted by $I_{1,\rm l}$. After integration, the oscillating terms gain a factor $\omega_{\rm z}^{-1}$ while the non-oscillating terms gain a factor $T$. We work in the region where the interrogation time is much longer than the oscillation period, $\omega_{\rm z}^{-1}\ll T$. Therefore, 
    \begin{align}
        \frac{I_{1,\rm l}}{I_{1,\rm o}} \propto \omega_{\rm z} T,  
    \end{align}
    where $I_{1,\rm o}$ is negligible in our setting. 
    
    \edit{As an easy example, we demonstrate that the oscillating term is indeed negligible by explicitly calculating the magnitude of the oscillating terms $\Delta_{1,\coh}$ for the case without decoherence. For simplicity, we consider the case where $\alpha$ is real. Substituting into~\cref{eqn:I_1_appendix,eqn:V_k_free} 
    \begin{align}
        \rho_{{\rm k},0,\qtm} = \frac{1}{1+C_{\rm i}} (\cos\theta \ket{\alpha} + e^{\iu\phi}\sin\theta\ket{-\alpha}) (\cos\theta \bra{\alpha} + e^{-\iu\phi}\sin\theta\bra{-\alpha}),
    \end{align}
    and 
    \begin{align}
        \rho_{{\rm k},0,\cls} = \cos^2\theta \ketbra{\alpha}{\alpha} +\sin^2\theta \ketbra{-\alpha}{-\alpha}, 
    \end{align}
    we obtain 
    \begin{align}
        I_{1,\qtm} = \frac{2C_{\rm g}\alpha}{(1+C_{\rm i})\omega_{\rm z}} \left( \cos2\theta \sin\omega_{\rm z}T + C_{\rm i} \tan\phi  (\cos\omega_{\rm z} T -1)\right) -C_{\rm k} T -C_{\rm r }T -C_{\rm k} \alpha^2 \frac{\sin 2\omega_{\rm z} T}{\omega_{\rm z} } - 2 C_{\rm k}  \frac{1-C_{\rm i}}{1+C_{\rm i}}\alpha^2 T, 
    \end{align}
    and 
    \begin{align}
        I_{1,\cls} = \frac{2C_{\rm g}\alpha}{\omega_{\rm z}} \cos2\theta \sin\omega_{\rm z} T - C_{\rm r} T - C_{\rm k} T + C_{\rm k} \alpha^2 \frac{\sin 2 \omega_{\rm z} T}{\omega_{\rm z}} - 2 C_{\rm k} \alpha^2  T.
    \end{align}
    When including the oscillating terms, $\Delta_{1,\coh}$ is given by
    \begin{align}
        \label{eqn:delta_1_free_full}
        \Delta_{1,\coh} = I_{1,\qtm} - I_{1,\cls} = \frac{C_{\rm i}}{1+C_{\rm i}} \left(\frac{2C_{\rm g}\alpha}{\omega_{\rm z}} \left( \tan\phi(\cos\omega_{\rm z} T-1)  - \cos2\theta \sin\omega_{\rm z} T\right)  + 4C_{\rm k} \alpha^2 T\right).
    \end{align}
    One should note that because $C_i\propto \cos\phi$, $\delta_{1,\coh}$ is finite even if $\phi=\frac{\pi}{2}$. 
    When excluding the oscillating terms, $\Delta_{1,\coh}$ is given by
    \begin{align}
        \label{eqn:delta_1_free_partial}
        \Delta_{1,\coh} = I_{1,\qtm} - I_{1,\cls} \approx 4C_{\rm k} \frac{C_{\rm i}}{1+C_{\rm i}}  \alpha^2 T.
    \end{align}
    Comparing~\cref{eqn:delta_1_free_full} to~\cref{eqn:delta_1_free_partial}, the oscillating terms we omitted in the main text are
    \begin{align}
        \label{eqn:omitted_term}
        \delta_{1,\coh}=\frac{2C_{\rm i}C_{\rm g}\alpha }{(1+C_{\rm i})\omega_{\rm z}} \left(\tan\phi(\cos\omega_{\rm z} T-1)- \cos2\theta \sin\omega_{\rm z} T \right).
    \end{align}
    We see that $\delta_{1,\coh}$ is upper bounded by 
    \begin{align}
        |\delta_{1,\coh}|\leq \frac{2C_{\rm i}C_{\rm g}\alpha }{(1+C_{\rm i})\omega_{\rm z}} \left(2|\tan\phi|+|\cos2\theta| \right).
    \end{align}
    By setting $\lambda_{\mathrm{Mg}}=468$~nm, $U_{\max,\mathrm{Mg}}=\edit{300} E_{\mathrm{r},\mathrm{Mg}}$, $\phi=\frac{\pi}{2}$, $\theta=\frac{\pi}{8}$ and $T=1$~s, ($d= \sqrt{2}z_\mathrm{s} \alpha$, i.e. $d=10$~nm corresponds to $\alpha=0.395$ for the $^{24}{\rm Mg}$ clock here), we obtain $\delta_{1,\coh} \lesssim 10^{-33}$~s, which is indeed negligible. 

    As can be seen in~\cref{eqn:omitted_term}, the oscillating terms we omitted is proportional to $C_{\rm g}$, therefore, they are due to the $C_{\rm g}(\hat{a}e^{-\iu\omega_{\rm z}t} + \textnormal{h.c.})$ term in~\cref{eqn:V_k_free} and thus the $mg\hat{z}$ term in~\cref{eqn:Hamiltonian_V_k}. Because the atoms confined in the optical lattice oscillates, the terms corresponding to the gravity also oscillates and fails to accumulate with respect to time. This fact combining with the small coefficient, the time dilation due to the gravity is not detectable.}
    
    Similarly, we split $ \left(\hat{V}_{\rm k}[t_2-t_1] \hat{V}_{\rm k}\right)\![t_1]$ in \cref{eqn:V_k_V_k_free} into oscillating terms with respect to at least one of $t_1$ and $t_2$ (first three lines) whose integration is denoted by $I_{2,\rm o}$ and non-oscillating terms (fourth line) whose integration is denoted by $I_{2,\rm q}$. After integration, the oscillating terms with respect to at least one of $t_1$ or $t_2$ gain either a factor $\omega_{\rm z}^{-2}$ or a factor $\omega_{\rm z}^{-1} T$, while the non-oscillating terms gain a factor $T^2$. Therefore, we obtain using $\omega_{\rm z}^{-1}s \ll T$
    \begin{align}
        \frac{I_{2,\rm q}}{I_{2,\rm o}} \propto \omega_{\rm z} T,  
    \end{align}
    where  $I_{2,\rm o}$ is again negligible. The integration thus results in \cref{eqn:delta_1_free}, \cref{eqn:delta_2_qtm_free,eqn:delta_2_cls_free}, as is discussed in~\cref{sec:FreeEvolution}. 
    \subsection{Amplitude Damping Channel}\label{apd:integration_amplitude}
    For amplitude damping channel, the time evolution of relevant operators are 
    \begin{align}
        \hat{V}_{\rm k}[t_1] & = (C_{\rm g} \hat{a}e^{-(\iu\omega_{\rm z} + \frac{1}{2}\Gamma_{\rm a} )t_1 }  + C_{\rm k} \hat{a}^2 e^{-(\iu 2\omega_{\rm z} + \Gamma_{\rm a}) t_1 } + \textnormal{h.c.})\nonumber \\
        & \ \ \ \ - (C_{\rm r} + C_{\rm k}(2 \hat{a}^\dagger \hat{a} e^{ - \Gamma_{\rm a} t_1 } + 1) ), \label{eqn:V_k_amplitude} \\
        \left(\hat{V}_{\rm k}[t_2-t_1]\hat{V}_{\rm k}\right)\![t_1] & =  ((C_{\rm g} \hat{a}e^{-(\iu\omega_{\rm z}+\frac{1}{2}\Gamma_{\rm a} ) t_2 } + C_{\rm k} \hat{a}^2 e^{-(\iu 2\omega_{\rm z} + \Gamma_{\rm a}) t_2 } )(C_{\rm g} \hat{a}e^{-(\iu\omega_{\rm z} + \frac{1}{2}\Gamma_{\rm a}  )t_1 } + C_{\rm k} \hat{a}^2 e^{-(\iu 2\omega_{\rm z} + \Gamma_{\rm a} )t_1 } ) + \textnormal{h.c.} )\nonumber \\
        & \ \ \ \ + C_{\rm g}^2 e^{-(\iu \omega_z + \frac{1}{2} \Gamma_{\rm a }  )(t_2-t_1)} (\hat{a}^\dagger \hat{a} e^{-\Gamma_{\rm a} t_1} + 1) + C_{\rm g}^2 e^{(\iu \omega_z-\frac{1}{2} \Gamma_{\rm a } )t_2 - (\iu \omega_z+\frac{1}{2} \Gamma_{\rm a } )t_1 } \hat{a}^\dagger \hat{a}  \nonumber \\
        & \ \ \ \ + C_{\rm g} C_{\rm k} e^{-(\iu \omega_z+ \frac{1}{2} \Gamma_{\rm a} ) t_2 + \iu 2 \omega_z t_1 } (\left.\hat{a}^\dagger\right.^2 \hat{a} e^{-\Gamma_{\rm a}  t_1}  + 2\hat{a}^\dagger  ) + C_{\rm g} C_{\rm k} e^{(\iu \omega_z - \frac{1}{2} \Gamma_{\rm a} ) t_2 - (\iu 2\omega_{\rm z}+\Gamma_{\rm a}) t_1 }\hat{a}^\dagger \hat{a}^2   \nonumber \\
        & \ \ \ \ + C_{\rm k} C_{\rm g} e^{-(2\iu \omega_{\rm z} +\Gamma_{\rm a}) t_2 + (\iu \omega_{\rm z} +\frac{1}{2} \Gamma_{\rm a} ) t_1 } ( \hat{a}^\dagger \hat{a}^2 e^{-\Gamma_{\rm a} t_1} + 2\hat{a}) +  C_{\rm k} C_{\rm g} e^{(2\iu \omega_{\rm z} -\Gamma_{\rm a}) t_2 - (\iu \omega_{\rm z} + \frac{1}{2} \Gamma_{\rm a} ) t_1 }  \left.\hat{a}^\dagger\right.^2 \hat{a} \nonumber \\
        & \ \ \ \ + C_{\rm k}^2 e^{-(\iu 2 \omega_{\rm z} +\Gamma_{\rm z} )(t_2-t_1)}(\left.\hat{a}^\dagger\right.^2 \hat{a}^2 e^{-2\Gamma_{\rm a} t_1 }+ 4 \hat{a}^\dagger \hat{a} e^{-\Gamma_{\rm a} t_1 } + 2) + C_{\rm k}^2 e^{(\iu 2 \omega_{\rm z} -\Gamma_{\rm z} )t_2-(\iu 2 \omega_{\rm z} + \Gamma_{\rm z} )t_1} \left.\hat{a}^\dagger\right.^2 \hat{a}^2 \nonumber \\ 
        & \ \ \ \ - C_{\rm g} e^{-(\iu\omega_{\rm z} +\frac{1}{2}\Gamma_{\rm a})t_2} (C_{\rm r} \hat{a} + C_{\rm k} (2 \hat{a}^\dagger \hat{a}^2 e^{-\Gamma_{\rm a} t_1} + 3\hat{a}) )\nonumber \\
        & \ \ \ \ - C_{\rm g} e^{(\iu\omega_{\rm z} -\frac{1}{2} \Gamma_{\rm a}) t_2} (C_{\rm r} \hat{a}^\dagger + C_{\rm k}(2 \left.\hat{a}^\dagger\right.^2 \hat{a} e^{-\Gamma_{\rm a} t_1} + \hat{a}^\dagger) ) \nonumber \\
        & \ \ \ \ - C_{\rm g}  e^{-(\iu\omega_{\rm z} +\frac{1}{2}\Gamma_{\rm a})t_1} (C_{\rm r} \hat{a} + C_{\rm k} (2 \hat{a}^\dagger \hat{a}^2 e^{-\Gamma_{\rm a} t_2} + \hat{a}) ) \nonumber \\
        & \ \ \ \ - C_{\rm g} e^{(\iu\omega_{\rm z} - \frac{1}{2}\Gamma_{\rm a}) t_1}(C_{\rm r} \hat{a}^\dagger + C_{\rm k} (2 \left.\hat{a}^\dagger\right.^2 \hat{a} e^{-\Gamma_{\rm a} t_2} +\hat{a}^\dagger(2 e^{-\Gamma_{\rm a} (t_2-t_1)}+ 1) )) \nonumber\\
        & \ \ \ \ - C_{\rm k}e^{-(\iu2\omega_{\rm z}+\Gamma_{\rm a}) t_2}(C_{\rm r} \hat{a}^2 +C_{\rm k} (2\hat{a}^\dagger \hat{a}^3 e^{-\Gamma_{\rm a} t_1} + 5\hat{a}^2 )) \nonumber \\
        & \ \ \ \ - C_{\rm k}e^{(\iu2\omega_{\rm z}-\Gamma_{\rm a}) t_2}(C_{\rm r} \left.\hat{a}^\dagger\right.^2 +C_{\rm k} (2\left.\hat{a}^\dagger\right.^3 \hat{a} e^{-\Gamma_{\rm a} t_1} + \left.\hat{a}^\dagger\right.^2 ))\nonumber \\ 
        & \ \ \ \ - C_{\rm k} e^{-(\iu2\omega_{\rm z} +\Gamma_{\rm a})t_1}(C_{\rm r} \hat{a}^2 + C_{\rm k}(2 \hat{a}^\dagger \hat{a}^3 e^{-\Gamma_{\rm a} t_2} + \hat{a}^2))\nonumber \\
        & \ \ \ \ - C_{\rm k} e^{(\iu2\omega_{\rm z} - \Gamma_{\rm a})t_1}(C_{\rm r} \left.\hat{a}^\dagger\right.^2 + C_{\rm k}(2 \left.\hat{a}^\dagger\right.^3 \hat{a} e^{-\Gamma_{\rm a} t_2} + \left.\hat{a}^\dagger\right.^2(4 e^{-\Gamma_{\rm a} (t_2-t_1)} + 1))) \nonumber \\
        & \ \ \ \ + (C_{\rm r}+C_{\rm k})(C_{\rm r}+C_{\rm k}(2 \hat{a}^\dagger \hat{a} e^{-\Gamma_{\rm a } t_1} + 2 \hat{a}^\dagger \hat{a} e^{-\Gamma_{\rm a } t_2}+1)) + 4 C_{\rm k}^2 e^{-\Gamma_{\rm a} t_2 } (\left.\hat{a}^\dagger\right.^2 \hat{a}^2 e^{-\Gamma_{\rm a} t_1} + \hat{a}^\dagger \hat{a}), \label{eqn:V_k_V_k_amplitude}
    \end{align}
    where we have used \cref{eqn:amplitude_a_adagger} for the evolution of polynomials of $\hat{a}$ and $\hat{a}^\dagger$. We can split $\hat{V}_{\rm k}[t_2-t_1]$ in \cref{eqn:V_k_amplitude} into oscillating terms (first line) whose integration over $t_1$ is denoted by $I_{1, \rm o}$ and non-oscillating terms (second line) whose integration is denoted by $I_{1,\rm l}$. After integration, the oscillating terms gain a factor $\omega_{\rm z}^{-1}$ while the non-oscillating terms gain either a factor $T$ or a factor $\Gamma_{\rm a}^{-1}$. In real experiments, one does not work in the regime where the quantum state has been destroyed, and thus set $\omega_{\rm z}^{-1} \ll \Gamma_{\rm a}^{-1} \simeq T $. Therefore, we obtain 
    \begin{align}
        \frac{I_{1,\rm l}}{I_{1,\rm o}} \propto \omega_{\rm z}T, 
    \end{align}
    therefore $I_{1,\rm o}$ is negligible. Similarly, we split $\left(\hat{V}_{\rm k}[t_2-t_1]\hat{V}_{\rm k}\right)\![t_1]$ in \cref{eqn:V_k_V_k_amplitude} into oscillating terms with respect to at least one of $t_1$ or $t_2$ (all but the last line) whose integration is denoted by $I_{2,\rm o}$ and non-oscillating terms (last line) whose integration is denoted by $I_{2, \rm q}$. After integration, the oscillating terms with respect to at least one of $t_1$ or $t_2$ gain either a factor $\omega_{\rm z}^{-2}$, a factor $\omega_{\rm z}^{-1} \Gamma_{\rm a}^{-1}$ or $\omega_{\rm z}^{-1} T$, while the non-oscillating terms gain a factor $T^2$, $\Gamma_{\rm a}^{-1} T$ or $\Gamma_{\rm a}$. Therefore, under our condition that $\omega_{\rm z}^{-1} \ll \Gamma_{\rm a}^{-1} \simeq T$
    \begin{align}
        \frac{I_{2,\rm q}}{I_{2,\rm o}} \propto \omega_{\rm z} T, 
    \end{align}
    therefore $I_{2,\rm o}$ is negligible. The integration results in \cref{eqn:delta_1_amplitude}, \cref{eqn:delta_2_qtm_amplitude,eqn:delta_2_cls_amplitude}, as is shown in~\cref{sec:NoiseTolerance}. 
    \subsection{Phase Damping Channel}\label{apd:integration_phase}
    For phase damping channel, the time evolution of relevant operators are
    \begin{align}
        \hat{V}_{\rm k} & = (C_{\rm g} \hat{a} e^{-(\iu \omega_{\rm z} + \frac{1}{2} \Gamma_{\rm p} )t_1} + C_{\rm k} \hat{a}^2 e^{-(\iu2\omega_{\rm z} + 2\Gamma_{\rm p} )t_1} + \textnormal{h.c.} ) \nonumber \\
        & \ \ \ \ -( C_{\rm r} + C_{\rm k}(2\hat{a}^\dagger \hat{a} + 1)), \label{eqn:V_k_phase}\\
        \left(\hat{V}_{\rm k}[t_2-t_1]\hat{V}_{\rm k} \right)\![t_1] & = (C_{\rm g}^2 \hat{a}^2 e^{-(\iu\omega_{\rm z}+\frac{1}{2}\Gamma_{\rm p})t_2 - (\iu\omega_{\rm z} + \frac{3}{2} \Gamma_{\rm p}) t_1} + C_{\rm k}^2 \hat{a}^4 e^{-(\iu 2\omega_{\rm z} + 2\Gamma_{\rm p}) t_2 - (\iu 2\omega_{\rm z} + 6\Gamma_{\rm p}) t_1} + \textnormal{h.c.}) \nonumber \\
        & \ \ \ \ +( C_{\rm g} C_{\rm k} \hat{a}^3 e^{-(\iu\omega_{\rm z} + \frac{1}{2}\Gamma_{\rm p})t_2 - (\iu2\omega_{\rm z} + 4\Gamma_{\rm p})t_1} + C_{\rm g} C_{\rm k} \hat{a}^3 e^{-(\iu2\omega_{\rm z}+2\Gamma_{\rm p})t_2 -(\iu\omega_{\rm z} + \frac{5}{2}\Gamma_{\rm p}) t_1} + \textnormal{h.c.}) \nonumber \\
        & \ \ \ \ + C_{\rm g}^2 (\hat{a}^\dagger \hat{a} + 1)e^{-(\iu\omega_{\rm z} +\frac{1}{2}\Gamma_{\rm p})(t_2 -t_1)} + C_{\rm g}^2 \hat{a}^\dagger \hat{a} e^{(\iu\omega_{\rm z} -\frac{1}{2}\Gamma_{\rm p})(t_2 -t_1)}\nonumber \\
        & \ \ \ \ + C_{\rm g} C_{\rm k}  e^{-(\iu\omega_{\rm z} + \frac{1}{2}\Gamma_{\rm p})t_2 + \iu 2 \omega_{\rm z} t_1}(\left.\hat{a}^\dagger\right.^2 \hat{a} + 2\hat{a}^\dagger) + C_{\rm g} C_{\rm k}  e^{(\iu\omega_{\rm z} - \frac{1}{2}\Gamma_{\rm p})t_2 - \iu 2 \omega_{\rm z} t_1} \hat{a}^\dagger \hat{a}^2 \nonumber \\
        & \ \ \ \ + C_{\rm k} C_{\rm g} e^{-(\iu2\omega_{\rm z} +2 \Gamma_{\rm p})t_2 + (\iu\omega_{\rm z} +\frac{3}{2}\Gamma_{\rm p})t_1 } (\hat{a}^\dagger \hat{a}^2 + 2\hat{a})+ C_{\rm k} C_{\rm g} e^{(2\iu\omega_{z} -2\Gamma_{\rm p} ) t_2 - (\iu\omega_{\rm z} - \frac{3}{2} \Gamma_{\rm p}) t_1} \left.\hat{a}^\dagger\right.^2 \hat{a} \nonumber \\
        & \ \ \ \ +   C_{\rm k}^2 e^{-(\iu 2\omega_{\rm z} + 2\Gamma_{\rm p})(t_2-t_1)} (\left.\hat{a}^\dagger\right.^2 \hat{a}^2 + 4 \hat{a}^\dagger a + 2) + C_{\rm k}^2 e^{(\iu 2\omega_{\rm z} - 2\Gamma_{\rm p})(t_2-t_1)} \left.\hat{a}^\dagger\right.^2 \hat{a}^2 \nonumber \\
        & \ \ \ \ - (C_{\rm r} + C_{\rm k})(C_{\rm g} \hat{a} e^{-(\iu \omega_{\rm z} + \frac{1}{2} \Gamma_{\rm p} )t_2} + C_{\rm k} \hat{a}^2 e^{-(\iu2\omega_{\rm z} + 2\Gamma_{\rm p} )t_2} + \textnormal{h.c.} ) \nonumber \\
        & \ \ \ \ - (C_{\rm r} + C_{\rm k})(C_{\rm g} \hat{a} e^{-(\iu \omega_{\rm z} + \frac{1}{2} \Gamma_{\rm p} )t_1} + C_{\rm k} \hat{a}^2 e^{-(\iu2\omega_{\rm z} + 2\Gamma_{\rm p} )t_1} + \textnormal{h.c.} ) \nonumber \\
        & \ \ \ \ - C_{\rm g} C_{\rm k} e^{-(\iu\omega_{\rm z} + \frac{1}{2}\Gamma_{\rm p})t_2} (\hat{a}^\dagger \hat{a}^2 +\hat{a} ) - C_{\rm g} C_{\rm k} e^{(\iu\omega_{\rm z}-\frac{1}{2}\Gamma_{\rm p})t_2} \left.\hat{a}^\dagger\right.^2 \hat{a} \nonumber \\
        & \ \ \ \ - C_{\rm g} C_{\rm k} e^{-(\iu\omega_{\rm z}+ \frac{1}{2}\Gamma_{\rm p})t_1} \hat{a}^\dagger \hat{a}^2  - C_{\rm g} C_{\rm k}  e^{(\iu\omega_{\rm z}-\frac{1}{2}\Gamma_{\rm p} t_1)} (\left.\hat{a}^\dagger\right.^2 \hat{a} + \hat{a}^\dagger ) \nonumber \\
        & \ \ \ \ - C_{\rm k}^2 e^{-(\iu2\omega_{\rm z}+2\Gamma_{\rm p} )t_2} (\hat{a}^\dagger \hat{a}^3 + 2 \hat{a}^2 ) - C_{\rm k}^2 e^{(\iu2\omega_{\rm z} - 2\Gamma_{\rm p} )t_2} \left.\hat{a}^\dagger\right.^3 \hat{a} \nonumber \\
        & \ \ \ \ - C_{\rm k}^2 e^{-(\iu2\omega_{\rm z} +2\Gamma_{\rm p} )t_1 } \hat{a}^\dagger \hat{a}^3 -  C_{\rm k}^2 e^{-(\iu2\omega_{\rm z} +2\Gamma_{\rm p} )t_1 } (\left. \hat{a}^\dagger\right.^3 \hat{a} +2 \left.\hat{a}^\dagger\right.^2) \nonumber \\
        & \ \ \ \ + (C_{\rm r} + C_{\rm k} (2\hat{a}^\dagger\hat{a}+1) )^2, \label{eqn:V_k_V_k_phase}
    \end{align}
    where we have used \cref{eqn:phase_a_adagger} for the evolution of polynomials of $\hat{a}$ and $\hat{a}^\dagger$. We split $\hat{V}_{\rm k}(t_1)$ in \cref{eqn:V_k_phase} into oscillating terms (first line) whose integration over $t_1$ is denoted by $I_{1,\rm o}$ and non-oscillating terms (second line) whose integration over $t_1$ is denoted by $I_{1,\rm l}$. After integration, the oscillating terms gain a factor of $\omega_{\rm z}^{-1}$ while the non-oscillating terms gain either a factor $T$ or a factor $\Gamma_{\rm p}^{-1}$. Because we work in the region where $\omega_{\rm z}^{-1}\ll \Gamma_{\rm p}^{-1} \simeq T$, we obtain
    \begin{align}
        \frac{I_{1,\rm l}}{I_{1,\rm o}} \propto \omega_{\rm z} T. 
    \end{align}
    Therefore $I_{1,\rm o}$ is negligible. Similarly, we split $\left(\hat{V}_{\rm k}[t_2-t_1]\hat{V}_{\rm k}\right)[t_1]$ in \cref{eqn:V_k_V_k_phase} into oscillating terms with respect to at least one of $t_1$ and $t_2$ (all but the last line) whose integration is $I_{2, \rm o}$ and non-oscillating terms (last line) whose integration is $I_{2,\rm q}$. After integration, the oscillating terms gain either a factor $\omega_{\rm z}^{-2}$, a factor $\omega_{\rm z}^{-1}\Gamma_{\rm p}^{-1}$ or a factor $\omega_{\rm z}^{-1} T$, while the non-oscillating terms gain a factor $T^2$, $\Gamma_{\rm p}^{-1} T$ or $\Gamma_{\rm p}^{-2}$. Using $\omega_{\rm z}^{-1} \ll \Gamma_{\rm p}^{-1} \simeq T$, we obtain
    \begin{align}
        \frac{I_{2,\rm q}}{I_{2,\rm o}} \propto \omega_{\rm z} T. 
    \end{align}
    As a result, we neglect $I_{2,\rm o}$. Integrating results in the same expressions as \cref{eqn:delta_1_free}, \cref{eqn:delta_2_qtm_free,eqn:delta_2_cls_free}. 
    \subsection{Diffusion Channel}\label{apd:integration_diffusion}
    For diffusion channel, the time evolution of relevant operators are 
    \begin{align}
        \hat{V}_{\rm k} & = (C_{\rm g} \hat{a}e^{-\iu\omega_{\rm z} t_1} + C_{\rm k} \hat{a}^2 e^{-\iu 2\omega_{\rm z} t_1} +\textnormal{h.c.} ) \nonumber\\
        & \ \ \ \ - (C_{\rm r} + C_{\rm k}(2\hat{a}^\dagger \hat{a} + 2\Gamma_{\rm d} t_1 + 1) ), \label{eqn:V_k_diffusion} \\
        \left(\hat{V}_{\rm k}[t_2-t_1] \hat{V}_{\rm k}\right)[t_1] & = ((C_{\rm g} \hat{a}e^{-\iu\omega_{\rm z}t_2 } + C_{\rm k} \hat{a}^2 e^{-\iu 2\omega_{\rm z} t_2 })(C_{\rm g} \hat{a}e^{-\iu\omega_{\rm z}t_1 } + C_{\rm k} \hat{a}^2 e^{-\iu 2\omega_{\rm z} t_1 }) +\textnormal{h.c.} ) \nonumber \\
        & \ \ \ \ + C_{\rm g}^2 e^{-\iu\omega_{\rm z} (t_2-t_1)} (\hat{a}^\dagger \hat{a} +\Gamma_{\rm d} t_1 + 1)+ C_{\rm g}^2  e^{\iu\omega_{\rm z} (t_2-t_1)}(\hat{a}^\dagger \hat{a} +\Gamma_{\rm d} t_1 ) \nonumber\\
        & \ \ \ \ + C_{\rm g} C_{\rm k} e^{-\iu\omega_{\rm z}t_2 + \iu2\omega_{\rm z} t_1 }(\left.\hat{a}^\dagger\right.^2 \hat{a} + 2(\Gamma_{\rm d} t_1 +1)  \hat{a}^\dagger) + C_{\rm g} C_{\rm k} e^{ \iu2\omega_{\rm z}t_2 - \iu\omega_{\rm z} t_1 }(\left.\hat{a}^\dagger\right.^2 \hat{a} + 2\Gamma_{\rm d} t_1 \hat{a}^\dagger) \nonumber \\
        & \ \ \ \ + C_{\rm g} C_{\rm k} e^{-\iu 2\omega_{\rm z} t_2+ \iu \omega_{\rm z}t_1)}(\hat{a}^\dagger \hat{a}^2 + 2(\Gamma_{\rm d} t_1 +1)\hat{a}) + C_{\rm g} C_{\rm k}  e^{\iu \omega_{\rm z} t_2 - \iu 2 \omega_{\rm z}t_1)}(\hat{a}^\dagger \hat{a}^2 + 2\Gamma_{\rm d}t_1 \hat{a} )\nonumber \\
        & \ \ \ \ + C_{\rm k}^2 e^{-\iu2\omega_{\rm z} (t_2-t_1)} (\left.\hat{a}^\dagger\right.^2 \hat{a}^2 + 4(\Gamma_{\rm d}t_1 +1)\hat{a}^\dagger \hat{a} + 2(\Gamma_{\rm d}t_1+1)^2) \nonumber \\
        & \ \ \ \ + C_{\rm k}^2 e^{\iu2\omega_{\rm z} (t_2-t_1)}(\left.\hat{a}^\dagger\right.^2 \hat{a}^2 + 4 \Gamma_{\rm d} t_1 \hat{a}^\dagger \hat{a} + 2\Gamma_{\rm d}^2 t_1^2) \nonumber \\
        & \ \ \ \ - (C_{\rm r}+C_{\rm k})(C_{\rm g} e^{-\iu\omega_{\rm z} t_2} \hat{a} + C_{\rm k}e^{-\iu 2\omega_{\rm z} t_2} \hat{a}^2 + \textnormal{h.c.}) \nonumber \\
        & \ \ \ \ - (C_{\rm r}+C_{\rm k}(2\Gamma_{\rm d} (t_2-t_1)+1))(C_{\rm g} e^{-\iu\omega_{\rm z} t_1} \hat{a} + C_{\rm k}e^{-\iu 2\omega_{\rm z} t_1} \hat{a}^2 + \textnormal{h.c.}) \nonumber \\ 
        & \ \ \ \ - C_{\rm g} C_{\rm k} e^{-\iu\omega_{\rm z} t_2} (2\hat{a}^\dagger \hat{a}^2 +2(2\Gamma_{\rm d} t_1 + 1)\hat{a}) - C_{\rm g} C_{\rm k} e^{-\iu\omega_{\rm z} t_1}( 2\hat{a}^\dagger \hat{a}^2 +4 \Gamma_{\rm d} t_1\hat{a}) \nonumber \\
        & \ \ \ \ - C_{\rm g} C_{\rm k} e^{\iu\omega_{\rm } t_2 } (2\left.\hat{a}^\dagger\right.^2\hat{a} + 4\Gamma_{\rm d}t_1 \hat{a}^\dagger ) - C_{\rm g} C_{\rm k}e^{\iu\omega_{\rm z} t_1} (2\left.\hat{a}^\dagger\right.^2 \hat{a} +2(2\Gamma_{\rm d} t_1 +1)\hat{a}^\dagger) \nonumber \\
        & \ \ \ \ - C_{\rm k}^2 e^{-\iu 2\omega_{\rm z} t_2}(2\hat{a}^\dagger \hat{a}^3 + 2(3\Gamma_{\rm d} t_1 + 2) \hat{a}^2 ) - C_{\rm k}^2 e^{-\iu 2\omega_{\rm z} t_1}(2\hat{a}^\dagger \hat{a}^3 + 6\Gamma_{\rm d} t_1  \hat{a}^2 ) \nonumber \\
        & \ \ \ \ - C_{\rm k}^2 e^{\iu 2\omega_{\rm z} t_2}(2\left.\hat{a}^\dagger\right.^3 \hat{a} + 6\Gamma_{\rm d} t_1  \left.\hat{a}^\dagger\right.^2 )  - C_{\rm k}^2 e^{\iu 2\omega_{\rm z} t_1}(2\left.\hat{a}^\dagger\right.^3 \hat{a} + 2(3\Gamma_{\rm d} t_1  + 2)\left.\hat{a}^\dagger\right.^2 )\nonumber\\
        & \ \ \ \ + (C_{\rm r} + C_{\rm k}(2\Gamma_{\rm d} (t_2-t_1) + 1 ) )(C_{\rm k} + C_{\rm r}( 2\hat{a}^\dagger \hat{a} +2\Gamma_{\rm d} t_1+1) ) \nonumber \\
        & \ \ \ \ + C_{\rm k}(C_{\rm r}+C_{\rm k}) (2\hat{a}^\dagger \hat{a} + 2\Gamma_{\rm d} t_1) + C_{\rm k}^2 (4\left.\hat{a}^\dagger\right.^2 \hat{a}^2 + 4(4\Gamma_{\rm d} t_1+1) \hat{a}^\dagger \hat{a} + 4 \Gamma_{\rm d} t_1(2\Gamma_{\rm d} t_1 +1) ), \label{eqn:V_k_V_k_diffusion}
    \end{align}
    where we have used \cref{eqn:diffusion_a_adagger} for the evolution of polynomials of $\hat{a}$ and $\hat{a}^\dagger$. We split $\hat{V}_{\rm k}(t_1)$ in \cref{eqn:V_k_diffusion} into oscillating terms (first line) whose integration over $t_1$ is denoted by $I_{1,\rm o}$ and non-oscillating terms (second line) whose integration over $t_1$ is denoted by $I_{1,\rm l}$. Recall that we work in the region where $\omega_{\rm z}^{-1} \ll \Gamma_{\rm d}^{-1} \simeq T$, which means ${\rm poly}(\Gamma_{\rm d} T) \simeq 1$. After integration, the oscillating terms gain a factor of ${\rm poly}(\Gamma_{\rm d}T) \cdot \omega_{\rm z}^{-1}$ while the non-oscillating terms gain a factor ${\rm poly}(\Gamma_{\rm d}T) \cdot T$. We thus obtain
    \begin{align}
        \frac{I_{1,\rm l}}{I_{1,\rm o}} \propto \omega_{\rm z} T. 
    \end{align}
    That means $I_{1,\rm o}$ is negligible. Similarly, we split $\left(\hat{V}_{\rm k}[t_2-t_1]\hat{V}_{\rm k}\right)[t_1]$ in \cref{eqn:V_k_V_k_diffusion} into oscillating terms with respect to at least one of $t_1$ and $t_2$ (all but the last two lines) whose integration is $I_{2, \rm o}$ and non-oscillating terms (last two lines) whose integration is $I_{2,\rm q}$. After integration, the oscillating terms gain either a factor ${\rm poly}(\Gamma_{\rm d}T) \cdot \omega_{\rm z}^{-2}$, a factor ${\rm poly}(\Gamma_{\rm d}T)\cdot \omega_{\rm z}^{-1}\Gamma_{\rm p}^{-1}$ or a factor $ {\rm poly}(\Gamma_{\rm d}T)\cdot \omega_{\rm z}^{-1} T$, while the non-oscillating terms gain a factor ${\rm poly}(\Gamma_{\rm d}T)\cdot T^2$. Using $\omega_{\rm z}^{-1} \ll \Gamma_{\rm d}^{-1} \simeq T$ and ${\rm poly}(\Gamma_{\rm d} T) \simeq 1$, we obtain
    \begin{align}
        \frac{I_{2,\rm q}}{I_{2,\rm o}} \propto \omega_{\rm z} T. 
    \end{align}
    We thus omit $I_{2,\rm o}$ in further calculations. It is direct but cumbersome to compute the integration explicitly, resulting in \cref{eqn:delta_1_diffusion}, \cref{eqn:delta_2_qtm_diffusion,eqn:delta_2_cls_diffusion}. 
\end{document}